\documentclass[pre,twocolumn,nobalancelastpage,amsmath,amssymb,showpacs,
nofootinbib, superscriptaddress]{revtex4-1}

\usepackage{graphics}      
\usepackage{graphicx}      
\usepackage{longtable}     
\usepackage{url}           
\usepackage{bm}            
\usepackage{xcolor}
\usepackage{amsmath}
\usepackage{dcolumn}
\usepackage{mathtools}
\usepackage{hyperref}

\begin{document}
\title{Shear-Driven Flow of Athermal, Frictionless, Spherocylinder Suspensions in Two Dimensions: Particle Rotations and Orientational Ordering}
\author{Theodore A. Marschall}
\affiliation{Department of Physics and Astronomy, University of Rochester, Rochester, NY 14627}
\author{Daniel Van Hoesen}
\affiliation{Department of Physics, Washington University,  St. Louis, MO 63130}
\author{S. Teitel}
\affiliation{Department of Physics and Astronomy, University of Rochester, Rochester, NY 14627}
\date{\today}

\begin{abstract}
We use numerical simulations to study the flow of a bidisperse mixture of athermal, frictionless, soft-core two dimensional spherocylinders driven by a uniform steady-state simple shear  applied at a fixed volume and a fixed finite strain rate $\dot\gamma$.  Energy dissipation is via a viscous drag with respect to a uniformly sheared host fluid, giving a simple model for flow in a non-Brownian suspension with Newtonian rheology.  Considering a range of packing fractions $\phi$ and particle asphericities $\alpha$ at small $\dot\gamma$, we study the angular rotation $\dot\theta_i$ and the nematic orientational ordering $\mathbf{S}_2$ of the particles induced by the shear flow, finding a non-monotonic behavior as the packing  $\phi$ is varied.  We interpret this non-monotonic behavior as a crossover from dilute systems at small  $\phi$,  where single-particle-like behavior occurs, to dense systems at large $\phi$,  where the geometry of the dense packing dominates and a random Poisson-like process for particle rotations results.  
We also argue that the  finite nematic ordering $\mathbf{S}_2$ is a consequence of the shearing serving as an ordering field, rather than a result of  long-ranged cooperative behavior among the particles.
We arrive at these conclusions by consideration of (i) the distribution of waiting times for a particle to rotate by $\pi$, (ii) the behavior of the system under pure, as compared to simple, shearing, (iii) the relaxation of the nematic order parameter $\mathbf{S}_2$ when perturbed away from the steady state, and (iv) by construction a numerical mean-field model for the rotational motion of a particle.
Our results also help to explain the singular behavior observed when taking the $\alpha\to 0$ limit approaching circular disks.
\end{abstract}
\maketitle

\section{Introduction}
\label{sec:intro}

In a system of athermal granular particles with only repulsive contact interactions, as the packing fraction of particles $\phi$ increases, the system undergoes a jamming transition \cite{OHern,LiuNagel} at a critical $\phi_J$.  For $\phi<\phi_J$ the system behaves similarly to a liquid, while for $\phi>\phi_J$ the system behaves like a rigid but disordered solid.  
One way to probe the jamming transition is through the application of a simple shear deformation to the system.  For an infinite system in the ``thermodynamic limit," if one applies a simple shear stress $\sigma$ no matter how small, then if the system is below $\phi_J$  the system responds with a simple shear flow, with a velocity profile that varies linearly in the direction transverse to the flow.  Above $\phi_J$, the application of a small shear stress causes the system to have an elastic shear distortion determined by the finite shear modulus of the solid phase; the system does not flow.  However, if $\sigma$ exceeds a critical yield stress $\sigma_0$, then plastic deformations cause the solid to flow.  The point where this yield stress $\sigma_0(\phi)$ vanishes upon decreasing $\phi$  then determines the shear-driven jamming transition $\phi_J$ \cite{OlssonTeitelPRL,OlssonTeitelPRE,VagbergOlssonTeitel}.  For frictionless particles, such as those considered in this work, $\sigma_0$ vanishes continuously \cite{OlssonTeitelPRL,OlssonTeitelPRE} as $\phi\to\phi_J$ from above.

Many numerical studies of the jamming transition, and granular materials more generally, have used spherically shaped particles for simplicity.  
It is therefore interesting to ask how behavior  is modified if the particles have shapes with a lower rotational symmetry \cite{Borzsonyi.Soft.2013}.  In a recent work \cite{MT1} we considered the shear-driven jamming of athermal, bidisperse, overdamped, frictionless, spherocylinders in two dimensions (2D), uniformly sheared at a fixed strain rate $\dot\gamma$.  In that work we  considered  the global rheology of the system, investigating how pressure, deviatoric shear stress, and macroscopic friction vary with particle packing fraction $\phi$, shear strain rate $\dot\gamma$ and particle asphericity $\alpha$. We determined the jamming packing fraction $\phi_J(\alpha)$ as a function of the spherocylinder asphericity, and the average number of contacts per particle at jamming, $Z_J(\alpha)$.  We also studied the probability for an inter-particle contact to form at a particular angle $\vartheta$ along the surface of the spherocylinder, and argued that the $\alpha\to 0$ limit approaching a circular particle was singular; we found that the total probability for a contact to form somewhere on one of the flat sides of the spherocylinder stays constant as $\alpha\to 0$, even as the length of those flat sides becomes a vanishing fraction of the total particle perimeter.

In the present work we continue our studies of this 2D spherocylinder model, but now concentrating on the rotational motion of particles and their orientational ordering.  As this work is a continuation of our work in  Ref.~\cite{MT1}, the introduction and  description of the model presented here are abbreviated.  We therefore refer the reader to Ref.~\cite{MT1} for a  discussion of the broader context of, and motivation for, our model, a more complete list of references, and more details of the derivation of our equations of motion.   Some of our results in the present work  have been presented previously \cite{MKOT}; here we broaden these prior investigations and present greater detail.

When sheared, aspherical particles are known to undergo orientational ordering due to the torques induced on the particles by the shear flow.  
Several numerical works focused on this shear-induced orientational ordering of ellipsoids \cite{Campbell} and rod-shaped particles \cite{Guo1,Guo2} of different aspect ratios in three dimensions (3D) approaching, but staying below, jamming.  They found that orientational order increased with increasing packing $\phi$, and that particles were preferentially oriented at a finite angle $\theta_2>0$ with respect to the direction of the shear flow.   Experiments and simulations of rod-shaped particles in 3D \cite{Borzsonyi1,Borzsonyi2,Wegner,Wegner2} found similar results, while also studying the rotation of particles in steady-state simple shear, and the transient approaches to the steady state.  Other experimental works have studied the transient behavior of orientational ordering and pressure $p$ of  ellipses in 2D under quasistatic shearing \cite{Farhadi,Wang}.  Numerical simulations, measuring rheological properties as well as orientational ordering 
in the hard-core limit below jamming, have been carried out for frictional 3D spherocylinders  sheared by biaxial compression \cite{Azema2010, Azema2012}, frictionless 3D spherocylinders  in steady-state simple shear \cite{Nagy}, and for both frictionless and frictional 2D ellipses in steady-state simple shear \cite{Trulsson}.  The rheology of 3D frictional and frictionless spherocylinders in steady simple shear has also recently been simulated \cite{Nath}.

In this work work we consider the uniform steady-state  simple shearing of a system of 2D spherocylinders, considering a broad range of particle asphericities, from moderately elongated to very nearly circular.  The above previous works \cite{Campbell, Guo1, Guo2, Borzsonyi1,Borzsonyi2,Wegner,Wegner2,Azema2010, Azema2012,Nagy,Trulsson,Nath}  modeled dry granular materials, in which energy is dissipated in particle collisions, rheology is Bagnoldian, and there may be microscopic inter-particle Coulombic friction.  In contrast, here we model particles in suspension, where the rheology is Newtonian at small strain rates below jamming.  We use a simple model that has been widely used in studies of the shear-driven jamming  of spherical and circular particles \cite{OlssonTeitelPRL,OlssonTeitelPRE,MT1,MKOT,Durian,Hatano,Heussinger,Andreotti,OT3,Wyart1,Vagberg.PRL.2014,Wyart2,Berthier}.
In this model, particles are frictionless with a soft-core, one-sided, harmonic repulsive interaction, and  
energy is dissipated by a viscous drag with respect to an affinely sheared host medium.  Particles obey an overdamped equation of motion and inertial effects are thus ignored.

Our simple model omits several physical processes that may be relevant to real physical suspensions, such as hydrodynamic forces \cite{hydro}, lubrication forces \cite{lub1,lub2,lub3},  inertial effects \cite{inertia}, and frictional contact interactions which have recently been proposed as a possible mechanism for  shear thickening \cite{DST0,DST1,DST2,DST3,DST4,DST5,DST6}.   
However, the greater simplicity of our model allows a more thorough investigation over a wide range of the parameter space, in particular  going to smaller values of the strain rate $\dot\gamma$ and smaller values of the particle asphericity $\alpha$.  
Our work is carried out in the spirit that it is useful to first understand the behavior of simple models before adding more realistic complexities.

The remainder of this paper is organized as follows.  In Sec.~\ref{sec:modelMethod} we define our model and give details of our numerical simulations.  In Sec.~\ref{sec:isolated} we consider the behavior of an isolated spherocylinder in an affinely sheared host medium, considering the rotational motion and the probability for the particle to be at a particular orientation.  Understanding the motion of an isolated  single particle  will help inform our understanding of the many particle system.

In Sec.~\ref{sec:ResultsRotation} we present our numerical results for the rotational motion of particles and their orientational ordering as the packing $\phi$ of particles increases through the jamming transition.  We compute the average  angular velocity of particles scaled by the strain rate, $\langle\dot\theta_i\rangle/\dot\gamma$, and the nematic orientational order parameter $\mathbf{S}_2$.  We addresses two basic questions in this section:  (1) What underlying physical processes are reflected in the observed non-monotonic behavior of both $\langle\dot\theta_i\rangle/\dot\gamma$ and the magnitude of the nematic order parameter $S_2$ as the packing $\phi$ increases, and (2) is the finite nematic ordering $\mathbf{S}_2$ a cooperative effect of multi-particle coherent motion, or is it a consequence of shearing acting like an ordering field?  We address these questions by considering (i)  the time dependence of particle rotations, (ii) the behavior of the system under pure, as opposed to simple, shearing, and (iii) the relaxation of $\mathbf{S}_2$ when it is perturbed away from its steady-state value, and (iv) by constructing a numerical mean-field model for the rotation of particles.  We also use these results  to explain the singular behavior we previously found \cite{MKOT} as the particle asphericity $\alpha\to 0$, and particles approach a circular shape.

In Sec.~\ref{sec:discus} we summarize our results.  We find that the non-montonic behavior of $S_2$ and $\langle\dot\theta_i\rangle/\dot\gamma$ can be viewed as a crossover from  a single particle-like behavior at small $\phi$, where the imposed simple shear results in a steady but non-uniform rotation of the particles, to a many particle behavior at large $\phi$, where the geometry of the dense packing and the decreasing free volume inhibits particle rotation, which becomes more of a random Poisson-like process.  We conclude that the orientational ordering is a consequence of the shear serving as an ordering field rather than due to cooperative behavior among the particles.  

Finally, in the  Appendices we consider several ancillary matters.  In Appendix~\ref{aOrientations} we  consider the distribution of particle orientations in steady-state shear flow and relate that distribution  to the orientation of the nematic order parameter.  In Appendix~\ref{sAtoZ} we present further analysis of the singular $\alpha\to 0$ limit, and explore how this limit is affected if we consider a system of particles polydisperse in  shape.  


\section{Model and Simulation Method}
\label{sec:modelMethod}

\begin{figure}
\centering
\includegraphics[width=2.5in]{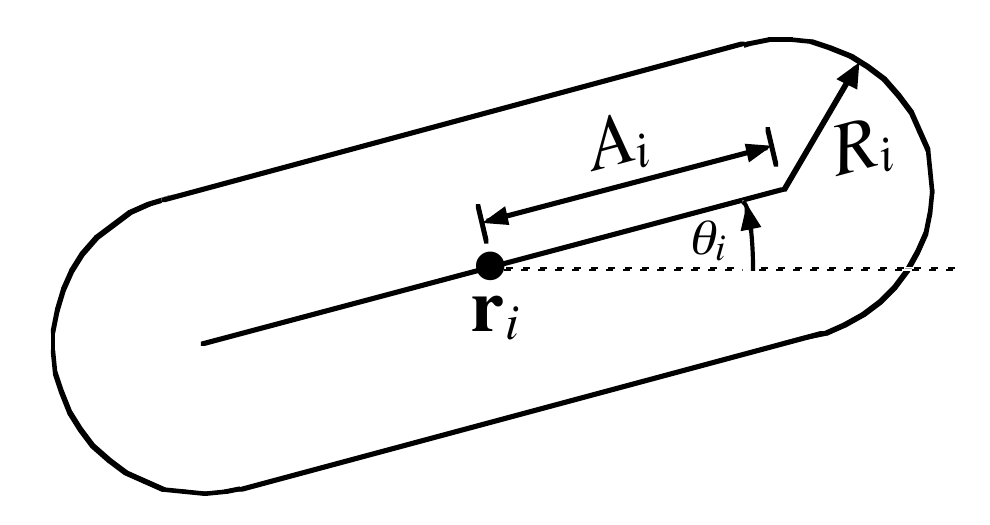}
\caption{ An isolated spherocylinder indicating the spine half-length $A_i$, end cap radius $R_i$, center of mass position $\mathbf{r}_i$, and angle of orientation $\theta_i$. }
\label{sphero} 
\end{figure}

Our model system is one of $N$ two dimensional,  athermal, frictionless spherocylinders, consisting of a rectangle with two semi-circular end caps, as illustrated in Fig.~\ref{sphero}.  The half-length of the rectangle of particle $i$ is $A_i$, the radius is $R_i$, and we define the asphericity $\alpha_i$ as,
\begin{equation}
\alpha_i=A_i/R_i
\end{equation}
so that $\alpha=0$ is a pure circular particle.  The ``spine" of the spherocylinder is the axis of length $2A_i$ that goes down the center of the rectangle.  For every point on the perimeter of the spherocylinder, the shortest distance from the spine is $R_i$.  The center of mass of the particle is $\mathbf{r}_i$ and the angle $\theta_i$ denotes the orientation of the spine with respect to the $\mathbf{\hat x}$ direction.  
Our system box has lengths $L_x$ and $L_y$ in the $\mathbf{\hat x}$ and $\mathbf{\hat y}$ directions, respectively.  We will in general take $L_x=L_y\equiv L$ unless otherwise noted.
If $\mathcal{A}_i$ is the area of spherocylinder $i$, the packing fraction $\phi$ is,
\begin{equation}
\phi=\frac{1}{L^2}\sum_{i=1}^N\mathcal{A}_i.
\end{equation}
Unless otherwise stated, all our particles have equal asphericity $\alpha$, and are bidisperse in size with equal numbers of big and small particles with length scales in the ratio $R_b/R_s=1.4$.

The dynamics of our model has been described in detail in Ref.~\cite{MT1}, here we  summarize the main features.    Periodic boundary conditions are taken along $\mathbf{\hat x}$, while Lees-Edward boundary conditions \cite{LeesEdwards} are taken along $\mathbf{\hat y}$ to introduce a simple shear strain $\gamma$.  We take $\gamma =\dot\gamma t$ to model simple shear flow in the $\mathbf{\hat x}$ direction at a fixed finite strain rate $\dot\gamma$.  Particles interact with each other via elastic contact interactions.  Energy dissipation is due to  a viscous drag between the particles and an affinely sheared host medium, 
\begin{equation}
\mathbf{v}_\mathrm{host}(\mathbf{r})=\dot\gamma y \mathbf{\hat x}, 
\end{equation}
modeling the behavior of particles in a uniform non-Brownian suspension.

Defining $r_{ij}$ as the shortest distance between the spines of spherocylinders $i$ and $j$ \cite{Pournin.GranulMat.2005}, and $d_{ij}=R_i+R_j$, two spherocylinders are in contact whenever $r_{ij}<d_{ij}$.  In this case there is a repulsive harmonic interaction between the particles with the force on $i$ being given by,
\begin{equation}
\mathbf{F}_{ij}^\mathrm{el}=\frac{k_e}{d_{ij}}\left(1-\frac{r_{ij}}{d_{ij}}\right)\mathbf{\hat n}_{ij},
\end{equation}
where $k_e$ is the particle stiffness and  $\mathbf{\hat n}_{ij}$ the unit vector pointing normally inwards to particle $i$ at the point of contact with $j$.  The force $\mathbf{F}_{ij}^\mathrm{el}$ acts at the contact point, which is located a distance $(R_i/d_{ij})r_{ij}$ from the spine of particle $i$, along the cord $r_{ij}$, and gives rise to a torque on particle $i$,
\begin{equation}
\boldsymbol{\tau}_{ij}^\mathrm{el}=\mathbf{\hat z} \tau_{ij}^\mathrm{el}=\mathbf{s}_{ij}\times\mathbf{F}_{ij}^\mathrm{el},
\end{equation}
where $\mathbf{s}_{ij}$ is the moment arm from the center of mass of $i$ to its point of contact with $j$.  The total elastic force and torque on particle $i$ are then
\begin{equation}
\mathbf{F}_i^\mathrm{el}=\sum_j \mathbf{F}_{ij}^\mathrm{el},\qquad
\tau_i^\mathrm{el}=\sum_j \tau_{ij}^\mathrm{el}
\end{equation}
where the sums are over all particles $j$ in contact with $i$.

The viscous drag between  particle $i$ and the host medium gives rise to a dissipative force,
\begin{equation}
\mathbf{F}_i^\mathrm{dis}=\int_i d^2r\,\mathbf{f}_i^\mathrm{dis}(\mathbf{r}),
\end{equation}
where the integral is over the area of particle $i$ and the dissipative force per unit area acting at position $\mathbf{r}$ on the particle is given by  the local velocity difference between the particle and the host medium,
\begin{equation}
\mathbf{f}_i^\mathrm{dis}(\mathbf{r})=-k_d[\mathbf{v}_i(\mathbf{r})-\mathbf{v}_\mathrm{host}(\mathbf{r})],
\end{equation}
where $k_d$ is a viscous damping coefficient and $\mathbf{v}_i(\mathbf{r})$ is the local velocity of the particle at position $\mathbf{r}$,
\begin{equation}
\mathbf{v}_i(\mathbf{r})=\mathbf{\dot r}_i+\dot\theta_i\mathbf{\hat z}\times (\mathbf{r}-\mathbf{r}_i).
\end{equation}
Here $\dot{\mathbf{r}}_i=d\mathbf{r}_i/dt$ is the center of mass velocity of the particle and $\dot\theta_i$ is its angular velocity about the center of mass.   The corresponding dissipative torque is,
\begin{equation}
\boldsymbol{\tau}_i^\mathrm{dis}=\mathbf{\hat z}\tau_i^\mathrm{dis}=\int_i d^2r\,(\mathbf{r}-\mathbf{r}_i)\times \mathbf{f}_i^\mathrm{dis}(\mathbf{r}).
\end{equation}

The above elastic and dissipative forces are the only forces included in our model; there are no inter-particle dissipative or frictional forces.  We will carry out our simulations in the overdampled (low particle mass) limit, where the total force and torque on each particle are damped to zero, 
\begin{equation}
\mathbf{F}_i^{\mathrm{el}} + \mathbf{F}_i^{\mathrm{dis}} = 0, 
\quad
\tau_i^{\mathrm{el}} + \tau_i^{\mathrm{dis}} = 0.
\end{equation} 
The resulting translational and rotational equations of motion for particle $i$ can then be written as \cite{MT1},
\begin{align}
\dot{\mathbf{r}}_i &=    \dot{\gamma}y_i{\mathbf{\hat x}}+\dfrac{\mathbf{F}_i^{\mathrm{el}}}{k_d \mathcal{A}_i},
\label{eq:ri_eom} \\
\dot{\theta}_i &= - \dot{\gamma} f(\theta_i)+ \dfrac{\tau_i^{\mathrm{el}}}{k_d  \mathcal{A}_i I_i},
\label{eq:theta_eom}
\end{align}
where $\mathcal{A}_i$ is the area of particle $i$, $I_i$ is the trace of the particle's moment of inertia tensor, and 
\begin{equation}
f(\theta)=\frac{1}{2}\left[1-\left({\Delta I_i}/{I_i}\right)\cos 2\theta\right],
\label{eftheta}
\end{equation}
where  $\Delta I_i$ is the absolute value of the difference of the two eigenvalues of the moment of inertia tensor.  We assume a uniform constant  mass density for both our small and big particles.

For our simulations we  take $2 R_s = 1$ as the unit of distance, $k_e = 1$ as the unit of energy, and $t_0 = (2 R_s)^2 k_d\mathcal{A}_s / k_e = 1$ as the unit of time.
For simplicity, we take the damping coeficient $k_d$ to vary with particle size, so that $k_d\mathcal{A}_i=1$ for all particles.
We numerically integrate the equations of motion (\ref{eq:ri_eom}) and (\ref{eq:theta_eom}) using a two-stage Heun method with a step size of $\Delta t = 0.02$.
Unless otherwise stated, we begin each shearing run in a finite energy configuration at the desired packing fraction $\phi$ with random initial particle positions and orientations.
To generate such initial configurations we place the spherocylinders in the system one-by-one, while rejecting and retrying any time a new placement would lead to an unphysical overlap where the spines of two spherocylinders  intersect.  In general we use $N=1024$ particles.  We have found this to be sufficiently large to avoid any significant finite size effects for the behaviors discussed in this work.
Most of our simulations typically extend to strains of at least $\gamma\approx 150$.  Discarding an   initial $\Delta\gamma\approx 20$ of the strain  from the averaging so as to eliminate transients effects, we find that our steady-state averages are generally insensitive to the particular starting configuration \cite{Vagberg.PRE.2011}.  See the Supplemental Material to Ref.~\cite{MKOT} for tests that these simulation parameters, in particular $N$ and $\Delta t$, are sufficient to obtain accurate results for particles with our smallest asphericity, $\alpha=0.001$.
Note that we restrict the strain coordinate $\gamma$ used in our Lees-Edwards boundary condition  to the range $\gamma\in \left(-L_x/2L_y, L_x/2L_y\right]$; whenever it exceeds this maximum it is reset by taking $\gamma \to \gamma - L_x/Ly$, allowing us to shear to arbitrarily large total strains.  

\section{Isolated Particles: Rotations and Orientational Ordering}
\label{sec:isolated}

\begin{figure}
\centering
\includegraphics[width=3.5in]{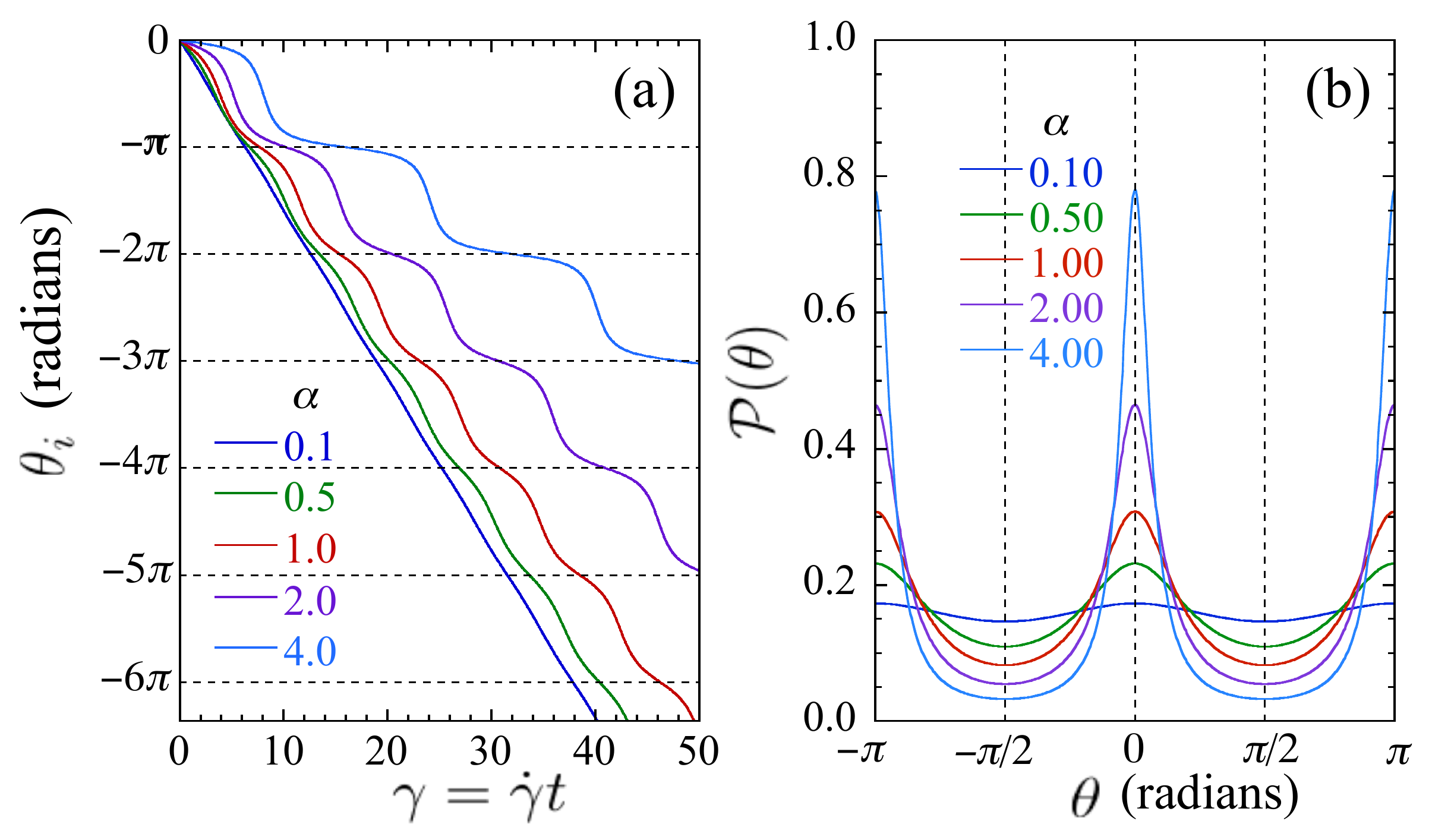}
\caption{For an isolated spherocylinder in a uniform shear flow, (a) orientation $\theta_i$ vs net shear strain $\gamma=\dot\gamma t$, and (b) probability density $\mathcal{P}(\theta)$ vs $\theta$ for the spherocylinder to be oriented at angle $\theta$.  
From bottom to top in (a) the curves are for spherocylinders with asphericity $\alpha=0.1$, 0.5, 1.0, 2.0 and 4.0, and similarly for the curves at $\theta=\pi$ in (b).
}
\label{theta-vs-g} 
\end{figure}

Although the main objective of this work is to study the behavior of many interacting particles, it is of interest to first consider the case of an isolated particle, for which $\mathbf{F}_i^\mathrm{el}=\boldsymbol{\tau}_i^\mathrm{el}=0$.  In this case Eq.~(\ref{eq:ri_eom}) gives that the particle flows with the local host velocity, $\dot{\mathbf{r}}_i=\dot\gamma y_i\mathbf{\hat x}$, while from Eq.~(\ref{eq:theta_eom}) the rotational motion obeys the deterministic equation, $\dot\theta_i=-\dot\gamma f(\theta_i)$, with $f(\theta)$ as in Eq.~(\ref{eftheta}).  Since in general $f(\theta)>0$, the particle will rotate continuously clockwise, but with a non-uniform angular velocity that is slowest at $\theta_i=0$ or $\pi$ where $f(\theta_i)$ is at its minimum, and fastest at $\theta_i=\pi/2$ or  $3\pi/2$ where $f(\theta_i)$ is at its maximum.  
This is analogous to the Jeffrey orbits of ellipsoids in a viscous fluid \cite{Jeffery.RSPA.1922}.
The particle will thus spend more time oriented at $\theta_i=0$, aligned parallel to the flow direction $\mathbf{\hat x}$.  We show this explicitly by integrating the equation of motion and plotting $\theta_i(t)$ vs $\gamma=\dot\gamma t$ in Fig.~\ref{theta-vs-g}(a) for spherocylinders of several different $\alpha$.

For such an isolated particle tumbling in the flow field of the host medium, we can compute the probability density for the particle's orientation to be at a particular angle $\theta$, 
\begin{align}
\mathcal{P}(\theta)&=\frac{1}{T}\int_0^T\!\!dt\,\delta (\theta_i(t)-\theta)\\[10pt]
&=\frac{1}{T}\int_0^{2\pi}\!\!d\theta_i\,\frac{\delta(\theta_i-\theta)}{|\dot\theta_i|}
=\dfrac{1}{T\dot\gamma f(\theta)},
\label{Piso}
\end{align}
where $T$ is the period of the rotation.  
We plot $\mathcal{P}(\theta)$ vs $\theta$ for spherocylinders with different $\alpha$ in Fig.~\ref{theta-vs-g}(b).
Normalization of $\mathcal{P}(\theta)$ then determines the period $T$ and thus gives for the average angular velocity,
\begin{equation}
-\dfrac{\langle\dot\theta_i\rangle}{\dot\gamma} = \dfrac{2\pi}{\dot\gamma T} = \frac{1}{2}\sqrt{1-(\Delta I_i/I_i)^2}.
\label{eomegasingle}
\end{equation}
For a circular particle one has $\Delta  I_i/I_i=0$ and so $-\langle\dot\theta\rangle/\dot\gamma=1/2$.  More generally, since $0\le \Delta I_i/I_i< 1$, one then  has $0<-\langle\dot\theta\rangle/\dot\gamma \le 1/2$.

\begin{figure}
\centering
\includegraphics[width=3.5in]{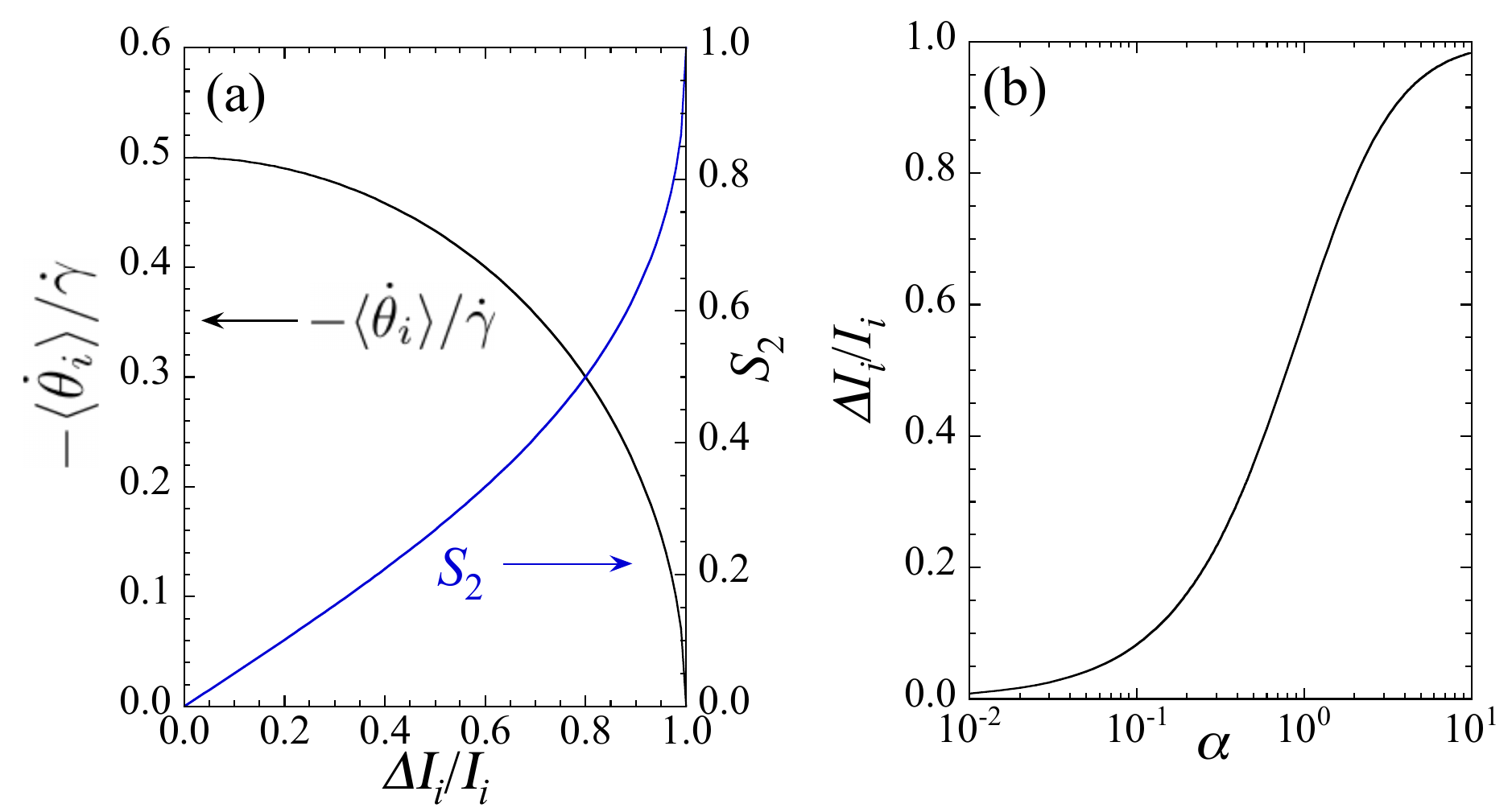}
\caption{(a) Average scaled angular velocity $-\langle\dot\theta_i\rangle/\dot\gamma$ and magnitude of the nematic order parameter $S_2$ vs $\Delta I_i/I_i$ for an isolated particle in a uniform shear flow. (b) Plot of $\Delta I_i/I_i$ vs $\alpha$ for spherocylinders of asphericity $\alpha$.
}
\label{S2-vs-C} 
\end{figure}

Since $\mathcal{P}(\theta+\pi)=\mathcal{P}(\theta)$, corresponding to the fact that the particle has neither a head nor a tail, orientational ordering will be nematic.  The direction of the nematic order parameter $\mathbf{S}_2$ is $\theta_2=0$, aligned with the flow, while the magnitude is given by,
\begin{equation}
S_2=\int_0^{2\pi}\!\!d\theta\,\mathcal{P}(\theta)\cos 2\theta = \dfrac{1-\sqrt{1-(\Delta I_i/I_i)^2}}{(\Delta I_i/I_i)}.
\label{eS2single}
\end{equation}
In Fig.~\ref{S2-vs-C}(a) we plot $-\langle \dot\theta\rangle/\dot\gamma$ and $S_2$ vs $\Delta I_i/I_i$ for an isolated particle, using Eqs.~(\ref{eomegasingle}) and (\ref{eS2single}).  
We see, not surprisingly, an anti-correlation between the two quantities; $-\langle\dot\theta\rangle/\dot\gamma$ decreases as the particle becomes more aspherical (i.e., as $\Delta I_i/I_i$ increases), while $S_2$ increases.
For spherocylinders of asphericity $\alpha$ we have,
\begin{equation}
\dfrac{\Delta I_i}{I_i}=\dfrac{2\alpha(4+3\pi\alpha+4\alpha^2)}{3\pi+24\alpha+6\pi\alpha^2+8\alpha^3},
\label{eDI}
\end{equation}
which we plot in Fig.~\ref{S2-vs-C}(b).

\begin{figure}
\centering
\includegraphics[width=3.5in]{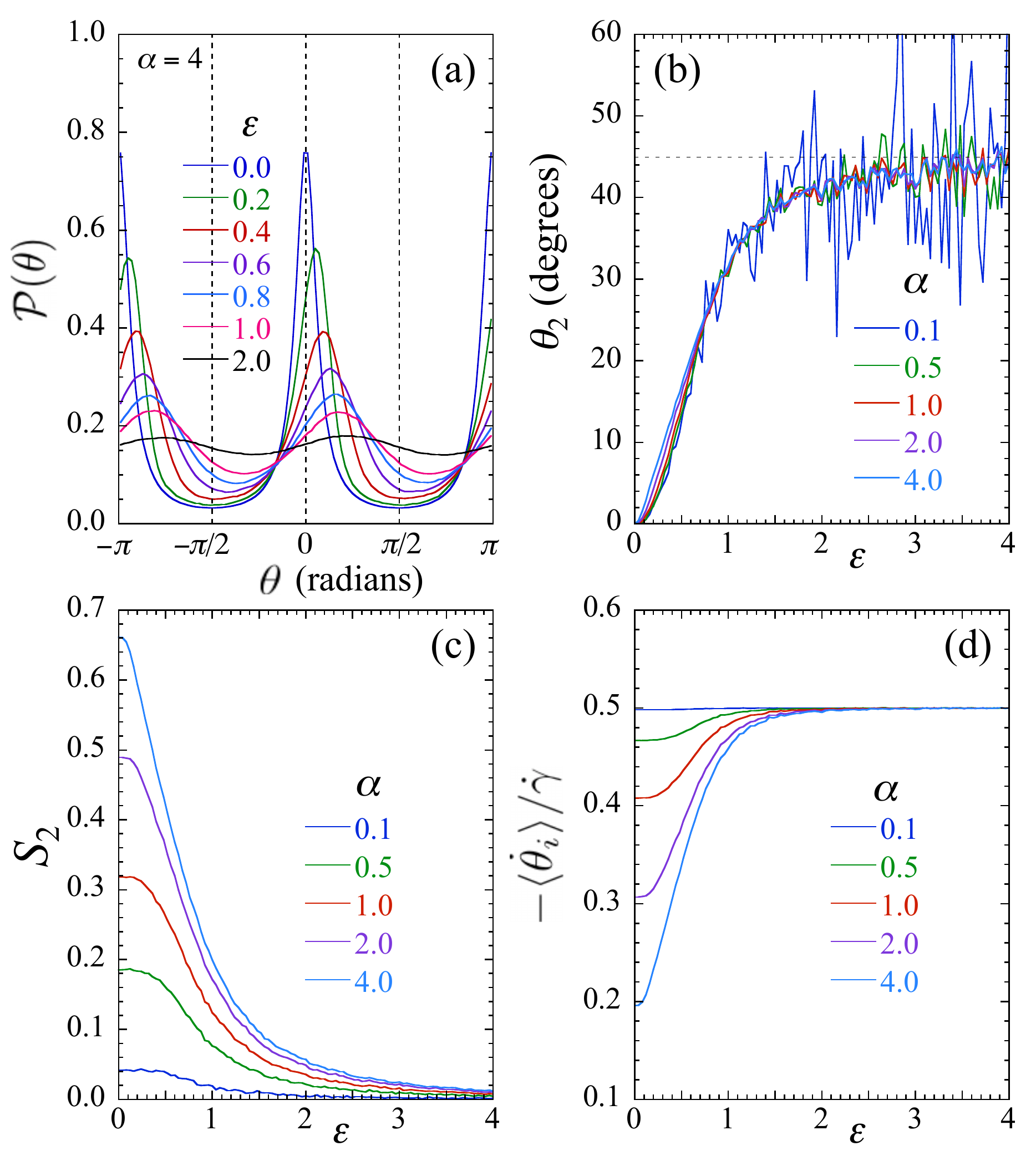}
\caption{(a) Probably density $\mathcal{P}(\theta)$ for a spherocylinder of asphericity $\alpha=4$ to be oriented at angle $\theta$, for various strengths $\varepsilon$ of uncorrelated random torque noise. (b) Orientation $\theta_2$ of the nematic order parameter, (c) magnitude $S_2$ of the nematic order parameter, and (d) scaled particle angular velocity $-\langle\dot\theta_i\rangle/\dot\gamma$ vs noise strength $\varepsilon$, for spherocylinders of various $\alpha$.
}
\label{noisy} 
\end{figure}

As the packing $\phi$ increases from zero, the above single particle behavior will be modified due to collisions that occur between particles, giving rise to elastic forces and torques.  It is interesting to consider a naive model in which, at small $\phi$, we regard these collisions as introducing uncorrelated random torques, as if the particle were at a finite temperature.  We therefore rewrite Eq.~(\ref{eq:theta_eom}) as,
\begin{equation}
\dfrac{\dot\theta_i}{\dot\gamma}=\dfrac{d\theta_i}{d\gamma}=-f(\theta_i) + \zeta(\gamma)
\label{enoisy}
\end{equation}
where $\zeta = \tau_i^\mathrm{el}/(k_d \mathcal{A}_i I_i\dot\gamma)$ and we assume,
\begin{equation}
\langle \zeta(\gamma)\rangle=0,\qquad \langle\zeta(\gamma)\zeta(\gamma^\prime)\rangle = \varepsilon^2\delta(\gamma-\gamma^\prime).
\end{equation}
 Numerically integrating Eq.~(\ref{enoisy}), in Fig.~\ref{noisy}(a) we plot the resulting probability density $\mathcal{P}(\theta)$ for a spherocylinder of $\alpha=4$, for various noise levels $\varepsilon$.  We see several significant changes from the noiseless $\varepsilon=0$ case.  As $\varepsilon$ increases, we see that the amplitude of the variation in $\mathcal{P}(\theta)$ decreases, and the location of the peak shifts from $\theta=0$ to larger $\theta >0$.  This indicates that the magnitude of the nematic order $S_2$ is decreasing and the nematic director becomes oriented at a finite positive angle with respect to the shear flow.  
 
 To quantify this observation, we compute the nematic order parameter as follows: For a particle in 2D, the magnitude $S_2$ and orientation $\theta_2$ of the nematic order parameter $\mathbf{S}_2$ are given by \cite{Torquato},
 \begin{equation}
 S_2=\max_{\theta_2}\left[\langle\cos(2[\theta-\theta_2])\rangle\right],
 \label{eS2iso0}
 \end{equation}
 where $\langle \dots\rangle$ denotes an average over time, or equivalently over strain $\gamma=\dot\gamma t$.   From this one can show,
 \begin{equation}
 S_2=\sqrt{\langle\cos 2\theta\rangle^2+\langle\sin 2\theta\rangle^2}
 \label{eS2iso1}
 \end{equation}
 and
 \begin{equation}
 \tan 2\theta_2 =\langle\sin 2\theta\rangle/\langle\cos 2\theta\rangle.
 \label{eS2iso2}
 \end{equation}
 
 In Fig.~\ref{noisy}(b) we plot $\theta_2$ vs noise level $\varepsilon$ for several different spherocylinder asphericities $\alpha$.  The values of $\theta_2$ for $\alpha=4$ coincide with the locations of the peaks in $\mathcal{P}(\theta)$ in Fig.~\ref{noisy}(a).  We see that there is no  strong dependence of $\theta_2$ on $\alpha$, except at small $\varepsilon$, and that $\theta_2$ saturates to $45^\circ$ as $\varepsilon$ gets large; $45^\circ$ corresponds to the eigen-direction of expansion of the affine strain rate tensor, and hence also the direction of minimal stress.
 
 In Fig.~\ref{noisy}(c) we plot $S_2$ vs $\varepsilon$ for different $\alpha$ and see that $S_2$ decays to zero as $\varepsilon$ increases; we find the large $\varepsilon$ tail of this decay to be well fit to an exponential $\sim \exp(-\varepsilon/\varepsilon_0)$, with $\varepsilon_0\approx 1.16$ for all $\alpha$.  Finally in Fig.~\ref{noisy}(d) we plot the scaled average angular velocity $-\langle\dot\theta_i\rangle/\dot\gamma$ vs $\varepsilon$ for different $\alpha$.  As $\varepsilon$ increases, $-\langle\dot\theta_i\rangle/\dot\gamma$ saturates to 1/2, the rotational velocity of the affinely sheared host medium, as well as the value expected for a circular particle.  We find the large $\varepsilon$ behavior to be well fit to the form $\sim \frac{1}{2}[1-c\exp(-\varepsilon/\varepsilon_0^\prime)]$, with $\varepsilon_0^\prime\approx 0.34$ for all $\alpha$.  As in Fig.~\ref{S2-vs-C}(a) we see that $S_2$ and $-\langle\dot\theta_i\rangle/\dot\gamma$ are anticorrelated; as one increases, the other decreases.

These results are easy to understand.  The nematic ordering, in the isolated particle limit, is determined by how long the particle spends at the preferred alignment $\theta=0$ or $\pi$, where $f(\theta)$ has its minimum.  When a particle oriented near $\theta=0$ receives a random kick directed counter-clockwise, the particle deflects to positive $\theta$, but then quickly relaxes back towards $\theta=0$ under the influence of the driving term $-f(\theta)$; however if the random kick is directed clockwise, the particle will rapidly rotate through $\pi$, before relaxing towards the preferred alignment at $\theta=\pi$.  This effect results in the particle spending more time at angles $\theta>0$ than at corresponding angles $\theta < 0$, and as a consequence $\theta_2$ becomes finite and positive, growing with the strength of the random kicks.  At the same time, the occurrence of clockwise directed random kicks serves to shorten the time the particle spends in the preferred aligned direction $\theta=0$ or $\pi$, resulting in an increase in the average angular velocity $-\langle\dot\theta\rangle/\dot\gamma$ and a decrease in the magnitude of the nematic ordering $S_2$.

In the following sections we explore what happens as the packing $\phi$ increases in a true model of $N$ interacting spherocylinders.  We will see that, as $\phi$ increases from small values, $\theta_2$ increases from zero in accord with the above naive model.  However we will see that  $S_2$ and $-\langle\dot\theta_i\rangle/\dot\gamma$ behave qualitatively the opposite of this naive model; as $\phi$ increases from small values, $S_2$ {\em increases} while $-\langle\dot\theta_i\rangle/\dot\gamma$ {\em decreases}.  As we will see in Sec.~\ref{sec:MF},  the reason for this difference is that, while our naive model above assumed the collisions provided no net torque $\langle\zeta\rangle=0$, in fact the collisions that occur due to  shearing create an orientation-dependent elastic torque  on a particle which on average is finite and counter-clockwise, thus slowing down the rotation of particles and increasing orientational ordering.

\section{Numerical results: Rotations and Orientational Ordering}
\label{sec:ResultsRotation}

\begin{figure}
\centering
\includegraphics[width=3.5in]{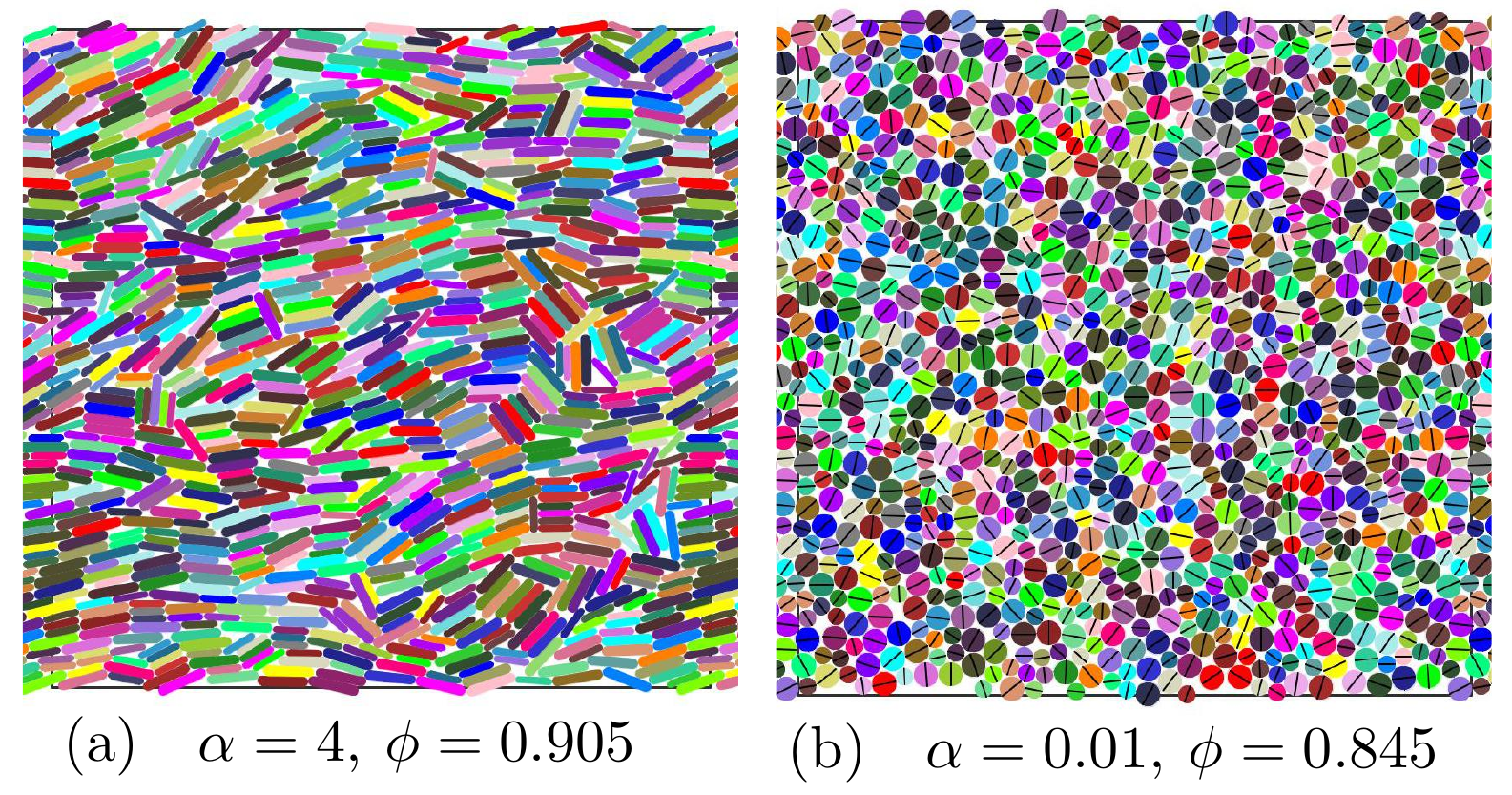}
\caption{Snapshot configurations in simple sheared steady-state with strain rate $\dot\gamma=10^{-6}$ for spherocylinders of asphericity (a)  $\alpha=4$ at packing $\phi=0.905$ near the jamming $\phi_J=0.906$, and (b) $\alpha=0.01$ at packing $\phi_J=0.845$.  For the nearly circular particles at $\alpha=0.01$, the black line bisecting each particle indicates the direction of the spherocylinder axis.  Colors are used to help distinguish different particles and have no other meaning.  Corresponding animations, showing the evolutions of these configurations under shearing, are available in the Supplemental Material \cite{SM}.
}
\label{configs} 
\end{figure}

At finite packing $\phi$, particles will come into contact, $\tau_i^\mathrm{el}$ will no longer be zero, and the  isolated particle behavior of the previous section will be modified.  Here we report on our numerical results for systems of particles with different asphericity from $\alpha=0.001$ to 4, for a range of packings $\phi$ from dilute, to jamming, and above.  We will  look in greater detail at the two specific cases of moderately elongated particles with $\alpha=4$, and nearly circular particles with $\alpha=0.01$.
In Fig.~\ref{configs} we show snapshots of typical steady-state configurations for these two cases, sheared at a rate $\dot\gamma=10^{-6}$.  For $\alpha=4$ we show a dense configuration at $\phi=0.905$, close to its jamming $\phi_J=0.906$; for $\alpha=0.01$ we show a configuration at its jamming $\phi_J=0.85$.

\begin{figure}
\centering
\includegraphics[width=3.5in]{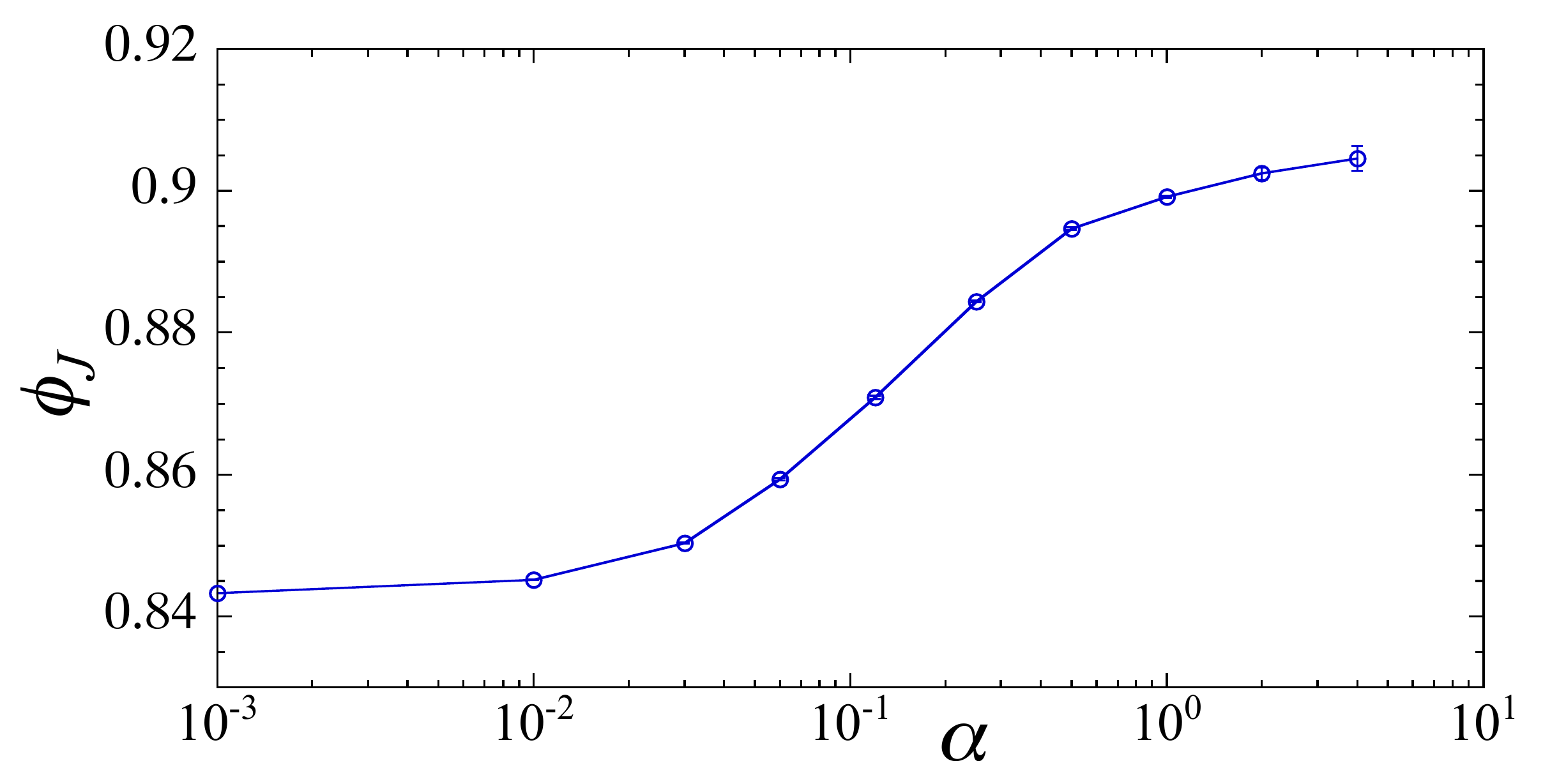}
\caption{Critical packing fraction $\phi_J$ for shear-driven jamming vs spherocylinder asphericity $\alpha$, from Ref.~\cite{MT1}.
}
\label{phiJ-vs-alpha} 
\end{figure}

When comparing results for systems of different  $\alpha$, we will  find it convenient to plot quantities in terms of a reduced  packing fraction, $\phi/\phi_J(\alpha)$, where $\phi_J(\alpha)$ is the shear-driven jamming packing fraction for particles of that particular value of $\alpha$.  For reference, in Fig.~\ref{phiJ-vs-alpha} we plot this $\phi_J$ vs $\alpha$, as we have determined in our earlier work \cite{MT1}.  Note that this $\phi_J(\alpha)$  monotonically increases with $\alpha$, for the range of $\alpha$ studied here.  This is in contrast to compression-driven jamming where $\phi_J(\alpha)$ reaches a maximum near $\alpha\approx 1$ and then decreases as $\alpha$ increases further \cite{MTCompress}.  This difference is because there is no nematic ordering for athermal isotropic compression \cite{MTCompress}, while (as we will see below) there is nematic ordering in the sheared system; the orientational ordering of the sheared system allows the particles to pack more efficiently and so results in a larger $\phi_J$ that continues to increases with increasing $\alpha$.

\subsection{Average Angular Velocity}
\label{sAV}

\begin{figure}
\centering
\includegraphics[width=3.5in]{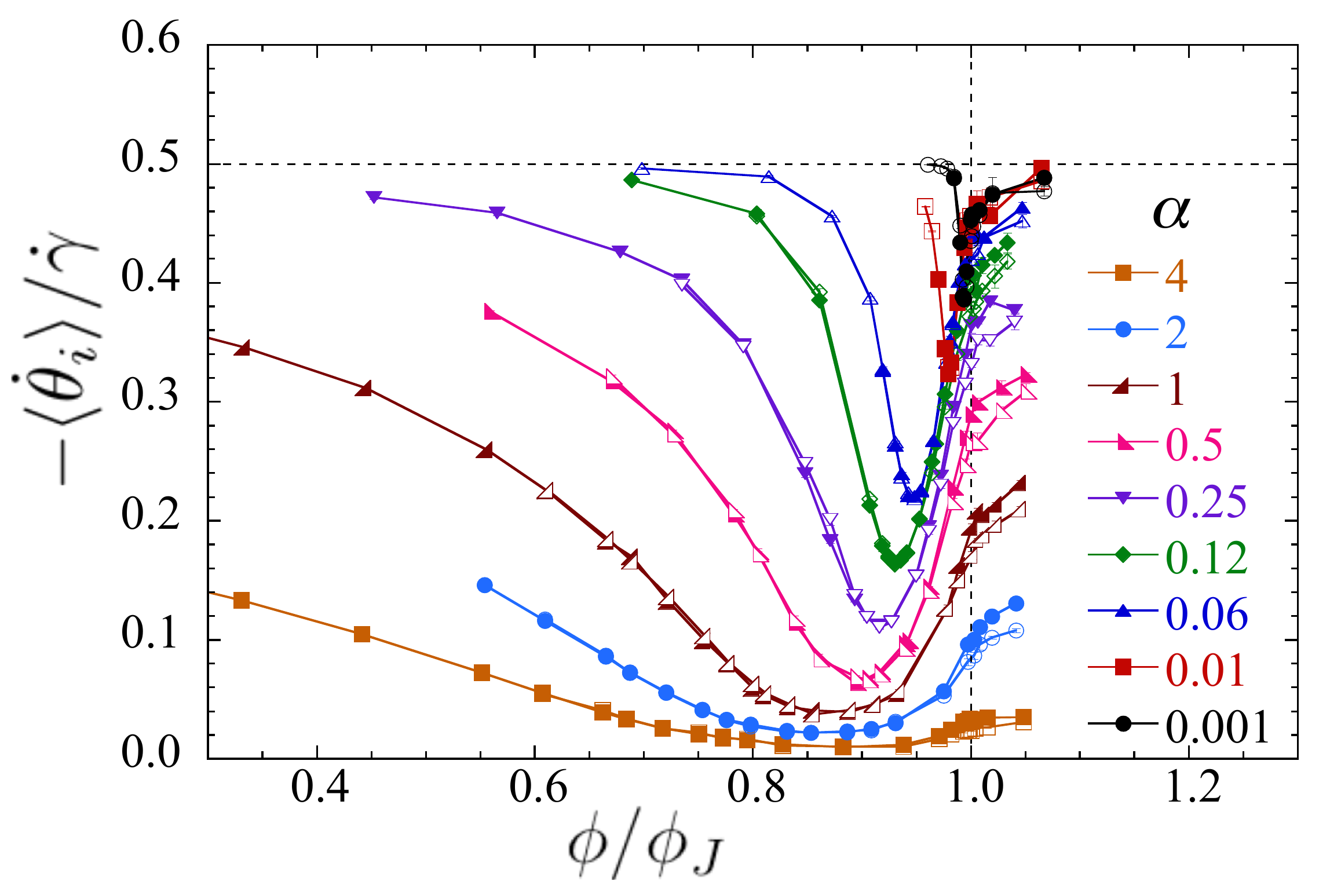}
\caption{Average particle angular velocity scaled by strain rate $-\langle\dot\theta_i\rangle/\dot\gamma$ vs reduced packing fraction $\phi/\phi_J$ for spherocylinders of different asphericity $\alpha$. For each $\alpha$ we show results for two different small strain rates $\dot\gamma_1$ (solid symbols) $<\dot\gamma_2$ (open symbols) (see Table~\ref{tab1} for values).   The vertical dashed line locates the jamming transition $\phi/\phi_J=1$.
The horizontal dashed line denotes the rotation $1/2$ of the affinely sheared host medium.
}
\label{dthdg-vs-phiophiJ} 
\end{figure}

We first consider the angular velocity of the particles' rotational motion.  For the coordinate system of our model, a counterclockwise rotation is a positive angular velocity, while a clockwise rotation is negative.  Since our particles have a net rotation that is clockwise, it is therefore convenient to consider $-\dot\theta_i$.  It will also be convenient to measure in dimensionless units, which we will find gives a finite value in the quasistatic limit $\dot\gamma\to 0$.  Hence, when we refer to the angular velocity of particle $i$, we will generally mean $-\dot\theta_i/\dot\gamma$.

From Eq.~(\ref{eq:theta_eom}) we can write for the average angular velocity of individual particles,
\begin{equation}
-\dfrac{\langle\dot\theta_i\rangle}{\dot\gamma} 
=\left\langle\dfrac{1}{N}\sum_{i=1}^N\left[f(\theta_i)-\dfrac{\tau_i^\mathrm{el}}{\dot\gamma k_d \mathcal{A}_i I_i}\right]\right\rangle,
\end{equation}
where $\langle\dots\rangle$ indicates an average over configurations in the steady state.   In an earlier letter \cite{MKOT} we plotted the resulting $-\langle\dot\theta_i\rangle/\dot\gamma$ vs the packing fraction $\phi$, for spherocylinders of different asphericity. 
In Fig.~\ref{dthdg-vs-phiophiJ} we reproduce those results for asphericities  $\alpha=0.001$ to 4, but now plotting vs the reduced packing fraction $\phi/\phi_J$, so as to more easily compare behaviors near the $\alpha$-dependent jamming transition.

For each $\alpha$ we show results at two different small strain rates, $\dot\gamma_1 <\dot\gamma_2$, in order to demonstrate that our results, except for the largest $\phi$ near and above jamming, are in the quasistatic limit where $\langle\dot\theta_i\rangle/\dot\gamma$ is independent of $\dot\gamma$.  The values of $\dot\gamma_1$ and $\dot\gamma_2$ used for each $\alpha$ are given in Table~\ref{tab1}.  That $-\langle\dot\theta_i\rangle/\dot\gamma>0$ indicates that the particles continuously rotate in a clockwise direction, and such rotation persists even in  dense configurations above jamming.  Here, and in subsequent plots, error bars represent one standard deviation of estimated statistical error; when error bars are not visible, they are smaller than the size of the symbol representing the data point.

\begin{figure}
\centering
\includegraphics[width=3.5in]{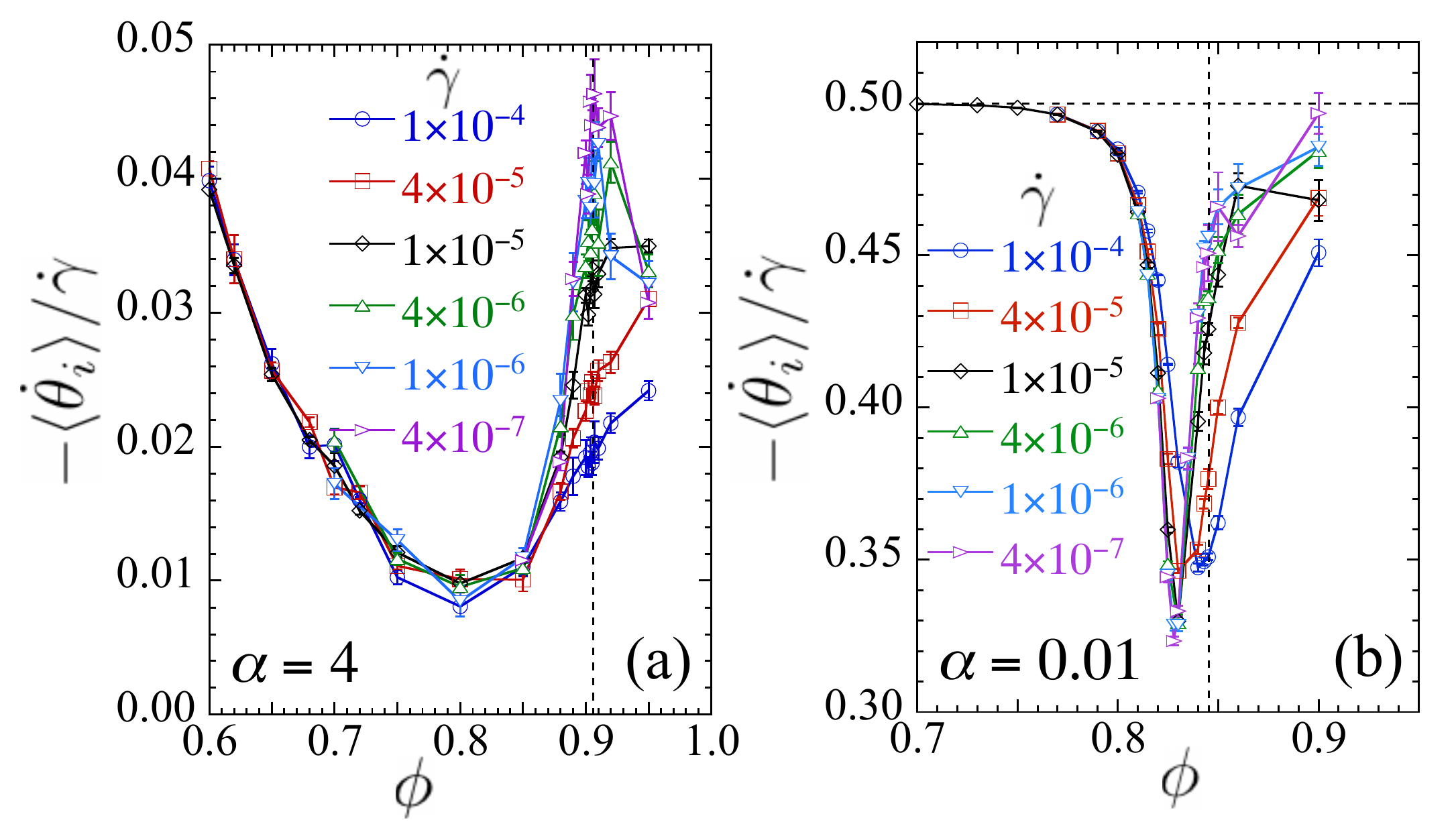}
\caption{Average particle angular velocity $-\langle\dot\theta_i\rangle/\dot\gamma$ vs packing $\phi$ for different strain rates $\dot\gamma$, for spherocylinders of asphericity (a) $\alpha=4$ and (b) $\alpha=0.01$.  Vertical dashed lines indicate the location of the jamming transitions, $\phi_J=0.906$ and $\phi_J=0.845$, respectively.
}
\label{av_v_phi-allgdot} 
\end{figure}

In Fig.~\ref{av_v_phi-allgdot} we similarly plot $-\langle\dot\theta_i\rangle/\dot\gamma$ vs $\phi$, but now showing results for multiple different strain rates $\dot\gamma$, for the two particular cases of moderately extended rods, with $\alpha=4$, and nearly circular particles, with $\alpha=0.01$.  We see, as mentioned above, that the $\dot\gamma$ dependence of the angular velocity increases as one approaches and goes above $\phi_J$, but seems to be approaching a finite limiting value as $\dot\gamma\to 0$.

 \begin{table}[h!]
\caption{Strain rate values used for data in Figs.~\ref{dthdg-vs-phiophiJ}, \ref{S2-vs-phiophiJ} and \ref{th2D-vs-phiophiJ}}
\begin{center}
\begin{tabular}{|c|c|c|}
\hline
$\alpha$ & $\dot\gamma_1$ & $\dot\gamma_2$  \\
\hline
0.001 & $1\times 10^{-7}$ & $4\times10^{-7}$ \\
0.01 & $4\times 10^{-7}$ & $1\times10^{-6}$  \\
$\alpha\ge0.06$ & $1\times 10^{-5}$ & $4\times 10^{-5}$ \\
\hline
\end{tabular}
\end{center}
\label{tab1}
\end{table}%

There are several obvious features to note  in Figs.~\ref{dthdg-vs-phiophiJ} and \ref{av_v_phi-allgdot}: (i) The angular velocity $-\langle \dot\theta_i\rangle/\dot\gamma$ is non-monotonic in $\phi$, initially decreasing as $\phi$ increases from the dilute limit, reaching a minimum at a $\phi_{\dot\theta\,\mathrm{min}}$ close to but below the jamming $\phi_J$, and then increasing again as $\phi$ further increases towards $\phi_J$ and goes above.  As $\alpha$ decreases, this variation in $-\langle \dot\theta_i\rangle/\dot\gamma$ gets squeezed into a narrower range of $\phi$, closer to $\phi_J$. One of our main objectives in this work will be to understand the physical origin of this non-monotonic behavior.  (ii) For small $\alpha$, at both small $\phi$ and large $\phi>\phi_J$, the angular velocity $-\langle \dot\theta_i\rangle/\dot\gamma\approx 1/2$,  the value expected for  perfectly circular particles.  However, even for the very nearly circular particles with $\alpha=0.001$, the dip in $-\langle \dot\theta_i\rangle/\dot\gamma$ at $\phi_{\dot\theta\,\mathrm{min}}$ remains sizable, about $20\%$ below 1/2.  The main result of our earlier Letter \cite{MKOT} was to argue that this dip remains finite in the  $\alpha\to 0$ limit approaching circular disks.  In this work we will provide further understanding of what causes this singular behavior as $\alpha\to 0$.
(iii) In the dilute limit at small $\phi$, the angular velocity $-\langle \dot\theta_i\rangle/\dot\gamma$ is decreasing as $\phi$ increases, which is the opposite of the behavior seen in Fig.~\ref{noisy}(d) for the noisy isolated particle model.  Thus one should not regard the elastic collisions in the dilute ``gas" limit as behaving simply like an effective temperature.

\begin{figure}
\centering
\includegraphics[width=3.5in]{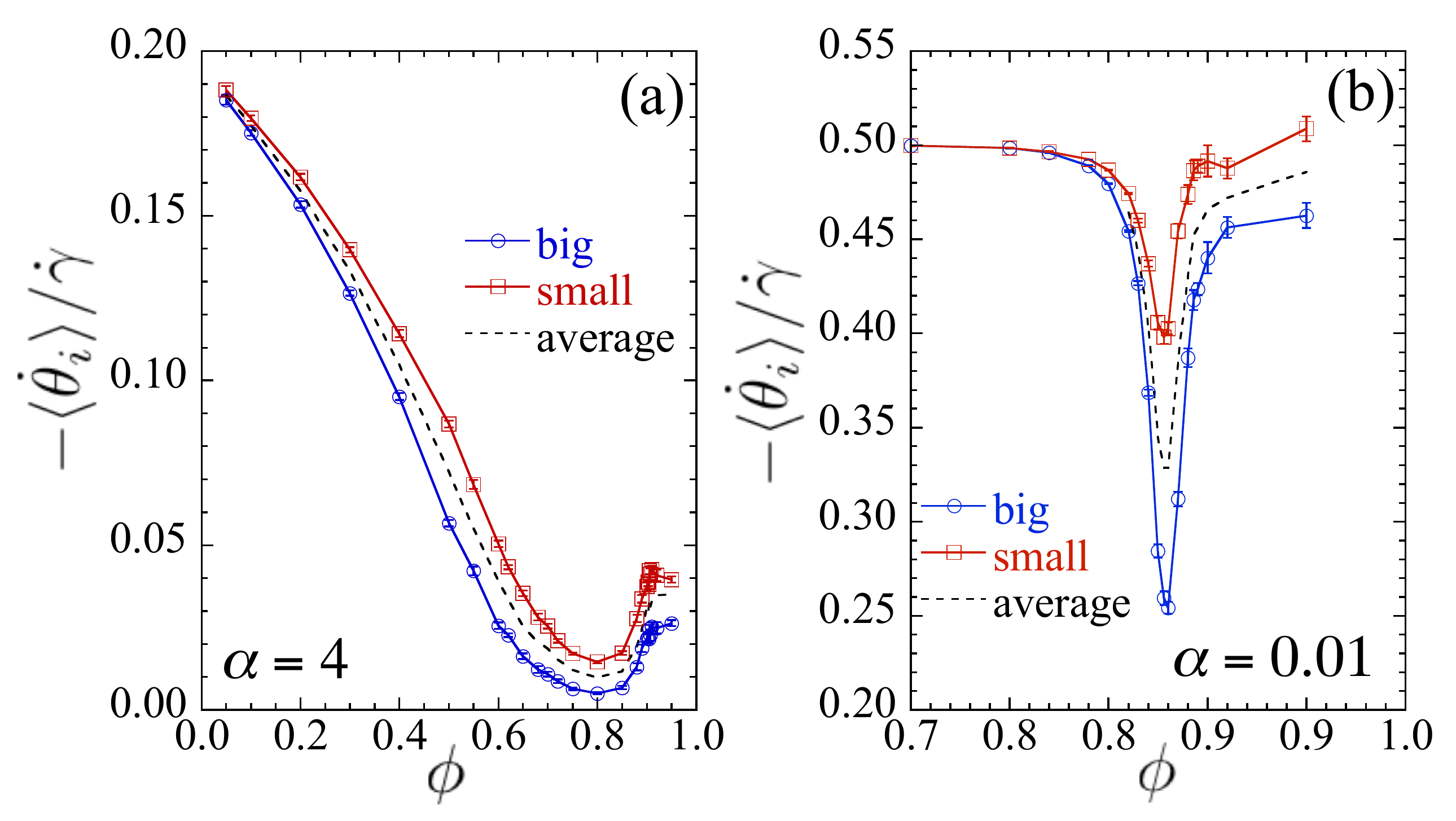}
\caption{Average angular velocity $-\langle\dot\theta_i\rangle/\dot\gamma$ vs $\phi$ for big and small particles separately, for spherocylinders with  (a) $\alpha=4$ at $\dot\gamma=10^{-5}$ and (b) $\alpha=0.01$ at $\dot\gamma=10^{-6}$.  The average over all particles is given by the dashed line.
}
\label{av-bs} 
\end{figure}

Finally, we make one last point concerning the angular velocity.  Since our system is bidisperse in particle size, one can separately compute the average angular velocity for big particles as compared to small particles.  In Figs.~\ref{av-bs}(a) and \ref{av-bs}(b) we plot these for spherocylinders with $\alpha=4$ and 0.01, respectively.  Not surprisingly, we see that big particles rotate more slowly than the average, while small particles rotate more quickly.  

\subsection{Nematic Orientational Ordering}
\label{sS2}

In this section we consider the orientational ordering of the interacting particles.  For a system in $d$ dimensions, the nematic order parameter $\mathbf{S}_2$ can be obtained from the traceless, symmetric, ordering tensor of an $N$ particle configuration,
\begin{equation}
\mathbf{T}=\dfrac{d}{(d-1)N}\sum_{i=1}^N\left[\boldsymbol{\hat\ell}_i\otimes\boldsymbol{\hat\ell}_i-\dfrac{1}{d}\mathbf{I}\right],
\end{equation}
where $\boldsymbol{\hat\ell}_i$ is a unit vector that lies along the spine of particle $i$, and $\mathbf{I}$ is the identity tensor. The magnitude $S_2$ of the nematic order parameter is given by the largest eigenvalue of $\mathbf{T}$, and the corresponding eigenvector $\boldsymbol{\hat\ell}_2$ gives the orientation of the nematic director.  
We will define the nematic order parameter as $\mathbf{S}_2=S_2\boldsymbol{\hat\ell}_2$. 
For our system in $d=2$ dimensions, the angle of $\boldsymbol{\hat\ell}_2$ with respect to the flow direction $\mathbf{\hat x}$ will define the orientation angle $\theta_2$ of the nematic director.

We  define the instantaneous nematic order parameter,  given by $S_2(\gamma)$ and $\theta_2(\gamma)$, in terms of the tensor $\mathbf{T}(\gamma)$ for the specific configuration of the system after a total strain $\gamma$.  We define the ensemble averaged nematic order parameter, given by $S_2$ and $\theta_2$, in terms of the ensemble averaged tensor $\langle\mathbf{T}\rangle$, which is an average over configurations in the steady state.  Note that while $\langle\mathbf{T}\rangle$ is a linear average over the instantaneous $\mathbf{T}(\gamma)$, the same is not in general true of $S_2$ and $\theta_2$ because of variations in the eigenvector directions of $\mathbf{T}(\gamma)$, due either to fluctuations about a steady-state, or to possible systematic variations of $\mathbf{T}(\gamma)$ with $\gamma$.

For a $d=2$ dimensional system, one can show that the above definitions for $S_2$ and $\theta_2$ are equivalent to  generalizations of Eqs.~(\ref{eS2iso0})-(\ref{eS2iso2}).  For a given  configuration after total strain $\gamma$ we have for the instantaneous order parameter,
\begin{equation}
S_2(\gamma)=\max_{\theta^\prime}\left[\frac{1}{N}\sum_{i=1}^N\cos (2[\theta_i-\theta^\prime])\right],
\label{eS2g0}
\end{equation}
with $\theta_2(\gamma)$ being the maximizing value of $\theta^\prime$.
From this one can show \cite{Torquato} that
\begin{equation}
S_2(\gamma)=\sqrt{\left[\frac{1}{N}\sum_{i=1}^N \cos (2\theta_i)\right]^2
+\left[\frac{1}{N}\sum_{i=1}^N \sin (2\theta_i)\right]^2}
\label{eS2g1}
\end{equation}
and
\begin{equation}
\tan [2\theta_2(\gamma)] = 
{\left[ \displaystyle{\frac{1}{N}\sum_{i=1}^N\sin (2\theta_i)}\right]}\bigg/
{\left[ \displaystyle{\frac{1}{N}\sum_{i=1}^N\cos (2\theta_i)}\right]}.
\label{eS2g2}
\end{equation}
The ensemble averaged order parameter, given by $S_2$ and  $\theta_2$, are similarly obtained, except by replacing the large square brackets $[\dots]$ in  Eqs.~(\ref{eS2g0})-(\ref{eS2g2}), which represent  sums over particles in a particular configuration,  by ensemble averages $\langle\dots\rangle$ over the many different configurations in the steady-state.


\subsubsection{Time Dependence of Nematic Ordering}
\label{sTNO}

The athermal shearing of aspherical rod-shaped particles has been compared to the thermalized shearing of nematic liquid crystals \cite{Borzsonyi1,Borzsonyi2,Wegner}.  In the latter case, several different types of behavior may occur depending on material parameters \cite{Jenkins,Larson2,Hess,Larson}.  The system may settle into a steady-state with constant $S_2$ and $\theta_2$; the system may ``tumble," with the orientation of the nematic director  $\theta_2$ rotating through $\pi$ over a well defined period; or the system might show ``wagging," in which $\theta_2$ has periodic variations back and forth within a fixed interval without rotating.   We thus wish to investigate whether such time varying behavior exists in our athermal system.  Given that we do find that individual particles continue to rotate even as the system gets dense, is there any coherent rotation of particles that would lead to a systematic variation of $\mathbf{S}_2(\gamma)$ with $\gamma$?
For our 2D spherocylinders we do indeed see both tumbling and wagging of the nematic director, however we believe that these occur only as a transient effect due  to poor equilibration of the rotational degrees of freedom, either because the density $\phi$ is so small that collisions are rare, or because $\alpha$ is so small that  small moment arms lead to small elastic torques and so take long times to reach proper equilibration. 

\begin{figure}
\centering
\includegraphics[width=3.5in]{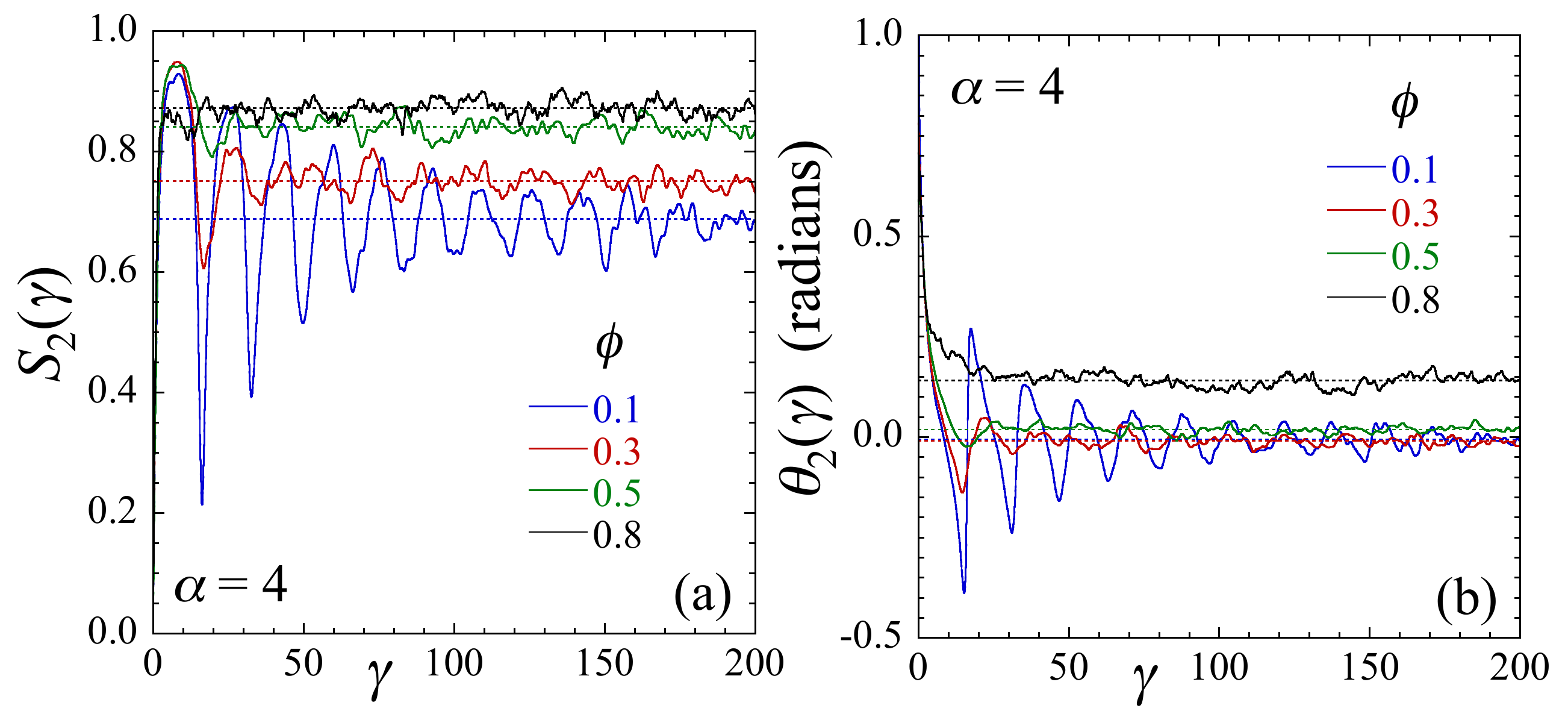}
\caption{For spherocylinders of asphericity $\alpha=4$ at $\dot\gamma=10^{-5}$: instantaneous (a) magnitude $S_2(\gamma)$ and (b) orientation $\theta_2(\gamma)$ of the nematic order parameter vs total strain $\gamma=\dot\gamma t$, for several different packing fractions $\phi$.  Horizontal dotted lines indicate the ensemble averaged values of $S_2$ and $\theta_2$.
}
\label{S2-v-g-a4} 
\end{figure}

In Fig.~\ref{S2-v-g-a4} we plot the instantaneous $S_2(\gamma)$ and $\theta_2(\gamma)$ vs total strain $\gamma=\dot\gamma t$, for spherocylinders of $\alpha=4$ at $\dot\gamma=10^{-5}$ for a few different packings $\phi$.  Our shearing starts from a random initial configuration for which $S_2(0)\approx 0$. For the very small $\phi=0.1$ we see damped oscillations in both $S_2(\gamma)$ and $\theta_2(\gamma)$ with a period $\Delta\gamma\approx 16.1$, almost equal to the period $16.04$ of  an isolated particle.  The behavior of $\theta_2(\gamma)$ identifies this as a wagging of the order parameter.  As $\gamma$ increases, the amplitude of these oscillations decays, but the periodicity remains.  For $\phi=0.3$, the behavior at small $\gamma$ is similar to that at $\phi=0.1$, but the amplitude of the oscillations dies out faster.  At larger $\gamma$ there is no longer any remnant of the initial periodic behavior, and $S_2(\gamma)$ and $\theta_2(\gamma)$ show  only random fluctuations about the ensemble averaged values $S_2$ and $\theta_2$. For larger $\phi$, the initial transient dies out even more quickly.

\begin{figure}
\centering
\includegraphics[width=3.5in]{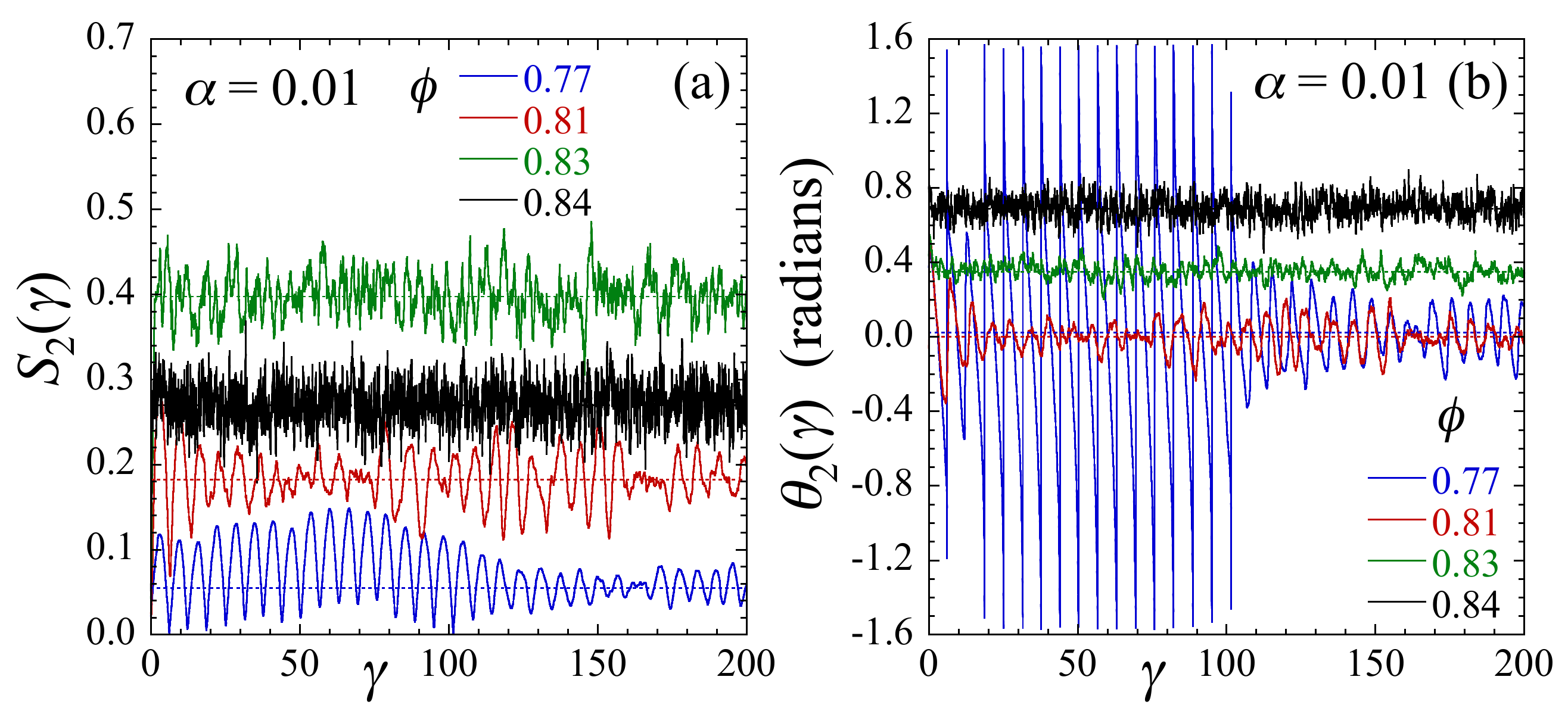}
\caption{For spherocylinders of asphericity $\alpha=0.01$ at $\dot\gamma=10^{-6}$: instantaneous (a) magnitude $S_2(\gamma)$ and (b) orientation $\theta_2(\gamma)$ of the nematic order parameter vs total strain $\gamma=\dot\gamma t$ for several different packing fractions $\phi$.  Horizontal dotted lines indicate the ensemble averaged values $S_2$ and $\theta_2$; for $\phi=0.77$ this average is taken only over the latter part of the run $\gamma>125$.
}
\label{S2-v-g-a01} 
\end{figure}

In Fig.~\ref{S2-v-g-a01} we show similar plots of $S_2(\gamma)$ and $\theta_2(\gamma)$, but now for particles of $\alpha=0.01$ at $\dot\gamma=10^{-6}$.  For the smallest $\phi=0.77$ shown we see strong oscillations in $S_2(\gamma)$, and $\theta_2(\gamma)$ initially makes full clockwise rotations with a period $\Delta\gamma\approx 6.7$, close to the  period $6.28$ for an isolated particle.  As $\gamma$ increases, the rotations become a wagging and the amplitude of the oscillations in $S_2(\gamma)$ decreases, but there remains a clear periodic behavior.  For $\phi=0.81$ there are no longer any initial rotations, but the wagging continues with a small erratic amplitude but definite periodicity out to the largest $\gamma$.  For $\phi=0.83$ and above, we see only random fluctuations about the ensemble averaged values.  We conclude from Figs.~\ref{S2-v-g-a4} and \ref{S2-v-g-a01} that the  rotating and wagging of the nematic order parameter $\mathbf{S}_2$ are only transient effects that should die out if the simulation is run long enough, rather than being stable periodic motions of the macroscopic order parameter.


\subsubsection{Ensemble Averaged Nematic Ordering}
\label{sec:NO}

\begin{figure}
\centering
\includegraphics[width=3.5in]{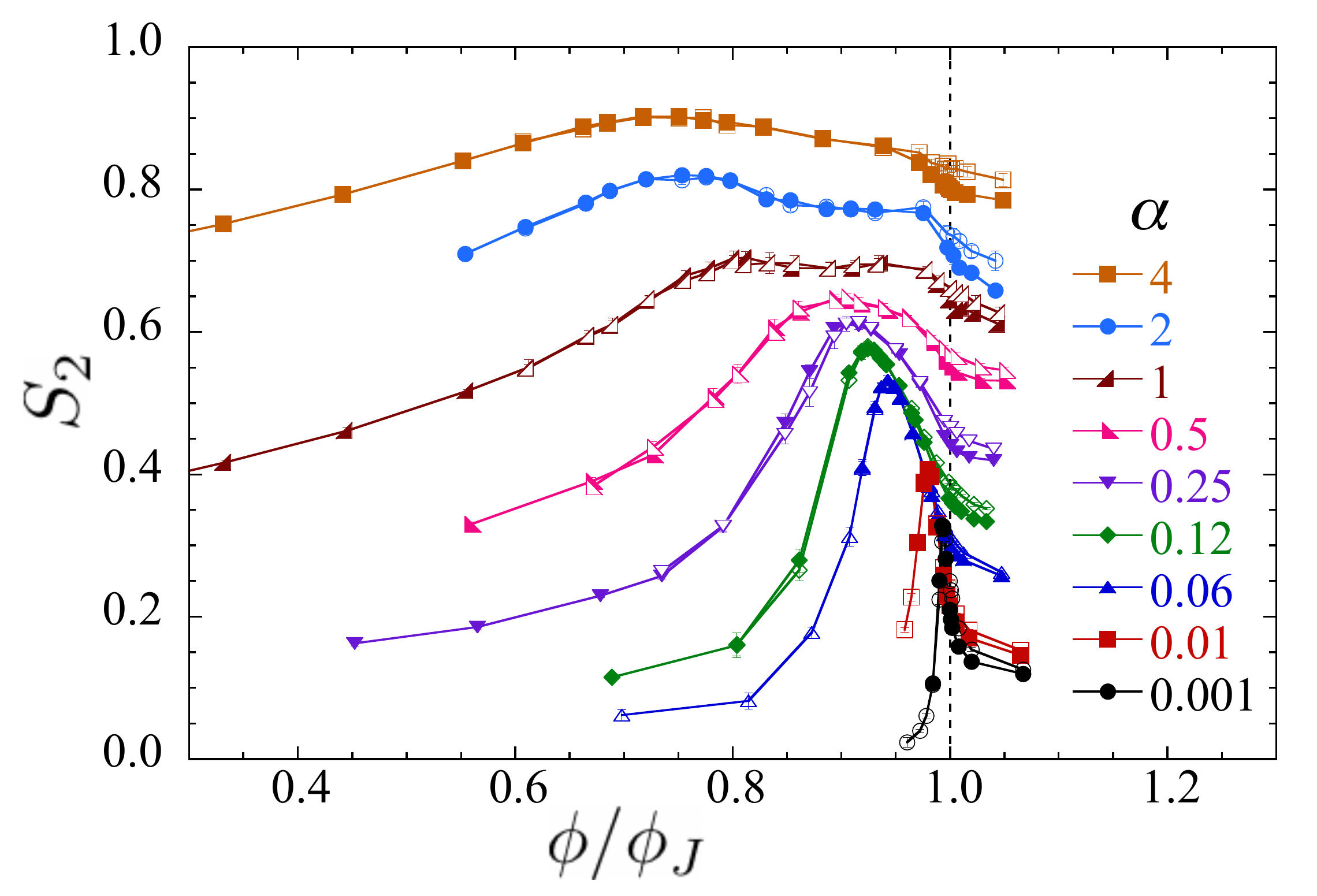}
\caption{Magnitude of the ensemble averaged nematic order parameter $S_2$ vs reduced packing fraction $\phi/\phi_J$ for spherocylinders of different asphericity $\alpha$.  For each $\alpha$ we show results for two different small strain rates $\dot\gamma_1$ (solid symbols) $< \dot\gamma_2$ (open symbols) (see Table~\ref{tab1} for values).  The vertical dashed line locates the jamming transition $\phi/\phi_J=1$.
}
\label{S2-vs-phiophiJ} 
\end{figure}

\begin{figure}
\centering
\includegraphics[width=3.5in]{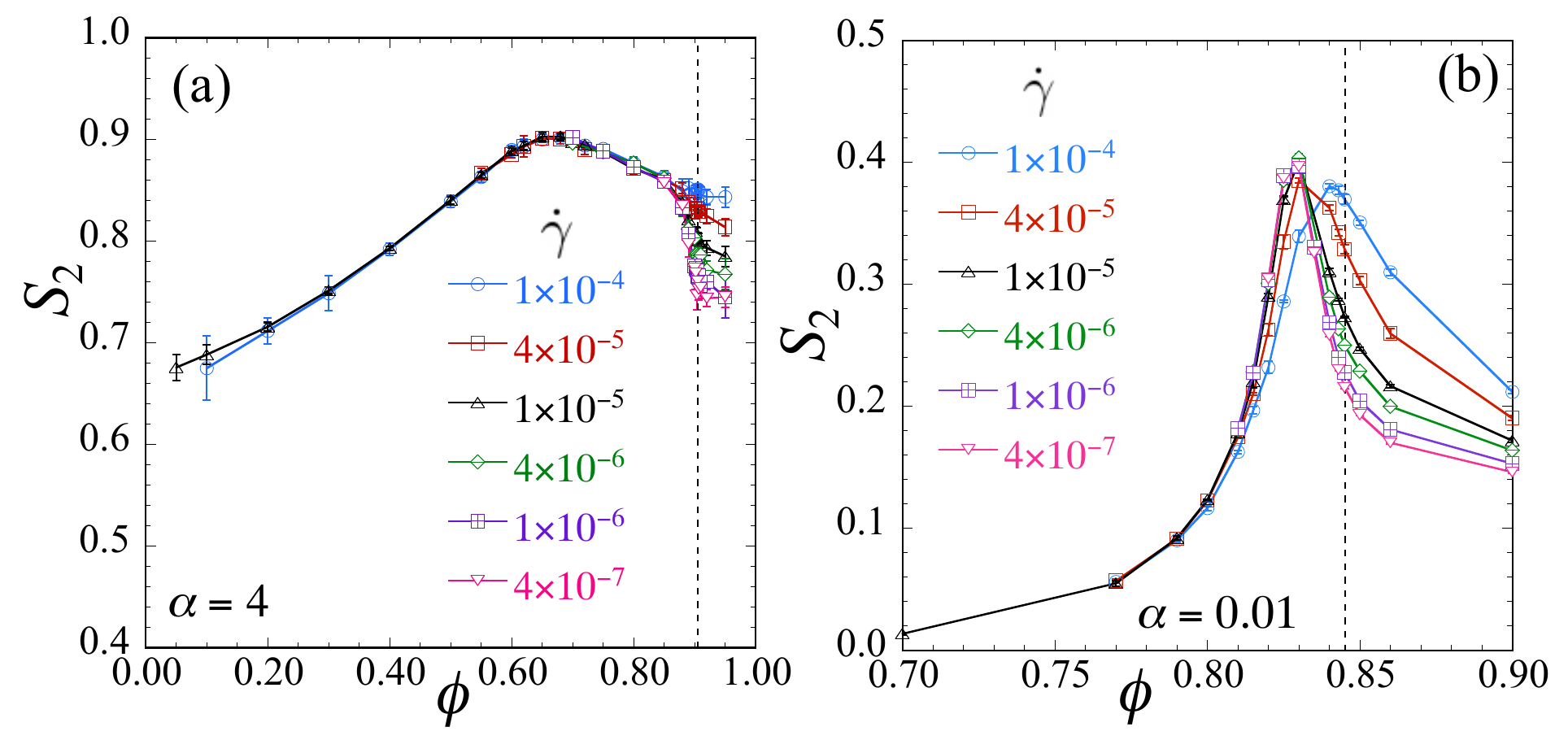}
\caption{Magnitude of the ensemble averaged nematic order parameter $S_2$ vs  packing fraction $\phi$ at different strain rates $\dot\gamma$, for spherocylinders of  asphericity (a) $\alpha=4$ and (b) $\alpha=0.01$.   Vertical dashed lines locate the jamming transitions, $\phi_J=0.906$ and $\phi_J=0.845$, respectively.
}
\label{S2_v_phi_a4-01} 
\end{figure}

Having argued in the preceding section that we expect no coherent time variation of the instantaneous nematic order parameter $\mathbf{S}_2(\gamma)$ in a well equilibrated system,
we turn now to consider the ensemble averaged nematic order parameter, given by  its magnitude $S_2$ and orientation angle $\theta_2$.  
In an earlier Letter \cite{MKOT} we plotted the ensemble averaged $S_2$ vs the packing $\phi$ for spherocylinders of different aspect ratios.
In Fig.~\ref{S2-vs-phiophiJ} we reproduce those results for asphericities $\alpha=0.001$ to 4, but now plotting vs the reduced packing fraction $\phi/\phi_J$.
For each $\alpha$ we show results at  two different strain rates $\dot\gamma_1<\dot\gamma_2$, whose values are given in Table~\ref{tab1}, to demonstrate that our results are in the quasistatic limit where $S_2$ becomes independent of $\dot\gamma$, except for the largest $\phi$ approaching and going above jamming.  In Fig.~\ref{S2_v_phi_a4-01} we similarly plot $S_2$ vs $\phi$, but now showing results for a wider range of strain rates $\dot\gamma$, for the two particular cases $\alpha=4$ and $\alpha=0.01$.  We see that the dependence of $S_2$ on $\dot\gamma$ is strongest near the jamming transition, but that $S_2$ appears to be approaching a finite limit as $\dot\gamma\to 0$.

\begin{figure}
\centering
\includegraphics[width=3.5in]{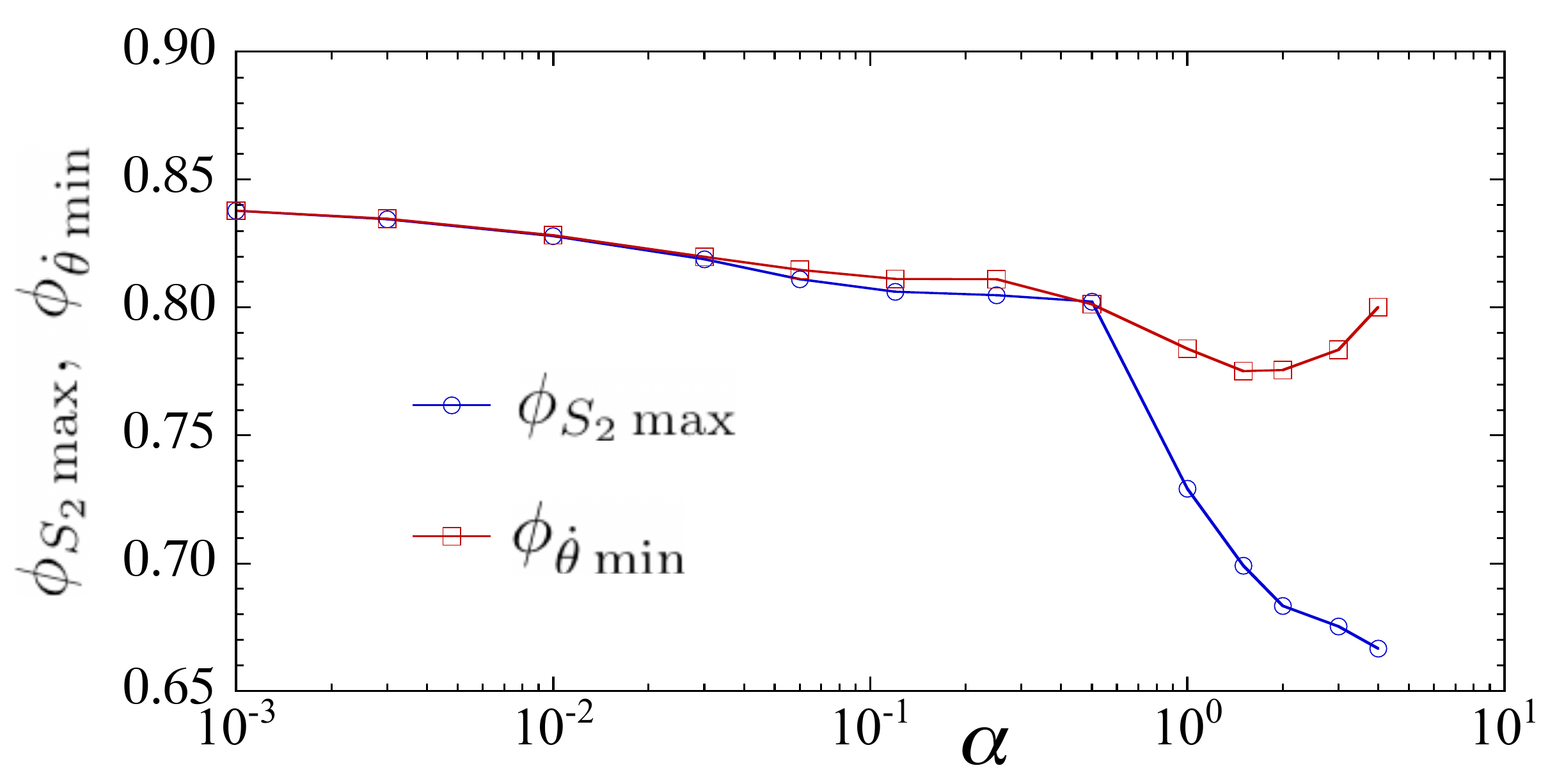}
\caption{Location $\phi_{S_2\,\mathrm{max}}$ of the maximum in the nematic order parameter $S_2$ of Fig.~\ref{S2-vs-phiophiJ}, and location $\phi_{\dot\theta\,\mathrm{min}}$ of the minimum in the angular velocity $-\langle\dot\theta_i\rangle/\dot\gamma$ of Fig.~\ref{dthdg-vs-phiophiJ}, vs particle asphericity $\alpha$.
}
\label{phi-extreme} 
\end{figure}

Similar to what we observed for the angular velocity $-\langle\dot\theta_i\rangle/\dot\gamma$ in Figs.~\ref{dthdg-vs-phiophiJ} and \ref{av_v_phi-allgdot}, 
our results for $S_2$ show several significant features:  (i) As was found for $-\langle\dot\theta_i\rangle/\dot\gamma$, $S_2$ is non-monotonic in $\phi$, reaching a maximum at $\phi_{S_2\,\mathrm{max}}$ somewhat below the jamming $\phi_J$.  As was found for an isolated particle in Fig.~\ref{S2-vs-C}(a), comparing Figs.~\ref{dthdg-vs-phiophiJ} and \ref{S2-vs-phiophiJ} we see  an anti-correlation between angular velocity and nematic ordering; roughly speaking, when  $-\langle\dot\theta_i\rangle/\dot\gamma$ decreases $S_2$ increases, and vice versa.  In Fig.~\ref{phi-extreme} we plot $\phi_{S_2\,\mathrm{max}}$, the location of the maximum in $S_2$, and $\phi_{\dot\theta\,\mathrm{min}}$, the location of the  minimum in $-\langle\dot\theta_i\rangle/\dot\gamma$, vs $\alpha$.  
We see that they are close and become roughly equal for $\alpha\lesssim 0.5$. 
(ii) As $\alpha$ decreases, the variation in $S_2$ gets squeezed into an increasingly narrow range of $\phi$, closer to $\phi_J$, and the degree of ordering $S_2$ decreases.  However, even for the very nearly circular particles with $\alpha=0.001$, the maximum value  $S_{2\,\mathrm{max}}=0.33$ remains relatively large.
This is another reflection of the singular $\alpha\to 0$ limit, discussed above in connection with the angular velocity $-\langle\dot\theta_i\rangle/\dot\gamma$, and  reported in our earlier letter \cite{MKOT}.
(iii) In the dilute limit at small $\phi$, we see $S_2$ is increasing as $\phi$ increases, which is the opposite of the behavior seen in Fig.~\ref{noisy}(c) for the noisy isolated particle.  Thus, as we concluded also from the behavior of $-\langle\dot\theta_i\rangle/\dot\gamma$, one cannot regard the elastic collisions in the dilute ``gas" limit as behaving similarly to an effective temperature.  In subsequent sections we will develop an understanding of the behaviors (i) and (ii).

\begin{figure}
\centering
\includegraphics[width=3.5in]{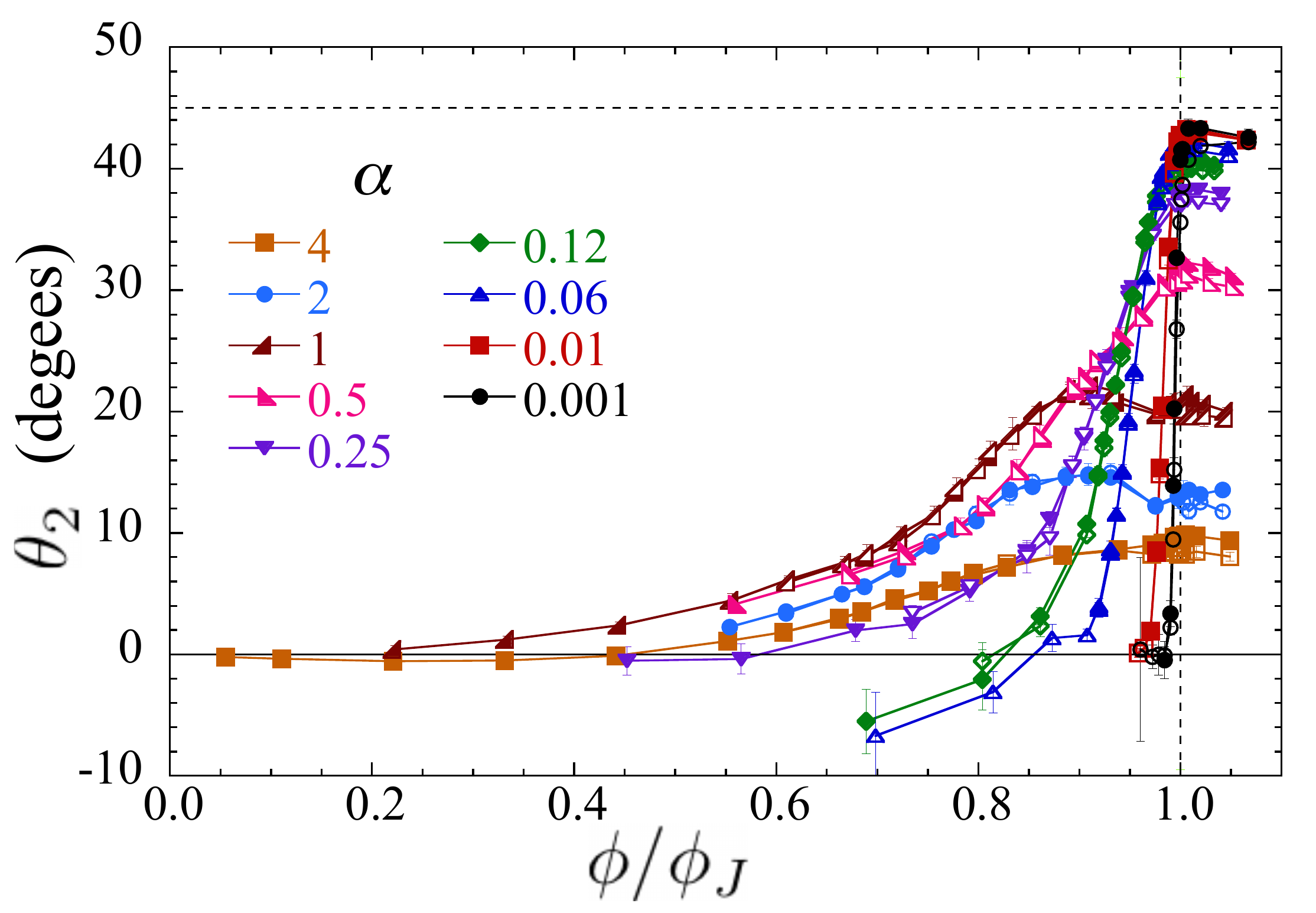}
\caption{Orientation of the ensemble averaged nematic order parameter $\theta_2$ vs reduced packing fraction $\phi/\phi_J$ for spherocylinders of different asphericity $\alpha$.  For each $\alpha$ we show results for two different small strain rates $\dot\gamma_1$ (solid symbols) $< \dot\gamma_2$ (open symbols) (see Table~\ref{tab1} for values).  The vertical dashed line locates the jamming transition $\phi/\phi_J=1$, the horizonal dashed line denotes $\theta_2=45^\circ$, while the  horizontal solid line denotes $\theta_2=0$.
}
\label{th2D-vs-phiophiJ} 
\end{figure}

Next we consider the orientation angle $\theta_2$ of the nematic director.  In Fig.~\ref{th2D-vs-phiophiJ} we plot $\theta_2$ vs the reduced packing $\phi/\phi_J$ for different asphericities $\alpha$, showing results for the two values of strain rate $\dot\gamma_1<\dot\gamma_2$ (see Table~\ref{tab1} for values).   For an isolated particle, $\theta_2=0$, indicating average alignment parallel to the flow direction $\mathbf{\hat x}$.  As $\phi$ increases from this small $\phi$ isolated particle limit, we see that $\theta_2$ initially goes negative.  Increasing $\phi$ further, $\theta_2$ increases, becomes positive, and upon approaching $\phi_J$ saturates to a value that increases towards $45^\circ$ as $\alpha$ decreases; as $\phi$ gets close to and goes above $\phi_J$, we see a slight decrease in $\theta_2$.   

\begin{figure}
\centering
\includegraphics[width=3.5in]{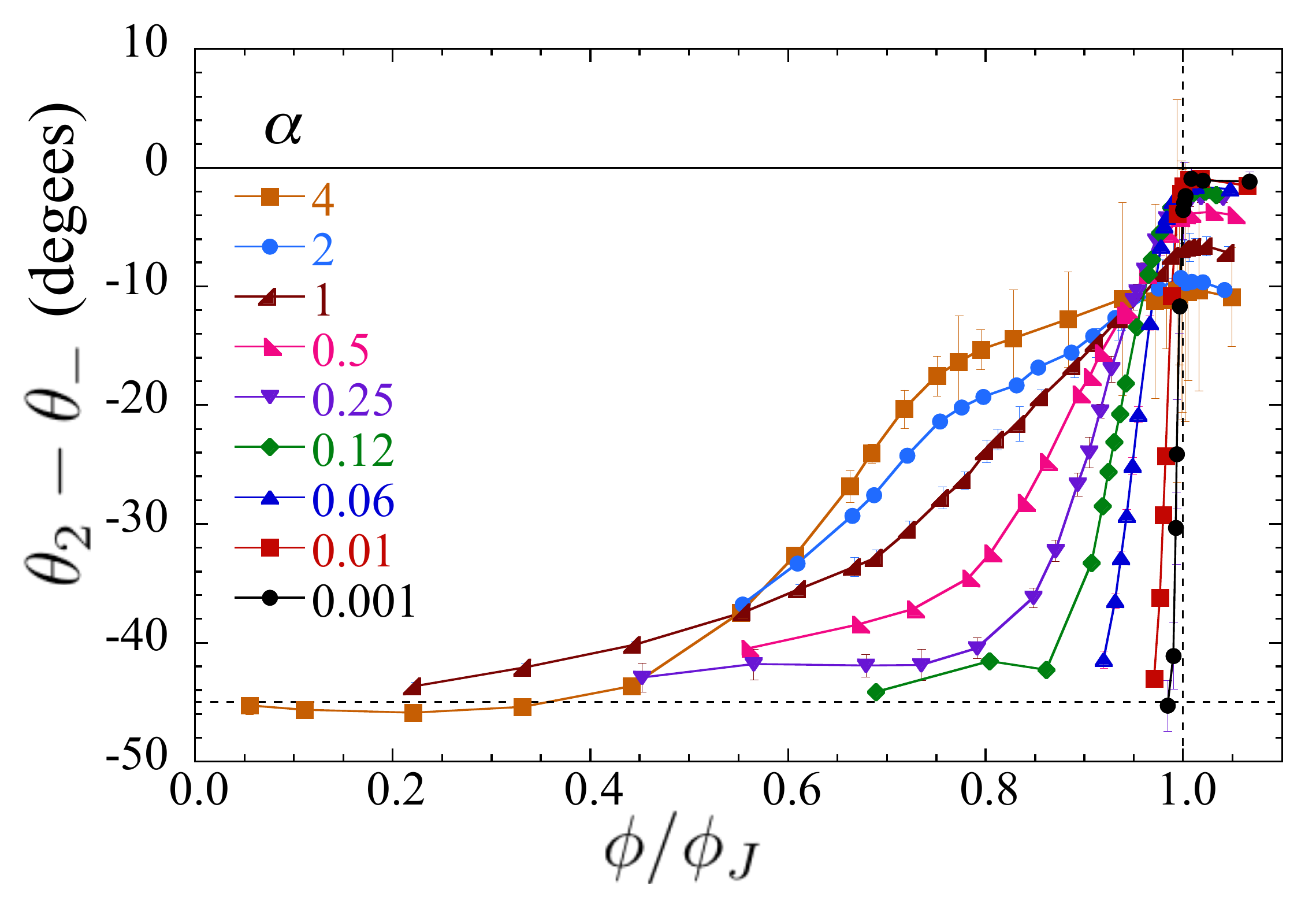}
\caption{Difference between nematic order parameter orientation $\theta_2$ and the orientation of the minimal stress eigenvector $\theta_-$,  vs reduced packing fraction $\phi/\phi_J$ for spherocylinders of different asphericity $\alpha$ at small strain rates $\dot\gamma_1$ (see Table~\ref{tab1} for values).  The vertical dashed line locates the jamming transition $\phi/\phi_J=1$, the horizonal dashed line denotes $\theta_2-\theta_-=-45^\circ$, and the  horizontal solid line denotes $\theta_2-\theta_-=0$.
}
\label{th2X-vs-phiophiJ} 
\end{figure}

While at very small packing $\phi$ the particles tend to align with the flow direction, one might think that, as the particle packing increases, the nematic director would align with the direction of minimal stress.  However we find that this is in general not so.  If $p$ is the pressure and $\sigma$ is the deviatoric shear stress, the orthogonal eigenvectors of the stress tensor, corresponding to eigenvalues $p\pm\sigma$, are oriented at angles $\theta_\pm$ with respect to the flow direction $\mathbf{\hat x}$.  In an earlier work \cite{MT1} we have computed the angle of the minimum stress eigenvector, $\theta_-$.  At small $\phi$ for any $\alpha$ we find $\theta_-\approx 45^\circ$, as it would be for a uniformly sheared continuum.  At dense $\phi$, near and above jamming, we find that $\theta_- \to 45^\circ$ as $\alpha\to 0$, but  otherwise decreases from $45^\circ$ as $\alpha$ increases.  In between, $\theta_-$ can vary non-monotonically as $\phi$ increases.
In Fig.~\ref{th2X-vs-phiophiJ} we plot $\theta_2-\theta_-$ vs $\phi$ for different $\alpha$, at the strain rate $\dot\gamma_1$ (see Table~\ref{tab1} for values).  We see that only for the smaller values $\alpha\lesssim 0.25$, and only approaching  $\phi_J$ and going above, do we find $\theta_2\approx \theta_-$, i.e. the nematic order parameter is aligning close to the minimum stress direction.  

In Appendix \ref{aOrientations} we discuss further properties of  particle orientations.  By considering the distribution of particle orientations $\mathcal{P}(\theta_i)$, we  show that the angle $\theta_2$ of the nematic order parameter  is  in general not equal to the most likely particle orientation, determined by the maximum in $\mathcal{P}(\theta_i)$, although the two are close.


\subsection{Time Dependence of Particle Rotations}
\label{stimedep}

A principle result of the preceding two sections is the observation that $-\langle\dot\theta_i\rangle/\dot\gamma$ and $S_2$  both vary non-monotonically as the packing $\phi$ increases.  In this section we provide a physical understanding of this behavior by demonstrating that the minimum in $-\langle\dot\theta_i\rangle/\dot\gamma$ represents a crossover from  small packings $\phi$, where particle rotations are qualitatively like the periodic rotations of an isolated particle (perturbed by inter-particle collisions), to  large packings $\phi$,  where the geometry of the dense packing becomes the dominant factor influencing rotations, which then behave similar to a random Poisson process.  We will show this by considering the distribution of strain intervals $\Delta\gamma$ between successive rotations of a particle by $\pi$.

\begin{figure}
\centering
\includegraphics[width=3.5in]{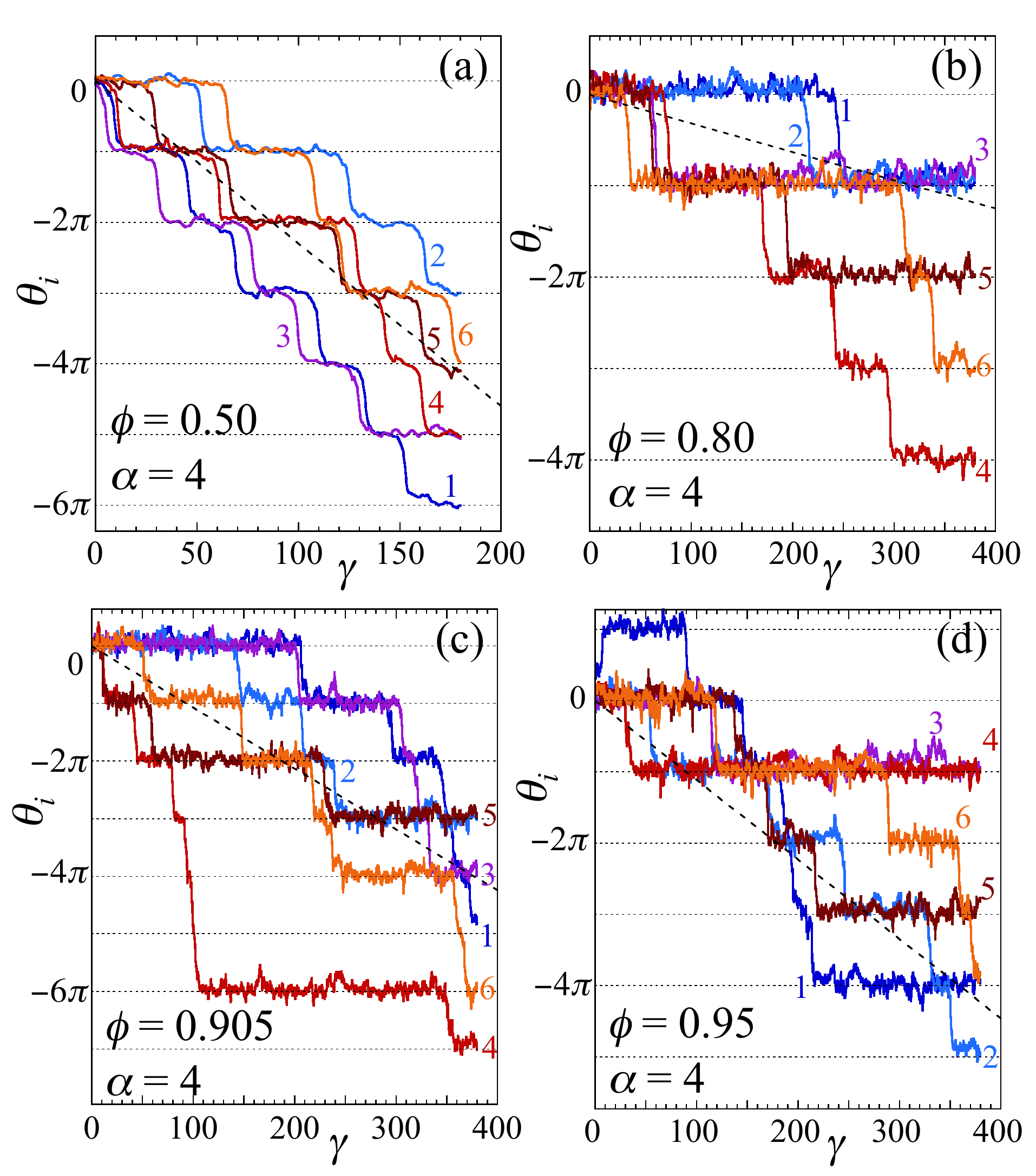}
\caption{For spherocylinders of asphericity $\alpha=4$ at strain rate $\dot\gamma=10^{-5}$, particle orientation $\theta_i$ vs net strain $\gamma=\dot\gamma t$ for six randomly selected particles at packings (a) $\phi=0.50$, (b) $\phi=0.80$, (c) $\phi=0.905\approx \phi_J$, and (d) $\phi=0.95$.  In each case particles 1, 2 and 3 are big particles, while 4, 5 and 6 are small particles.  The diagonal dashed lines indicate the average rotation, $\theta_i=[\langle\dot\theta_i\rangle/\dot\gamma]\gamma$.
}
\label{theta-a4} 
\end{figure}

In Sec.~\ref{sAV} we discussed  the average angular velocity of individual particle rotations, $-\langle\dot\theta_i\rangle/\dot\gamma$.  Now we  consider  the time evolution of a particle's rotation.  We consider first the case of elongated particles with $\alpha=4$.  In Fig.~\ref{theta-a4} we plot $\theta_i(\gamma)$ vs $\gamma=\dot\gamma t$ for six randomly selected particles, three big and three small, at several different packing fractions $\phi$ and $\dot\gamma=10^{-5}$.  The average motion, $\theta_i = [\langle\dot\theta_i\rangle/\dot\gamma]\gamma$, is indicated by the dashed diagonal line. Comparing Fig.~\ref{theta-a4} with the corresponding curve for a isolated particle shown in Fig.~\ref{theta-vs-g}(a), we see a general similarity:  There are plateaus near integer values $\theta_i=-n\pi$, separated by regions where $\theta_i$ rapidly transitions by an amount $-\pi$, representing a clockwise flipping of the orientation of the particle.  Upon further inspection, however, there are two important differences.  For the case of the isolated particle in Fig.~\ref{theta-vs-g}(a), the plateaus show a small downwards slope due to the finite angular velocity $\dot\theta_i/\dot\gamma =d\theta_i/d\gamma =- f(0)=-[1-(\Delta I_i/I_i)]/2$ when the particle is oriented parallel to the flow.  In Fig.~\ref{theta-a4} however, the plateaus appear on average to be mostly flat.  For the isolated particle, the jumps in $\theta_i$ by $-\pi$, as the particle flips orientation, occur in a perfectly periodic fashion.  In Fig.~\ref{theta-a4} however, the timing between such jumps appears to be more random.  In the densest system at $\phi=0.95 >\phi_J$, shown in Fig.~ \ref{theta-a4}(d), we also see that particle 1 makes a counterclockwise flip of $+\pi$ at small $\gamma$; However for $\alpha=4$ these counterclockwise flips are rare events, occurring infrequently for $\phi=0.95$, and even less so for smaller $\phi$, over the length of our simulations.

In  Fig.~\ref{theta-a4} we see that the average value of $\theta_i$ on these plateaus lies slightly above the values $-n\pi$ at the larger values of $\phi$; the particles are thus at some small finite angle $[\theta_i \text{ modulo } \pi]>0$ with respect to the flow direction.  This is a consequence of the increasing orientation angle of the nematic director $\theta_2$ as $\phi$ increases, as shown in Fig.~\ref{th2D-vs-phiophiJ}. We also see that the fluctuations about the plateaus tend to increase as $\phi$ increases.   This is a consequence of the broadening of the distribution of orientations $\mathcal{P}(\theta_i)$ as $\phi$ increases, as shown in  Appendix~\ref{aOrientations}.  

\begin{figure}
\centering
\includegraphics[width=3.5in]{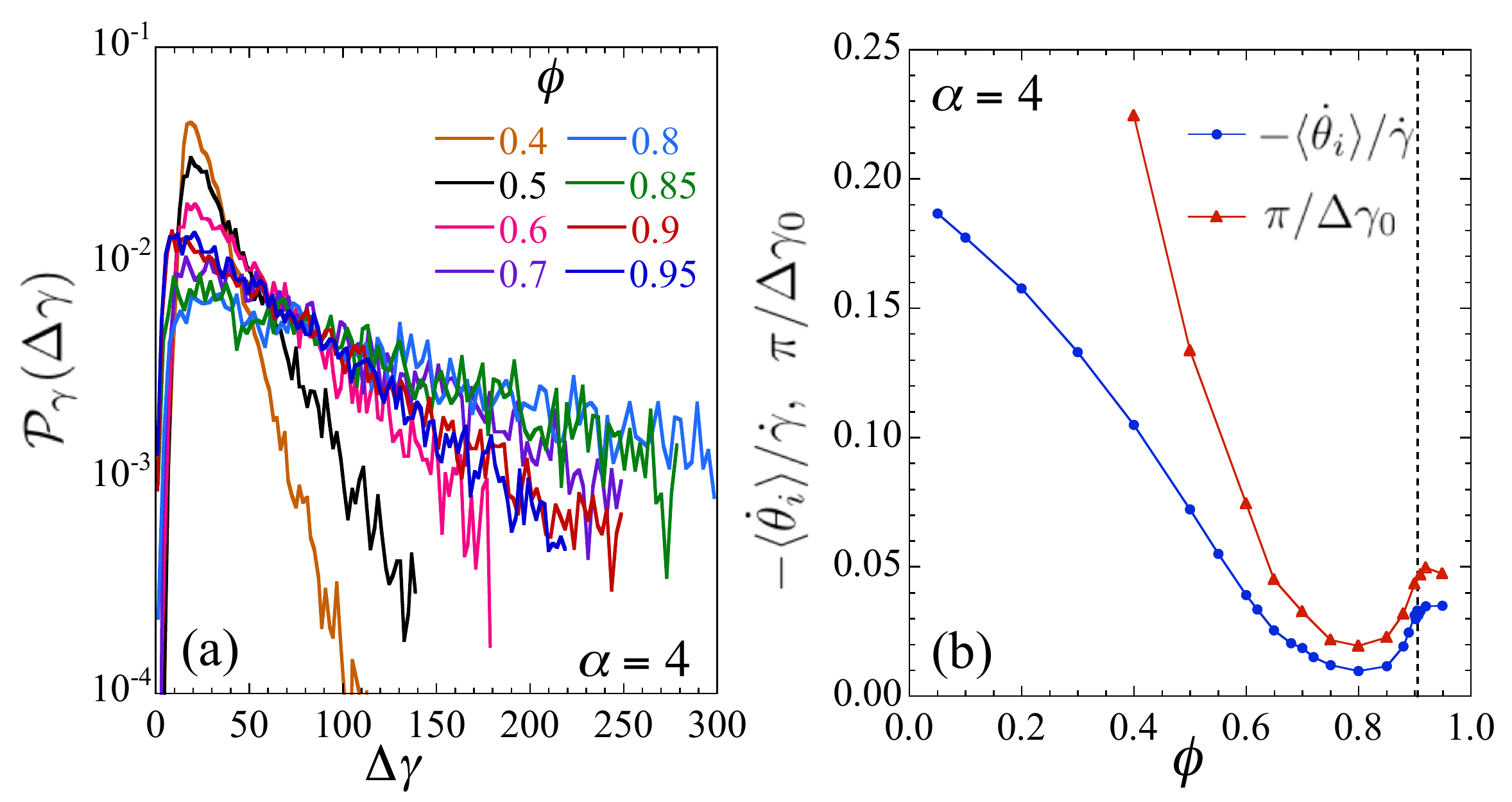}
\caption{For spherocylinders of asphericity $\alpha=4$ at strain rate $\dot\gamma=10^{-5}$: (a) Distribution $\mathcal{P}_\gamma(\Delta\gamma)$ of the strain interval $\Delta\gamma=\dot\gamma\Delta t$ between successive clockwise rotations of a particle by $\pi$ for different packings $\phi$.
(b) With $\Delta\gamma_0$ obtained from fitting the exponentially decaying large $\Delta\gamma$ tail of  $\mathcal{P}_\gamma$ to $\exp[-\Delta\gamma/\Delta\gamma_0]$, a comparison of $\pi/\Delta\gamma_0$ vs the average particle angular velocity $-\langle\dot\theta_i\rangle/\dot\gamma$.  The vertical dashed line locates the jamming $\phi_J$.
}
\label{flipHist-a4} 
\end{figure}

Measuring the strain $\Delta\gamma$ between two successive rotational flips of a particle by $-\pi$, we plot the distribution  $\mathcal{P}_\gamma(\Delta\gamma)$ vs $\Delta\gamma$ for different $\phi$ at fixed $\dot\gamma=10^{-5}$ in Fig.~\ref{flipHist-a4}(a).  For the smaller values of $\phi$ we find that $\mathcal{P}_\gamma$ peaks at the value $\Delta\gamma\approx 16$, which is the same as the strain interval between the periodic flips by $-\pi$ for an isolated particle, as seen in Fig.~\ref{theta-vs-g}(a); however as $\phi$ increases, the distribution broadens and is increasingly skewed towards values on the large $\Delta\gamma$ side of the peak.  As $\phi$ increases further, we see that the location of the peak in $\mathcal{P}_\gamma$ steadily shifts to smaller values of $\Delta\gamma$ and the large $\Delta\gamma$ tail of the distribution becomes exponential, as seen by the roughly linear decrease of the distributions on our semi-log plot.  This exponential waiting time between flips, $\Delta t=\Delta\gamma/\dot\gamma$, suggests that at large $\phi$ particle flips are a Poisson-like process, and that, aside from an initial waiting time corresponding to the rise of $\mathcal{P}_\gamma$ to its peak, the time until the next particle flip is independent of how long the particle has spent since its last flip.  
Thus, unlike  the case of an isolated particle for which the particle undergoes periodic rotation  with a non-uniform angular velocity, here our results  suggest a scenario in which,  as the particle density increases, the reduced free volume  between particles blocks particle rotations, leaving 
particles to spend most of their time having small angular deflections about a fixed value.  Then, after some random strain $\Delta\gamma$, a local rearrangement appears that allows the particle to rotate rapidly through $\Delta\theta_i=-\pi$.  The exponential distribution of the waiting times implies that the appearance of such local rearrangements are uncorrelated, except for a minimal waiting time.

Fitting the large $\Delta\gamma$ tail of the distribution to $\mathcal{P}_\gamma\propto\exp[-\Delta\gamma/\Delta\gamma_0]$, we determine the  rate of particle flips $1/\Delta\gamma_0$.  This rate, which is just the slope of the linearly decreasing distributions in the semi-log plot of Fig.~\ref{flipHist-a4}(a), is seen to be non-monotonic in $\phi$, reaching a minimum value near $\phi\approx 0.80$. 
In Fig.~\ref{flipHist-a4}(b) we plot this rate as $\pi/\Delta\gamma_0$ vs $\phi$ and compare it to the average angular velocity $-\langle\dot\theta_i\rangle/\dot\gamma$, shown previously in Fig.~\ref{av_v_phi-allgdot}(a).  If the $\mathcal{P}_\gamma$ were exactly  exponential distributions, these two curves would be equal.  But $\mathcal{P}_\gamma$ is not precisely exponential, due to the waiting time represented by the rise of $\mathcal{P}_\gamma$ to its peak value.  Because of this waiting time we expect $\langle\Delta\gamma\rangle > \Delta\gamma_0$, and so $-\langle\dot\theta_i\rangle/\dot\gamma=\pi/\langle\Delta\gamma\rangle$ will lie below $\pi/\Delta\gamma_0$, as we indeed find to be the case.  Nevertheless we see that at the larger $\phi$, $\pi/\Delta\gamma_0$ behaves qualitatively the same as $-\langle\dot\theta_i\rangle/\dot\gamma$, with a similar minimum around $\phi_{\dot\theta\,\mathrm{min}}\approx 0.80$; the difference between the two curves becomes greatest as $\phi$ decreases below the minimum.  

We thus form the following picture.  At small $\phi$ particles behave similarly to isolated particles, with the typical strain $\Delta\gamma$ between particle flips being roughly equal to that of an isolated particle, but with random fluctuations due to particle collisions; these fluctuations are skewed to larger $\Delta\gamma$ thus causing the decrease in $-\langle\dot\theta_i\rangle/\dot\gamma$.  The average $\langle\Delta\gamma\rangle$ at these small $\phi$ is significantly different from the $\Delta\gamma_0$ that describes the large $\Delta\gamma$ tail of the distribution.  As $\phi$ increases however, the flips become more of a Poisson-like process in which the average time until the next particle flip is independent of the time since the last flip.  The exponential part of the distribution $\mathcal{P}_\gamma$ dominates the behavior and $\Delta\gamma_0$ gives a qualitative explanation for the average angular velocity $-\langle\dot\theta_i\rangle/\dot\gamma$ in the range of $\phi$ approaching the minimum $\phi_{\dot\theta\,\mathrm{min}}$ and going above.

Note, although we described the rotations by $\pi$ in Figs.~\ref{theta-a4}(c) and \ref{theta-a4}(d) as ``rapid,"  this is meant as rapid relative to the strain interval $\Delta\gamma$ between successive particle rotations.  Upon closer examination, a particle rotation  takes place over a typical strain scale of $\delta\gamma\sim 5$; this is roughly the strain needed for particles of tip-to-tip length $5D_s$, in neighboring rows parallel to the flow direction, to slide past one another.  Thus the entire configuration has undergone substantial change over the time it takes the particle to rotate; moreover, although as we will argue later there is no long range coherence in particle motion, there are strong correlations in particle motion on short length scales.  It is  therefore not obvious to visually identify the 
particular configurational fluctuations that facilitate particle rotations.


\begin{figure}
\centering
\includegraphics[width=3.5in]{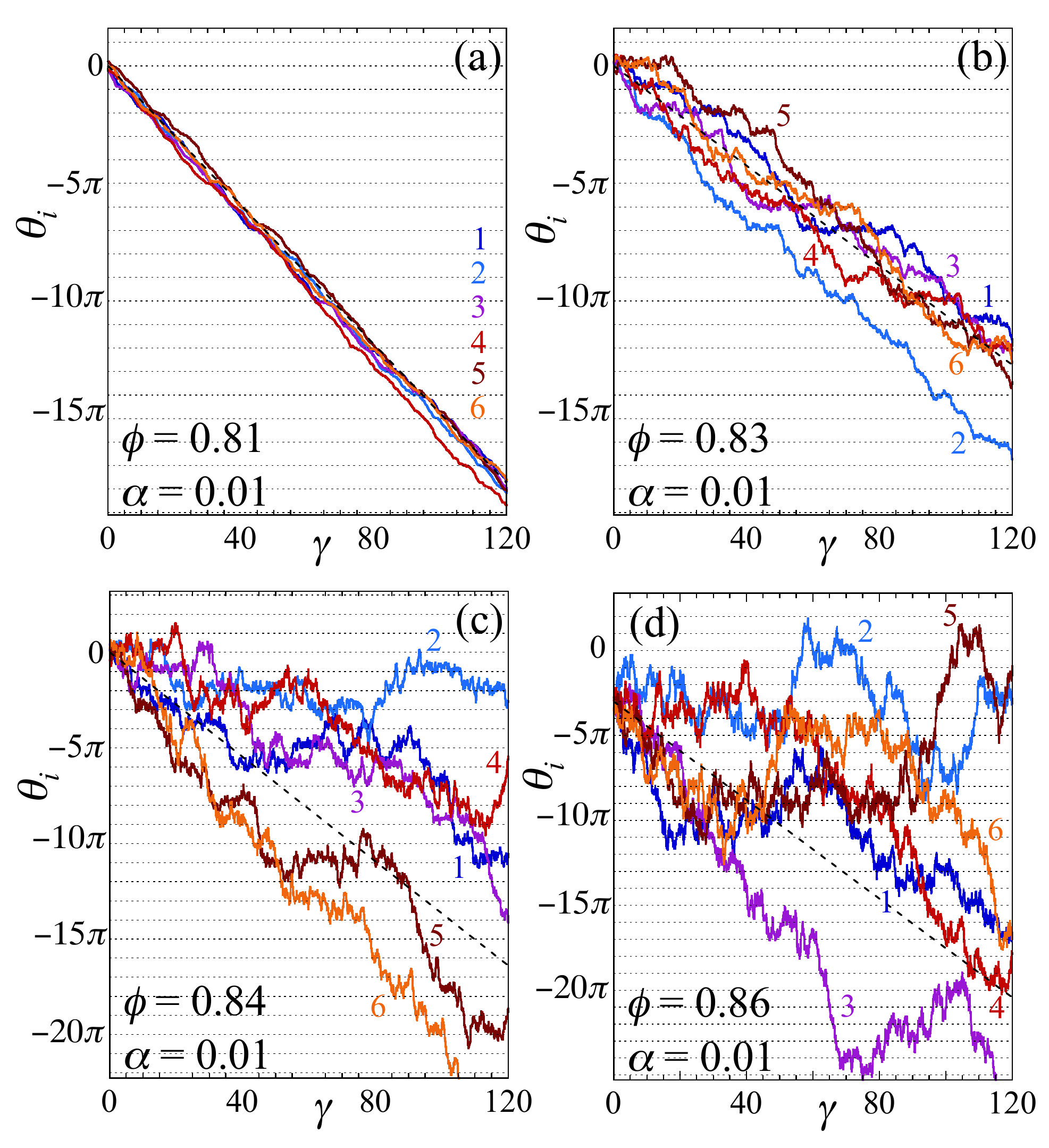}
\caption{For spherocylinders of asphericity $\alpha=0.01$ at strain rate $\dot\gamma=10^{-6}$, particle orientation $\theta_i$ vs net strain $\gamma=\dot\gamma t$ for six randomly selected particles at packings (a) $\phi=0.81$, (b) $\phi=0.83$, (c) $\phi=0.84\approx \phi_J=0.845$, and (d) $\phi=0.86$.  In each case particles 1, 2 and 3 are big particles, while 4, 5 and 6 are small particles.  The dashed lines indicate the average rotation, $\theta_i=[\langle\dot\theta_i\rangle/\dot\gamma]\gamma$.
}
\label{theta-a01} 
\end{figure}

Next we consider the case of nearly circular particles with $\alpha=0.01$.  For an isolated particle, $\Delta I_i/I_i=0.0085$ is so small that a plot of $\theta_i$ vs $\gamma$ would look like a straight line of slope $-1/2$; no plateaus are observable to the eye.  In Fig.~\ref{theta-a01} we plot $\theta_i(\gamma)$ vs $\gamma=\dot\gamma t$ for  six randomly selected particles, three big and three small, at several different packing fractions $\phi$ and $\dot\gamma=10^{-6}$.  The average motion, $\theta_i = [\langle\dot\theta_i\rangle/\dot\gamma]\gamma$, is indicated by the dashed diagonal line.   For $\phi=0.81$, below the minimum  in $-\langle\dot\theta_i\rangle/\dot\gamma$ at $\phi_{\dot\theta\,\mathrm{min}}$ (see Fig.~\ref{av_v_phi-allgdot}(b)), we see in  Fig.~\ref{theta-a01}(a) small fluctuations about the isolated particle behavior.  For $\phi=0.83\approx \phi_{\dot\theta\,\mathrm{min}}$ in  Fig.~\ref{theta-a01}(b), near the minimum in $-\langle\dot\theta_i\rangle/\dot\gamma$, we see larger fluctuations, some small isolated plateaus where particles stay at a fixed orientation, but for the most part particles are rotating nearly uniformly.  However, for $\phi=0.84$ in Fig.~ \ref{theta-a01}(c), just below the jamming $\phi_J=0.845$, and for $\phi=0.86$ in  Fig.~\ref{theta-a01}(d), above $\phi_J$, we see dramatically different behavior.  Fluctuations are now extremely large, and rotation is highly non-uniform.  Compared to Fig.~\ref{theta-a4} for $\alpha=4$, here it is hard to identify clear plateaus, and there is considerable counterclockwise rotation (where $\theta_i$ increases with increasing $\gamma$) in addition to clockwise rotation  (where $\theta_i$ decreases with increasing $\gamma$).

Nevertheless, we can still carry out an analysis of flipping times in analogy with what we did for $\alpha=4$ in Fig.~\ref{flipHist-a4}.  If we denote by $\gamma_1$ the strain at which a given particle trajectory first passes through $\theta_i=-(n+1/2)\pi$ upon rotating clockwise, and by $\gamma_2$  the strain at which it next passes through $\theta_i=-(n+3/2)\pi$, then $\Delta\gamma_-=\gamma_2-\gamma_1$ can be taken as the net strain displacement over which the particle has flipped its orientation, rotating clockwise through an angle $\pi$.  In a similar way we can determine $\Delta\gamma_+$, the net strain displacement for the particle to flip its orientation rotating counterclockwise through an angle $\pi$.  

\begin{figure}
\centering
\includegraphics[width=3.5in]{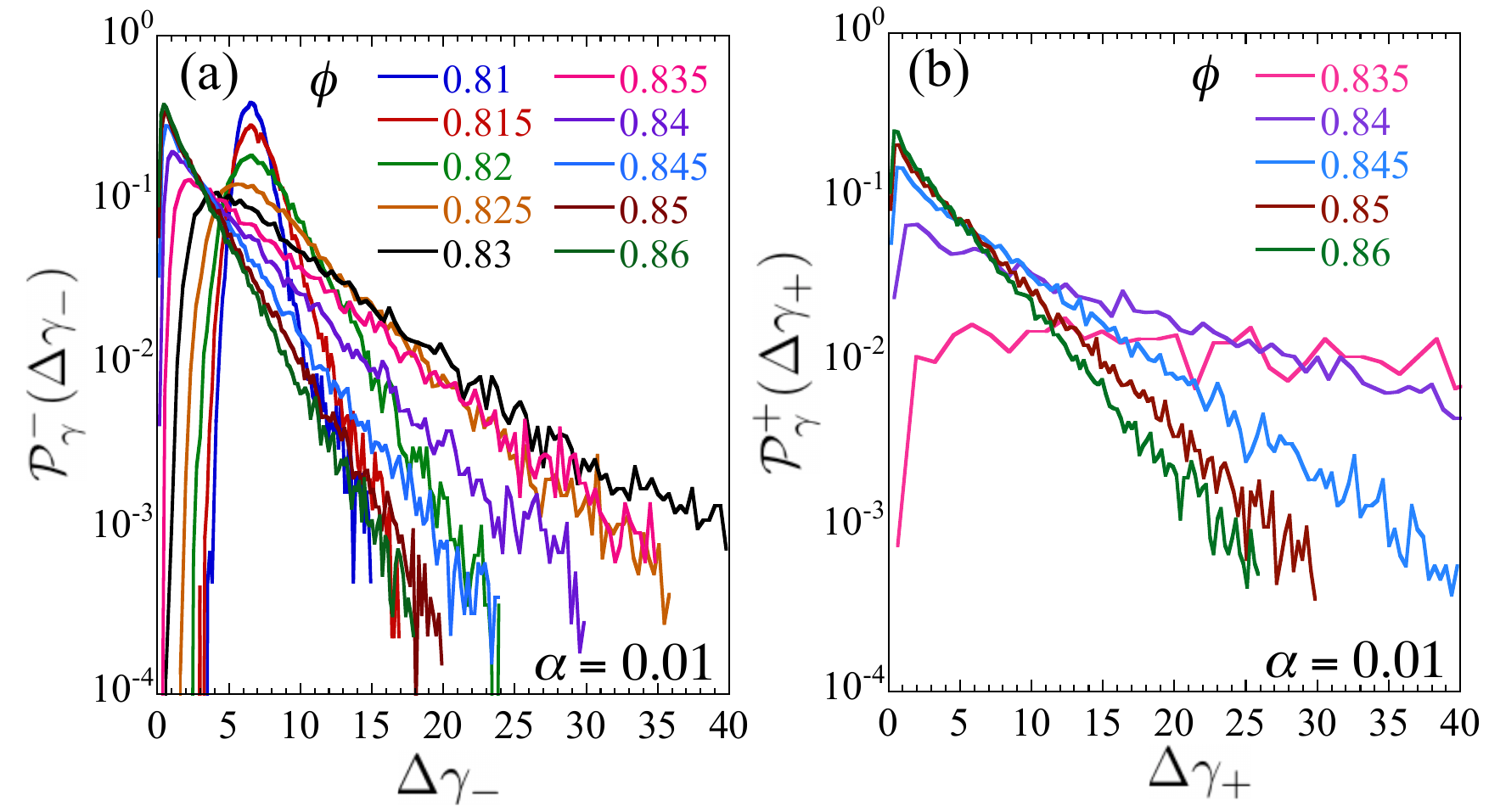}
\caption{For spherocylinders of asphericity $\alpha=0.01$ at strain rate $\dot\gamma=10^{-6}$: Distributions (a) $\mathcal{P}^-_\gamma(\Delta\gamma_-)$ for the strain interval $\Delta\gamma_-$ between successive clockwise rotations of a particle by $\pi$ for different packings $\phi$, and 
(b) $\mathcal{P}^+_\gamma(\Delta\gamma_+)$ for the strain interval $\Delta\gamma_+$ between successive counterclockwise rotations of a particle by $\pi$ for different packings $\phi$, 
}
\label{flipHist-a01} 
\end{figure}

In Figs.~\ref{flipHist-a01}(a) and \ref{flipHist-a01}(b) we plot the distributions $\mathcal{P}_{\gamma}^{-}(\Delta\gamma_-)$ for clockwise flips, and $\mathcal{P}^{+}_{\gamma}(\Delta\gamma_+)$ for counterclockwise flips, respectively, for different packings $\phi$ at $\dot\gamma=10^{-6}$.  Despite the qualitative differences in the trajectories $\theta_i(\gamma)$ for $\alpha=0.01$, shown in Fig.~\ref{theta-a01}, from those for $\alpha=4$, shown in Fig.~\ref{theta-a4}, the distribution $\mathcal{P}_\gamma^-$ for $\alpha=0.01$ shows the same qualitative behavior as the $\mathcal{P}_\gamma$ found for $\alpha=4$ in Fig.~\ref{flipHist-a4}(a).  For small $\phi\lesssim 0.82$, the peak in $\mathcal{P}_\gamma^-$ lies close to $\Delta\gamma_-\approx 6.3$, which is the same as the strain interval between the periodic rotations by $\pi$ of an isolated particle.  However as $\phi$ increases, approaching the minimum in $-\langle\dot\theta_i\rangle/\dot\gamma$ at  $\phi_{\dot\theta\,\mathrm{min}}\approx 0.83$, the distribution broadens and an exponential tail appears on the large $\Delta\gamma_-$ side of the peak.  As $\phi$ increases above $0.83$ the location of the peak in $\mathcal{P}_\gamma^-$ shifts towards smaller $\Delta\gamma_-$ and the exponential tails grow, until at our largest values of $\phi$ the distribution $\mathcal{P}_\gamma^-$ is almost a pure exponential.  Fitting to the large $\Delta\gamma_-$ tail of $\mathcal{P}_\gamma^-$ we determine the exponential rate $1/\Delta\gamma_{0-}$, which is just the slope of the linearly decreasing distributions in the semi-log plot of Fig.~\ref{flipHist-a01}(a).  We see that this rate is non-monotonic, having its smallest value at $\phi\approx 0.83\approx \phi_{\dot\theta\,\mathrm{min}}$ where the average angular velocity $-\langle\dot\theta_i\rangle/\dot\gamma$ is minimum.

\begin{figure}
\centering
\includegraphics[width=3.5in]{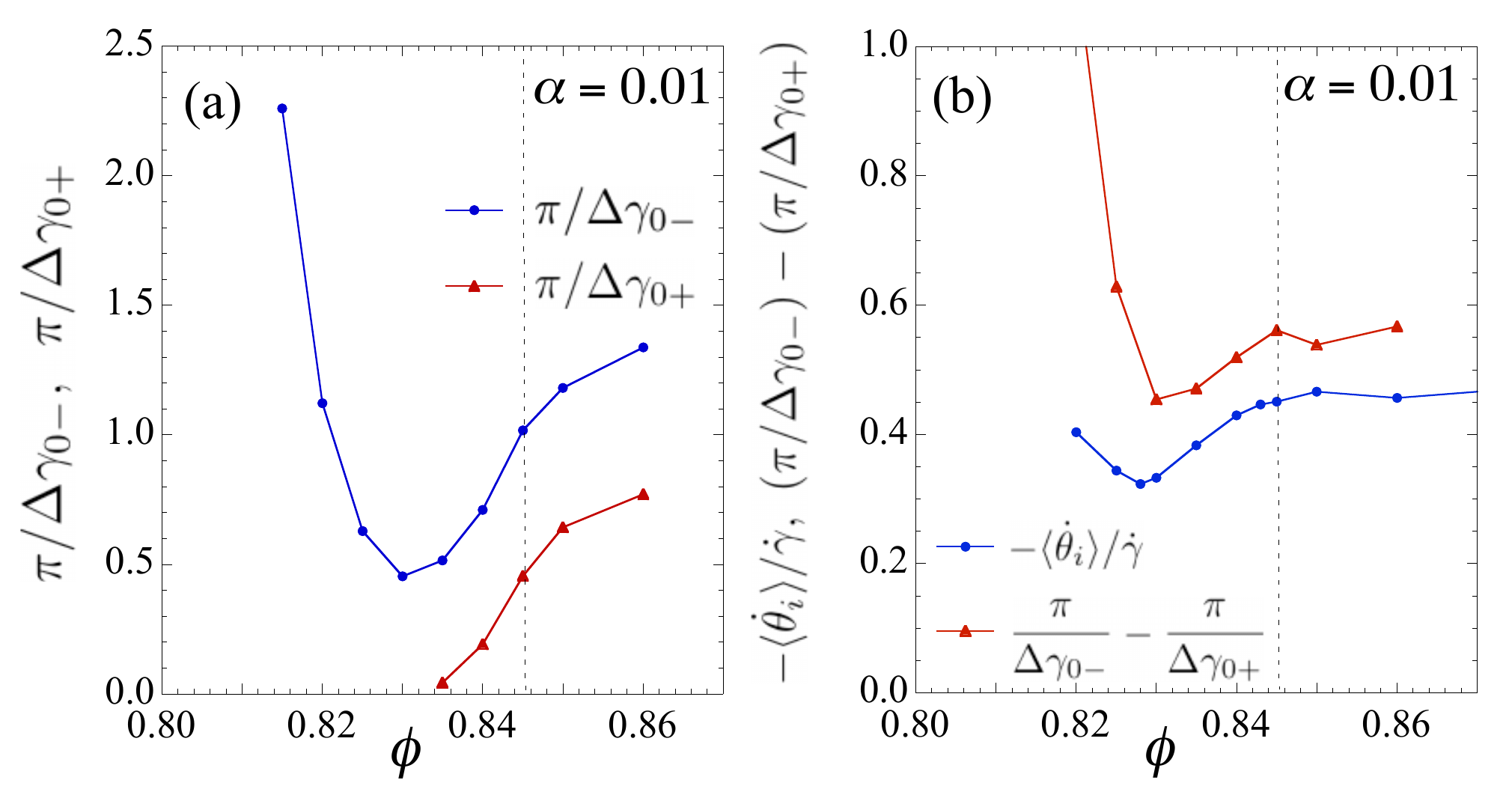}
\caption{For spherocylinders of asphericity $\alpha=0.01$ at strain rate $\dot\gamma=10^{-6}$: (a) Rates $\pi/\Delta\gamma_{0-}$ and $\pi/\Delta\gamma_{0+}$ characterizing the exponential tails of the distributions $\mathcal{P}_\gamma^-$ and $\mathcal{P}_\gamma^+$ for the wait times for clockwise and counterclockwise rotations of a particle by $\pi$, and (b) average particle angular velocity $-\langle\dot\theta_i\rangle/\dot\gamma$ compared to $(\pi/\Delta\gamma_{0-})-(\pi/\Delta\gamma_{0+})$ vs packing $\phi$.  The dashed vertical line locates the jamming $\phi_J$.
}
\label{flipRate-a01} 
\end{figure}

For counterclockwise rotations, we see that the
distributions of $\mathcal{P}_\gamma^+$, shown  in Fig.~\ref{flipHist-a01}(b), are close to exponential, with a rate that rapidly decreases as $\phi$ decreases from above jamming towards the $\phi_{\dot\theta\,\mathrm{min}}\approx0.83$ that locates the minimum in $-\langle\dot\theta_i\rangle/\dot\gamma$.  For $\phi<0.835$, counterclockwise rotations are so rare over the length of our simulation runs that  we are unable to determine the distribution $\mathcal{P}_\gamma^+$ at such small $\phi$.  For $\phi\ge0.835$ we fit  the large $\Delta\gamma_+$ tails of $\mathcal{P}_\gamma^+$ to determine the exponential rate $1/\Delta\gamma_{0+}$.  In Fig.~\ref{flipRate-a01}(a) we plot the clockwise and counterclockwise rates as $\pi/\Delta\gamma_{0-}$ and $\pi/\Delta\gamma_{0+}$ vs $\phi$.  As found for $\pi/\Delta\gamma_0$ for $\alpha=4$ in Fig.~\ref{flipHist-a4}(b), we see that $\pi/\Delta\gamma_{0-}$ has a minimum at $\phi=0.83\approx \phi_{\dot\theta\,\mathrm{min}}$ where $-\langle\dot\theta_i\rangle/\dot\gamma$ is minimum.  In contrast, $\pi/\Delta\gamma_{0+}$ is getting small, and perhaps vanishing,  as $\phi\to0.83$ from above.  

If the distributions $\mathcal{P}_\gamma^-$ and $\mathcal{P}_\gamma^+$ were exactly exponential, then the average angular velocity would just be $(\pi/\Delta\gamma_{0-})-(\pi/\Delta\gamma_{0+})$.  In Fig.~\ref{flipRate-a01}(b) we compare this quantity with the exactly computed $-\langle\dot\theta_i\rangle/\dot\gamma$, plotting both vs the packing $\phi$.  As for the case of spherocylinders with $\alpha=4$, shown in Fig.~\ref{flipHist-a4}(b), we see that these two curves qualitatively agree upon approaching the minimum at $\phi_{\dot\theta\,\mathrm{min}}=0.83$ and going above, but they quickly separate as $\phi$ decreases below 0.83.  As with $\alpha=4$, the difference between the two curves results from the fact that the distributions $\mathcal{P}_\gamma^-$ and $\mathcal{P}_\gamma^+$ are not  exactly exponential, with $\langle\Delta\gamma_{\pm}\rangle > \Delta\gamma_{0\pm}$ due to the rise of the distributions to their peak at a finite $\Delta\gamma_\pm$; this difference becomes most pronounced at the smaller $\phi < 0.83$.

Our analysis of spherocylinders with both $\alpha=4$ and $\alpha=0.01$ thus points to a common scenario.  The minimum in $-\langle\dot\theta_i\rangle/\dot\gamma$ at $\phi_{\dot\theta\,\mathrm{min}}$ results from a crossover between two different types of behavior as $\phi$ varies.  For $\phi\ll\phi_{\dot\theta\,\mathrm{min}}$, particles behave qualitatively like isolated particles.  While an isolated particle will have perfectly periodic rotations by $\pi$ given by a strain period of  $\Delta\bar\gamma=2\pi/\sqrt{1-(\Delta I_i/I_i)^2}$ (see Eq.~(\ref{eomegasingle})), the interacting particles will have a distribution of $\Delta\gamma$ that peaks near $\Delta\bar\gamma$ but has a finite width, with a skew to the large $\Delta\gamma$ side of the peak; the width of the distribution and the skew increase as $\phi$ increases, giving a decreasing $-\langle\dot\theta_i\rangle/\dot\gamma$.  This effect is presumably a result of the reduction in free volume  between the particles as $\phi$ increases, thereby inhibiting rotations.
For $\phi\gtrsim\phi_{\dot\theta\,\mathrm{min}}$, however, the distribution peak shifts down towards zero, and the distribution becomes increasingly exponential, as $\phi$ increases.  This exponential distribution suggests that rotations by $\pi$ become a Poisson-like process;  particles in general fluctuate about fixed orientations, while flips with a $\pi$ rotation occur at uncorrelated random times set by a rate $1/\Delta\gamma_0$.  The time until the next flip is largely independent of the time since the last flip, except for a minimum waiting time.  As $\phi$ increases above $\phi_{\dot\theta\,\mathrm{min}}$, the flipping rate $1/\Delta\gamma_0$ increases and so $-\langle\dot\theta_i\rangle/\dot\gamma$ increases.  

\subsection{Pure vs Simple Shearing}
\label{sPure}

In this section we present another analysis that again suggests that the non-monotonic behavior of $-\langle\dot\theta_i\rangle/\dot\gamma$ and $S_2$, as $\phi$ increases, results from a crossover from single particle like behavior to behavior dominated by the geometry of the dense packing.  Our analysis here focuses on the magnitude of the nematic order parameter $S_2$.  Our results  will also offer an explanation for the singular behavior reported in our earlier Letter \cite{MKOT}, in which we found for simple shearing  that as $\alpha\to 0$, and particles approach a circular shape, $S_2$ vanishes for $\phi<\phi_J$ but $S_2$ remains finite at and just above $\phi_J$.

All the results elsewhere in this paper involve the  behavior of our system under  simple shearing.  Here, however, we consider the behavior of our system under pure shearing.  As we discuss below, the behavior of an isolated single particle is dramatically different under pure vs simple shearing.  We will find that the behavior of $S_2$ of our many particle system is similarly qualitatively different for pure vs simple shearing at small packings, but that they are qualitatively the same at large packings, thus suggesting the crossover described above.

In our model, dissipation arises due to a viscous drag between the local velocity of the particle and the local velocity $\mathbf{v}_\mathrm{host}(\mathbf{r})$ of the suspending host medium.  For simple shear in the $\mathbf{\hat x}$ direction, $\mathbf{v}_\mathrm{host}(\mathbf{r})=\dot\gamma y\mathbf{\hat x}$.  For a more general linear deformation of the host medium we can write,
\begin{equation}
\mathbf{v}_\mathrm{host}(\mathbf{r})=\dot{\boldsymbol{\Gamma}}\cdot\mathbf{r},
\end{equation}
with $\dot{\boldsymbol{\Gamma}}$ the strain rate tensor.
For simple shear we can write,
\begin{equation}
\dot{\boldsymbol{\Gamma}}_\mathrm{ss}=
\left[
\begin{array}{cc}
0 & \dot\gamma\\
0 & 0
\end{array}
\right]
=
\left[
\begin{array}{cc}
0 & \dot\gamma/2\\
\dot\gamma/2 & 0
\end{array}
\right]
+\left[
\begin{array}{cc}
0 & \dot\gamma/2\\
-\dot\gamma/2 & 0
\end{array}
\right].
\label{epures}
\end{equation}
The first term on the right most side of Eq.~(\ref{epures}) represents a pure shear distortion, in which the host medium is expanded in the $\mathbf{\hat x}+\mathbf{\hat y}$ direction, while being compressed in the $\mathbf{\hat x}-\mathbf{\hat y}$ direction, both at a rate $\dot\gamma/2$, so as to preserve the system area.  The second term represents a clockwise rotation $(-\dot\gamma/2)\mathbf{\hat z}\times \mathbf{r}$,  with angular velocity $-\dot\gamma/2$.  Thus a simple shear can be viewed as the sum of a pure shear and a rotation.
It is this rotational part which gives rise to the constant term  $1/2$ in the angular driving function $f(\theta)$ of Eq.~(\ref{eftheta}), while the pure shear part gives rise to the $\cos 2\theta$ term.  It is the rotational part that 
drives the continuous rotation of particles under simple shear, resulting in the finite $-\langle\omega_{zi}\rangle/\dot\gamma >0$ found in steady-state, as seen in Fig.~\ref{dthdg-vs-phiophiJ}.  Studying pure shear thus allows us to study the orientational ordering of the system in the absence of the rotational drive.

For our pure shear simulations we  choose $\mathbf{\hat x}$ as the expansive direction and $\mathbf{\hat y}$ as the compressive direction, using periodic boundary conditions in both directions. In this case, the translational and rotational equations of motion for pure shear become,
\begin{equation}
\dot{\mathbf{r}}_i=\dfrac{\dot\gamma}{2}[x_i\mathbf{\hat x}-y_i\mathbf{\hat y}] +\dfrac{\mathbf{F}^\mathrm{el}_i}{k_d\mathcal{A}_i},
\end{equation}
\begin{equation}
\dot\theta_i=-\dfrac{\dot\gamma}{2}\dfrac{\Delta I_i}{I_i}\sin 2\theta_i + \dfrac{\tau_i^\mathrm{el}}{k_d \mathcal{A}_i I_i}.
\end{equation}
For an isolated particle, where $\tau_i^\mathrm{el}=0$, one can solve the rotational equation of motion analytically,
\begin{equation}
|\tan\theta_i(t)| = \mathrm{e}^{-\dot\gamma t \Delta I_i/I_i}|\tan \theta_i(0)|.
\label{ethetatps}
\end{equation}
An isolated particle will   relax  exponentially to $\theta_i=0$ or $\pi$ with a relaxation time $t_\mathrm{relax}$ set by a total strain $\gamma_\mathrm{relax}=\dot\gamma t_\mathrm{relax} = I_i/\Delta I_i$.  Unlike simple shearing, there is no continuing rotation of the particle.
Thus, for an isolated particle under pure shearing, we  find   perfect nematic ordering with $S_2=1$ and $\theta_2=0$ for particles of {\em any} asphericity $\alpha$.  This is in contrast to the behavior under simple shearing where, due to continuing particle rotation, Eq.~(\ref{eS2single}) gives  $S_2 < 1$.  

This difference between pure and simple shearing  is most dramatic for the case of a nearly circular particle with small $\alpha$.
For small $\alpha$, Eq.~(\ref{eDI}) gives a small $\Delta I_i/I_i\sim \alpha$.  For pure shearing, an isolated particle will relax to perfect ordered alignment with the minimal stress direction, $S_2=1$ and $\theta_2=0$, although the 
relaxation strain  to achieve that ordered state, $\gamma_\mathrm{relax}=I_i/\Delta I_i\sim 1/\alpha$, grows large as $\alpha$ decreases.  For simple shearing, however, an isolated particle with small $\alpha$ will continue to rotate, with a nearly uniform angular velocity $\dot\theta_i\approx -\dot\gamma/2$, so that Eq.~(\ref{eS2single}) gives $S_2\sim \Delta I_i/I_i\sim \alpha$, which thus vanishes as $\alpha$ decreases to zero.

To investigate the response to pure shear at a finite packing $\phi$, in particular near and above jamming, we carry out numerical simulations. 
Unlike simple shear, where the system lengths $L_x$ and $L_y$ remain constant as the system strains, under pure shear these lengths change with the total strain $\gamma$ according to $L_x(\gamma)=L_x(0)\mathrm{e}^{\gamma/2}$ and $L_y(\gamma)=L_y(0)\mathrm{e}^{-\gamma/2}$.
Thus a practical limitation of pure shear simulations is  that, unlike for simple shear, there is a limit to the total strain $\gamma$ that can be applied to a finite numerical system before the system collapses to a narrow height of order one particle length.  Therefore, to increase the total possible strain $\gamma$, we use systems with an initial system aspect ratio of $L_y(0)/L_x(0)=\beta$, and shear to a  strain $\gamma$ such that $L_y(\gamma)/L_x(\gamma)=1/\beta$, thus allowing a maximum  strain of $\gamma_\mathrm{max}=2\ln \beta$.  The value of $\beta$ and the number of particles $N$ are varied with $\alpha$, so that the final system height after the maximal strain is comparable to the fixed system length of our simple shear simulations.  In particular, for $\alpha\le0.01$ we use $\beta=12$ and $N=4096$; for $0.01<\alpha<4$ we use $\beta=16$ and $N=8192$; for $\alpha=4$ we use $\beta=20$ and $N=16384$.
All our results below use a fixed strain rate $\dot\gamma=10^{-6}$, and start from  random initial configurations, constructed in the same manner as for our simple shear simulations.  

\begin{figure}
\centering
\includegraphics[width=3.5in]{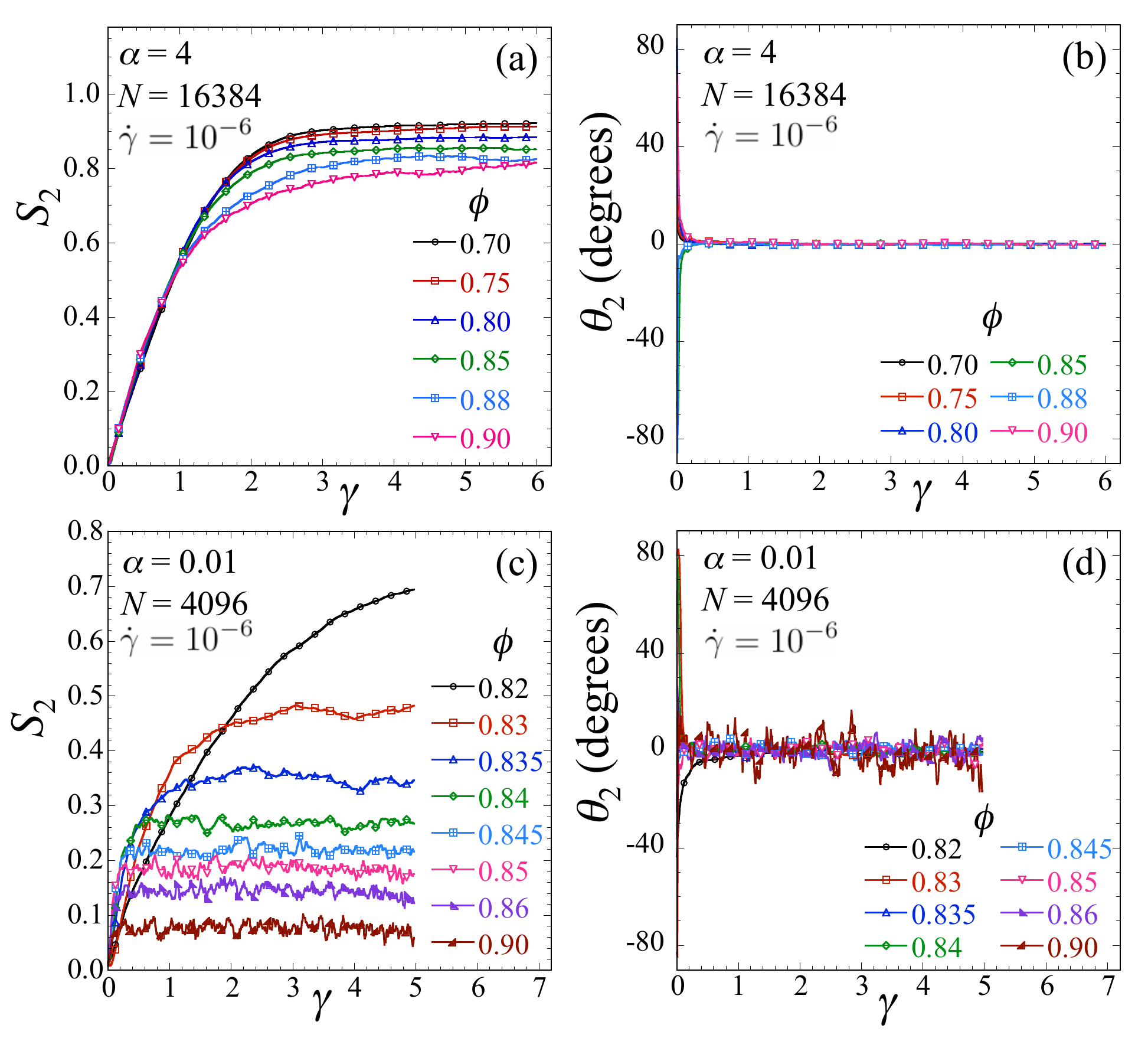}
\caption{For a pure shear deformation, (a) and (c) show the magnitude of the nematic order parameter $S_2$ vs total strain $\gamma=\dot\gamma t$ at different packing fractions $\phi$, for particles of asphericity $\alpha=4$ and 0.01, respectively; (b) and (d) show the corresponding orientation $\theta_2$ of the nematic order parameter.  Results are for a strain rate $\dot\gamma=10^{-6}$ with the number of particles $N$ as indicated in each panel.  Solid lines connect data points; symbols are shown only on a dilute set of the data points, so as to aid identification of the different curves.  The jamming packing fraction is $\phi_J=0.906$ for $\alpha=4$ and $\phi_J=0.845$ for $\alpha=0.01$.
}
\label{S2-vs-gamma} 
\end{figure}

In Fig.~\ref{S2-vs-gamma}(a) we plot $S_2$ vs strain $\gamma$ at several different packings $\phi$, for our elongated particles with $\alpha=4$.  We see that as $\gamma$ increases, $S_2$ rises from its near zero value in the initial random configuration and saturates to a constant steady-state value at large $\gamma$.  As $\phi$ increases, this steady-state value of $S_2$ decreases, as the decreasing free volume associated with the increasing particle density blocks particles from perfect alignment.  In Fig.~\ref{S2-vs-gamma}(b) we plot the corresponding orientation of the nematic order parameter $\theta_2$ vs $\gamma$.  We see that $\theta_2$ starts at some finite value, depending on the small, randomly directed, residual $\mathbf{S}_2$ in the initial random configuration, and then rapidly decays to $\theta_2=0$ as $\gamma$ increases.  Thus, as expected, the pure shearing orders the particles with a nematic order parameter oriented parallel to the minimal stress direction.  Our results in Figs.~\ref{S2-vs-gamma}(a) and \ref{S2-vs-gamma}(b) are from a single pure shear run at each $\phi$.

In Figs.~\ref{S2-vs-gamma}(c) and \ref{S2-vs-gamma}(d) we show corresponding results for $S_2$ and $\theta_2$ vs $\gamma$ for the case of nearly circular particles with $\alpha=0.01$.  Again we see that $S_2$ increases from zero to saturate at a steady-state value as $\gamma$ increases. Unlike the very slow relaxation $\gamma_\mathrm{relax}\sim 1/\alpha$ we expect for an isolated particle, here we see that relaxation to the steady-state is relatively rapid at large packings $\phi$; the frequent collisions between particles at large densities act to quickly equilibrate the system.  However as $\phi$ decreases, the relaxation strain $\gamma_\mathrm{relax}$ increases, and at our smallest packing $\phi=0.82$, $S_2$ fails to saturate to the steady-state value within our maximum strain $\gamma_\mathrm{max}=2\ln 12\approx 5$.  We previously reported similar results for $\alpha=0.001$ in the Supplemental Material to Ref.~\cite{MKOT}.  Our results in Figs.~\ref{S2-vs-gamma}(c) and \ref{S2-vs-gamma}(d) are from the average of two independent runs at each $\phi$.

We note that similar simulations have been carried out by Az{\'e}ma and Radja{\"i} in Ref.~\cite{Azema2010} for {\em frictional} 2D spherocylinders near the jamming packing, but using a constant lateral pressure rather than a constant volume, and shearing only to much smaller total strains than we do here.  They similarly find that particles orient parallel to the minimal stress direction as they are sheared, but they seem to reach the large strain steady-state only for relatively small particle asphericities.

\begin{figure}
\centering
\includegraphics[width=3.5in]{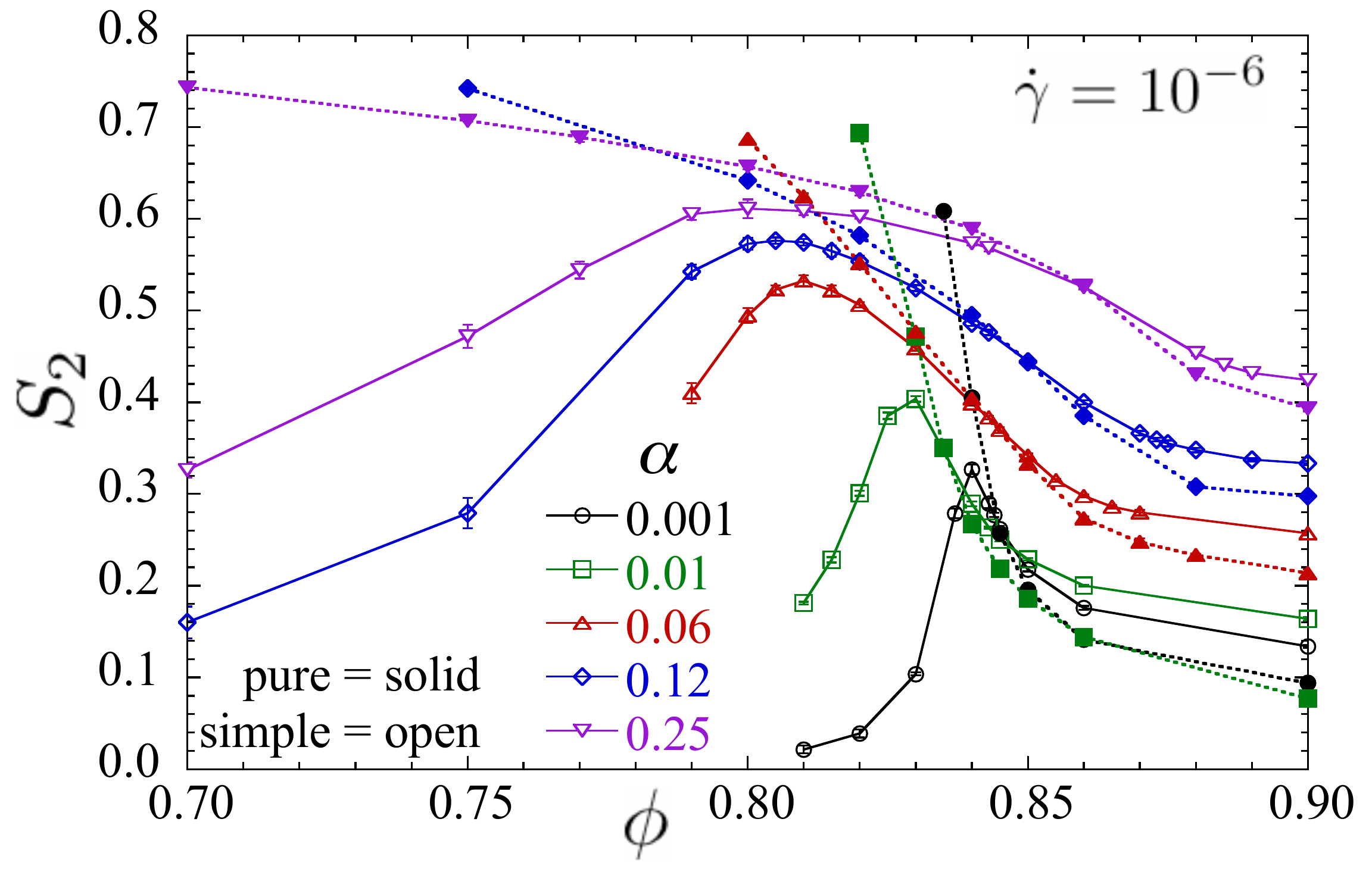}
\caption{Magnitude of the steady-state nematic order parameter $S_2$ vs packing $\phi$ for pure shear (solid symbols, dotted lines) compared to simple shear (open symbols, solid lines), for several small values of particle asphericity $\alpha$.  For pure shear the strain rate is $\dot\gamma=10^{-6}$.  For simple shear $\dot\gamma=10^{-6}$ for $\alpha=0.001$ and 0.01; for larger $\alpha$ a larger $\dot\gamma$ is used, but one that is still in the quasistatic limit where $S_2$ becomes independent of $\dot\gamma$.
}
\label{S2-pure-simple} 
\end{figure}

In Fig.~\ref{S2-pure-simple} we plot the pure shear steady-state value of $S_2$ vs $\phi$ (solid symbols, dotted lines) at several of our smaller $\alpha$, showing only results where $S_2(\gamma)$ has saturated to the large $\gamma$ steady-state value.  We see that as $\phi$ decreases, $S_2$ monotonically increases.  Based on the behavior of an isolated particle, given by Eq.~(\ref{ethetatps}), we believe that  $S_2$ will continue to increase and approach unity as $\phi\to 0$, however we cannot see this explicitly since we would need larger strains $\gamma$ to reach the steady-state as $\phi$ decreases.

For comparison, we also show in Fig.~\ref{S2-pure-simple} our results for the steady-state value of $S_2$ vs $\phi$ obtained from simple shearing (open symbols, solid lines).  For $\alpha=0.001$ and 0.01 we show results for $\dot\gamma=10^{-6}$, the same rate as we used in the pure shear simulations.  For $\alpha=0.06$ we use $\dot\gamma=4\times 10^{-6}$ and for $\alpha>0.06$ we use $\dot\gamma=10^{-5}$; however, in these cases the results of Fig.~\ref{S2-vs-phiophiJ} show that these larger $\dot\gamma$ have already reached the quasistatic limit, where $S_2$ becomes independent of $\dot\gamma$, for the range of $\phi$ of interest.  

While at the largest $\phi$ we see that $S_2$ from pure shearing is somewhat smaller than that from simple shearing, the two are qualitatively similar, and remain so as $\phi$ decreases.  However as $\phi$ approaches and decreases below $\phi_{S_2\,\mathrm{max}}$, the location of the peak in $S_2$ for simple shearing, we see that $S_2$ for pure shearing continues to increase while $S_2$ for simple shearing reaches its maximum and then decreases.  Thus above $\phi_{S_2\,\mathrm{max}}$ pure and simple shearing induce qualitatively similar orientational ordering, while below $\phi_{S_2\,\mathrm{max}}$ they become dramatically different.  

The non-monotonic behavior of $S_2$ under simple shearing can thus be understood as a competition between rotational drive and free volume.  At large $\phi$, the small free volume inhibits particles from aligning.  As $\phi$ decreases, the free volume increases allowing a better particle alignment and a larger $S_2$.  In such dense configurations, particles undergoing  simple shear still rotate with a finite $\langle\dot\theta_i\rangle/\dot\gamma$, however, according to the results of Sec.~\ref{stimedep}, these rotations occur randomly as  a Poisson-like process with the average rotation rate being determined by the long waiting time tails of the distribution (see Figs.~\ref{flipHist-a4}(a) and \ref{flipHist-a01}(a)); particle orientations are driven primarily by the interactions with other particles.  As $\phi$ decreases below $\phi_{S_2\,\mathrm{max}}$, the rotational drive of the simple shear becomes dominant, and particle rotation becomes more similar to the periodic rotations of an isolated particle, but with random perturbations due to particle collisions (see Sec.~\ref{stimedep}, particularly Figs.~\ref{flipHist-a4}(a) and \ref{flipHist-a01}(a)).  In this case, the particle rotations act to reduce the orientational ordering (and destroy it as $\alpha\to 0$), and $S_2$ decreases; this is unlike the case of pure shearing where there is no such rotational driving term [i.e. the second term on the right hand side of Eq.~(\ref{epures})] and $S_2$ continues to increases as $\phi$ decreases.  

The above scenario also helps to understand the singular $\alpha\to 0$ behavior under simple shearing,  discussed in our recent Letter \cite{MKOT}, in which as  particles approach a circular shape, $S_2$ vanishes for $\phi<\phi_J$ but $S_2$ remains finite at and just above $\phi_J$.
Such singular behavior is suggested in Fig.~\ref{S2-pure-simple} where we see that, for nearly circular particles with $\alpha=0.001$ undergoing simple shearing, the peak value of $S_{2\,\mathrm{max}}\approx 0.3$ remains relatively large, even though the fraction of the particle perimeter occupied by the two flat sides is only $0.064\%$.  In Appendix~\ref{sAtoZ} we present further analysis to determine the $\phi$ dependence of both $S_2$ and $-\langle \dot\theta_i\rangle/\dot\gamma$ in the $\alpha\to 0$ limit (see Fig.~\ref{S2-av-alpha-to-0}).

For nearly circular particles with small $\alpha$, 
at small $\phi$ well below  $\phi_{S_2\,\mathrm{max}}$, the rotational drive causes the particles to rotate almost uniformly with $-\langle\dot\theta_i\rangle/\dot\gamma\approx 1/2$, which by Eqs.~(\ref{eS2single}) and (\ref{eDI}) results in a small $S_2\propto \alpha$.  Particle collisions that give significant torques that increase $S_2$ only occur as the particle density increases to $\phi_{S_2\,\mathrm{max}}$, which itself increases  to the $\alpha=0$ jamming fraction $\phi_J^{(0)}$ as $\alpha\to 0$ \cite{MKOT}.  Thus we expect that as $\alpha\to 0$, $S_2\propto\alpha\to 0$ for all $\phi<\phi_J^{(0)}$.  Above $\phi_J^{(0)}$, however, particle interactions dominate over the rotational drive, and $S_2$ behaves as it would under pure shearing, with a finite $S_2$ that decreases  as $\phi$ increases.  Moreover, as $\alpha\to 0$, we found in Fig.~\ref{th2D-vs-phiophiJ} that  the orientation of the the nematic order parameter becomes $\theta_2\approx 45^\circ$ above $\phi_J^{(0)}$, hence $\mathbf{S}_2$ is aligning along the minimal stress direction (see also Fig.~\ref{th2X-vs-phiophiJ}), again just as it does under pure shearing.  Thus the singular behavior of $S_2$ as $\alpha\to 0$ for simple shearing is due to a sharp transition from the domination by rotational drive at $\phi<\phi_J$, to domination by geometric effects of the dense packings at $\phi>\phi_J$.

We have thus explained the non-monotonic behavior we have found for $S_2$ in terms of the competition between rotation and free volume.  However, recent simulations by Trulsson \cite{Trulsson}, on the simple shearing of 2D  ellipses, found that the non-monotonic behavior of $S_2$, seen for frictionless particles as $\phi$ increases, goes away once inter-particle frictional forces are added.  Instead of $S_2$ decreasing as $\phi$ increases above some $\phi_{S_2\,\mathrm{max}}$, for frictional particles $S_2$ seems to saturate to a constant value as $\phi$ increases.  However Trulsson simulates in the hard-core particle limit, and so all his simulations take place for $\phi\lesssim\phi_J(\mu_p)$, where $\phi_J(\mu_p)$ is the jamming packing fraction for particles with inter-particle frictional coefficient $\mu_p$.  For frictional particles, the additional frictional forces act to stabilize particle packings at smaller densities than the geometric jamming limit found for frictionless particles \cite{Makse,Otsuki}, and so $\phi_J(\mu_p)<\phi_J(\mu_p=0)$.  The difference between $\phi_J(\mu_p)$ and $\phi_J(\mu_p=0)$ increases as $\alpha$ increases \cite{Trulsson}.  Whereas for simple shear-driven jamming $\phi_J(\mu_p=0)$ seems to monotonically increase as $\alpha$ increases, $\phi_J(\mu_p)$ initially increases, reaches a maximum, and then decreases; the  difference in $\phi_J$ between the frictionless and the frictional cases becomes more dramatic as $\mu_p$ increases (see Fig.~6 of Ref.~\cite{Trulsson}).  Thus Trulsson's simulations do not probe the large density limit approaching geometric random close packing,  and so might not reach the dense limit where  free volume effects are dominating the behavior of $S_2$.  Fixed volume simulations with soft-core frictional particles, allowing one to investigate the range of $\phi$ above $\phi_J(\mu_p)$, might thus help to clarify  the situation.


\subsection{Relaxation to the Steady-State}
\label{sRSS}

In this section we address a second issue concerning the nematic orientational ordering of aspherical particles in simple shear flow.
Since there is a finite orientational order $\mathbf{S}_2$ even for an isolated single particle, is the finite $\mathbf{S}_2$ observed in the many particle system just a consequence of shearing acting like an ordering field? Or is the macroscopic $\mathbf{S}_2$ in the many particle system a consequence of cooperative behavior among the particles, as in an equilibrium ordering transition?
In this section we investigate this question by considering the relaxation of the system when perturbed away from the steady-state.

In Sec.~\ref{sTNO} we argued that the nematic order parameter $\mathbf{S}_2$ does not show any coherent time-dependent behavior, but rather has a constant value in the sheared steady-state.  However, if $\mathbf{S}_2$ is perturbed away from this steady-state value by a coherent rotation of all particles, it will relax back to the steady-state.  
In Ref.~\cite{Wegner} Wegner et al. suggested, by analogy with behavior in ordered  nematic liquid crystals, that the relaxation of $\mathbf{S}_2$ should obey a macroscopic equation of motion that can be written in the form,
\begin{equation}
\dot\theta_2=-\dot\gamma C(1-\kappa\cos 2\theta_2).
\label{dth2dt}
\end{equation}
If such an equation holds, it would suggest that $\mathbf{S}_2$ reflects a macroscopic ordering resulting from the coherent interaction of many particles.

The macroscopic equation~(\ref{dth2dt}) is similar to Eq.~(\ref{eq:theta_eom}) for the rotation of an isolated particle, except now it is assumed that  $\kappa >1$.  This gives a stable steady-state equilibrium value of $\theta^\mathrm{ss}_{2} = \frac{1}{2}\arccos(1/\kappa)$ and an unstable equilibrium value ($\dot\theta_2=0$) at $\theta_2=-\theta_2^\mathrm{ss}$.  One can then rewrite Eq.~(\ref{dth2dt}) as,
\begin{equation}
\dot\theta_2=-\dot\gamma C \left( 1-\dfrac{\cos 2\theta_2}{\cos 2\theta_2^\mathrm{ss}}\right).
\label{dth2dt2}
\end{equation}
Defining $\theta_2\in (-\pi/2,\pi/2]$, the above equation of motion predicts that when 
$|\theta_2| <\theta_2^\mathrm{ss}$, then 
$\mathbf{S}_2$ will relax to the steady state by rotating counter-clockwise to approach $\theta_2^\mathrm{ss}$; however, when $\theta_2$ lies outside this interval, 
$\mathbf{S}_2$ will relax to the steady state by rotating clockwise to approach $\theta_2^\mathrm{ss}$.  

\begin{figure}
\centering
\includegraphics[width=3.5in]{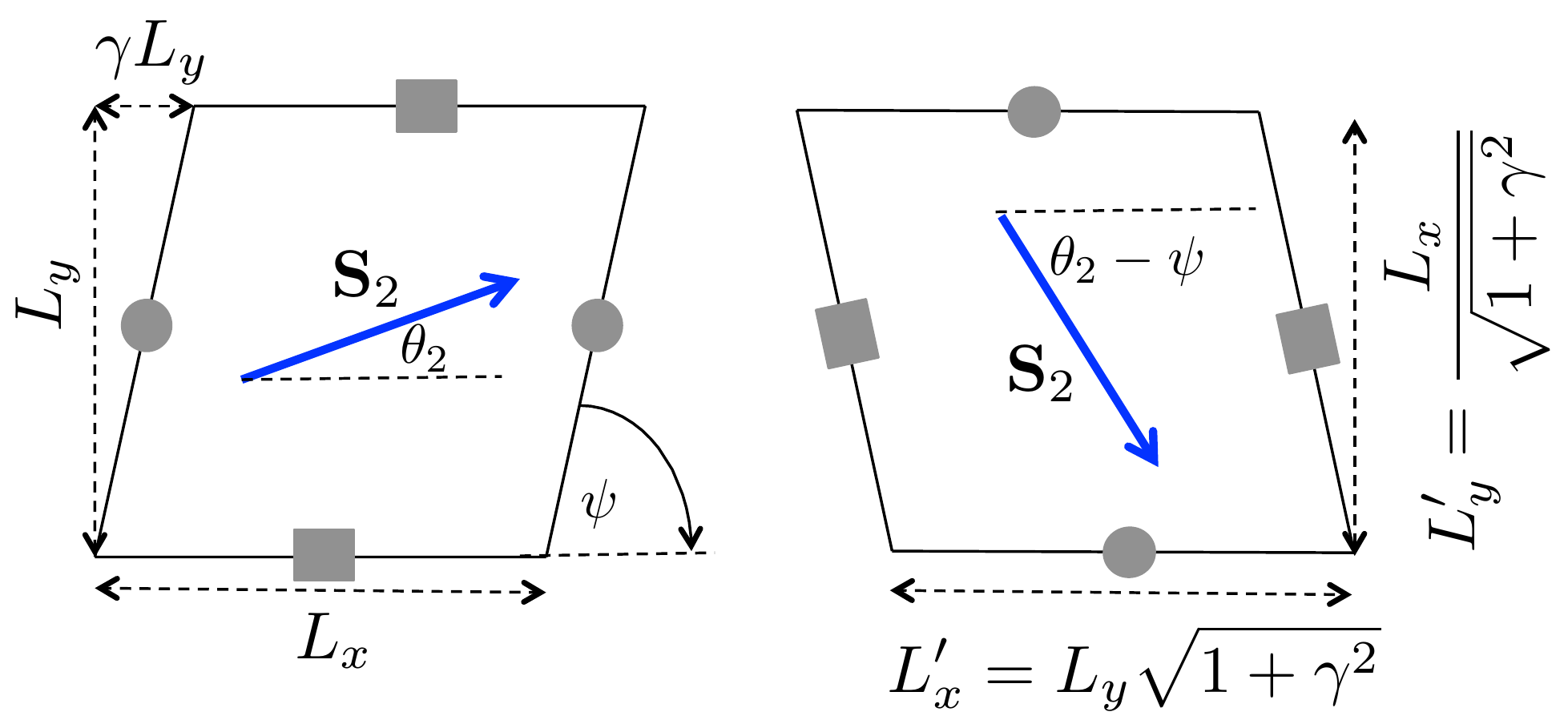}
\caption{Schematic of the procedure to construct a configuration in which the nematic order parameter $\mathbf{S}_2$ is rotated clockwise by an angle $\psi$.  Start with a configuration with a net shear strain $\gamma=\cot\psi$ (left figure) and rotate by $\psi$ to create the new configuration (right figure).  Under this transformation the configuration boundary conditions are preserved, as indicated by the shaded circles and squares on the various sides of the system boundary, but the  system aspect ratio changes, $L_y/L_x \to L_x/[L_y(1+\gamma^2)]$.
}
\label{flipS2} 
\end{figure}

To test this prediction we prepare numerical samples in which the steady-state $\mathbf{S}_2$ is rotated clockwise by a predetermined amount, and then measure the relaxation of $S_2$ and $\theta_2$ back to the steady-state as the system is sheared.  To create these samples with rotated $\mathbf{S}_2$ we use the method illustrated in Fig.~\ref{flipS2}.  A system with shear strain $\gamma$, sampled from our steady-state ensemble, is rotated clockwise by the angle $\psi=\mathrm{cot}^{-1}\gamma$, so that the two sides of the system boundary which were previously slanted now become the horizontal sides parallel to the flow direction.  We then continue to shear the system in the horizontal direction.

Such a rotation preserves the boundary conditions of the original configuration; the periodic boundary condition previously obeyed at the slanted sides now becomes the Lees-Edwards boundary condition at the new horizontal sides, and vice versa, as illustrated by the shaded circles and squares on the various sides in Fig.~\ref{flipS2}.  If the original configuration had a length $L_x$ and a height $L_y$, the new rotated configuration has length $L_y\sqrt{1+\gamma^2}$ and height $L_x/\sqrt{1+\gamma^2}$.  If the original $\mathbf{S}_2$ was at an angle $\theta_2$, close to but not necessarily exactly equal to $\theta_2^\mathrm{ss}$ because of fluctuations, the new $\mathbf{S}_2$ will be at an angle $\theta_2-\psi$.  By choosing different strains $\gamma$ at which to make this system rotation, we wind up with configurations in which the original steady-state $\mathbf{S}_2$ has been rotated by various angles $\psi=\cot^{-1}\gamma$.  To avoid a too elongated system when we rotate at a  large $\gamma$ (so as to produce a small rotation angle $\psi$), we start with an initial system in which $L_x>L_y$, instead of our usual $L_x=L_y$.

\begin{figure}
\centering
\includegraphics[width=3.5in]{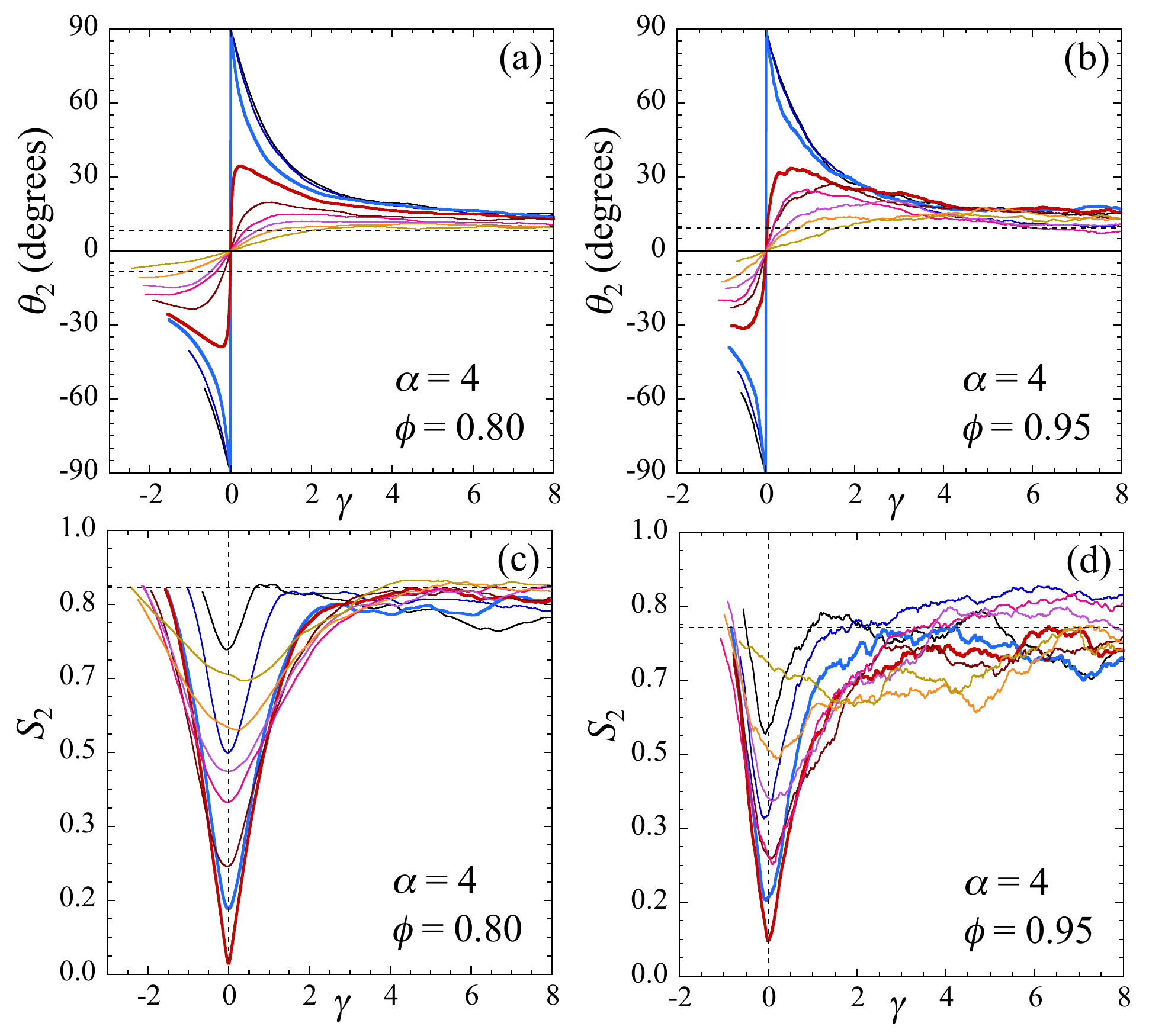}
\caption{For spherocylinders of asphericity $\alpha=4$ at strain rate $\dot\gamma=10^{-5}$: (a) and (b) instantaneous angle $\theta_2$, and (c) and (d) instantaneous magnitude $S_2$  of the nematic order parameter $\mathbf{S}_2$, vs shear strain $\gamma=\dot\gamma t$, after a rotation of a configuration in the steady-state  by different angles $\psi$ as illustrated in Fig.~\ref{flipS2}.  (a) and (c) are for $\phi=0.80$ near the minimum in $-\langle\dot\theta_i\rangle/\dot\gamma$, while (b) and (d) are for $\phi=0.95$ above the jamming $\phi_J=0.906$.
In (a) and (b) the left most point on each curve gives the initial value  $\theta_2^\mathrm{init}$ after the system rotation; the horizontal dashed lines give the ensemble averaged steady state values of $\pm\theta_2^\mathrm{ss}$.  In (c) and (d) the horizontal dashed line gives the ensemble averaged steady state value of $S_2$.  For ease of comparison, the strain axis has been shifted for each curve so that the point where $\theta_2=0$ or $90^\circ$ occurs at $\gamma=0$.  The two thicker curves denote (i) the largest of our $\theta_2^\mathrm{init}$ that results in a pure clockwise relaxation to the steady-state, and (ii)  the smallest of our $\theta_2^\mathrm{init}$ that results in a mostly counter-clockwise relaxation.
}
\label{relax-a4} 
\end{figure}

We first consider the relaxation of a system of moderately elongated spherocylinders with asphericity $\alpha=4$.
Using a system sheared at a strain rate $\dot\gamma=10^{-5}$, 
Fig.~\ref{relax-a4} shows the relaxation  of the rotated nematic order parameter $\mathbf{S}_2$ back to the steady state.
In Figs.~\ref{relax-a4}(a) and \ref{relax-a4}(b) we show the relaxation of the orientation $\theta_2$ vs net strain $\gamma=\dot\gamma t$, at packing fractions $\phi=0.80$ and $\phi=0.95$, respectively; $\phi=0.80$ is the packing that gives the minimum in $-\langle\dot\theta_i\rangle/\dot\gamma$, while $\phi=0.95$ is above the jamming $\phi_J=0.906$.  Figures~\ref{relax-a4}(c) and \ref{relax-a4}(d) show the corresponding relaxation of the magnitude $S_2$.  For each $\phi$ we show results for rotations through several different angles $\psi$, giving different initial values of $\theta_2^\mathrm{init}=\theta_2^\mathrm{ss}-\psi$.  For ease of comparison, for each curve the strain axis has been shifted so that the point where $\theta_2=0$ occurs at $\gamma=0$; this  also corresponds to the point where $|d\theta_2/d\gamma|$ is largest (for the cases with the smallest $\theta_2^\mathrm{init}$, where particles relax by a pure clockwise rotation, this point corresponds to  where $\theta_2$, consistent with our definition of $\theta_2\in (-\pi/2,\pi/2]$, takes a discontinuous jump from $-90^\circ$ to $+90^\circ$).

Denoting the values of $\pm\theta_2^\mathrm{ss}$ by horizontal dashed lines, in Figs.~\ref{relax-a4}(a) and \ref{relax-a4}(b) we see that for $\theta_2^\mathrm{init}$ sufficiently more negative than $-\theta_2^\mathrm{ss}$, the order parameter angle $\theta_2$ does relax back to the steady state by rotating clockwise, in agreement with Eq.~(\ref{dth2dt2}).  Similarly, for  $-\theta_2^\mathrm{ss}<\theta_2^\mathrm{init}<0$ we see that $\theta_2$ relaxes by rotating counter-clockwise, again in agreement with Eq.~(\ref{dth2dt}).  However there exists a region of $\theta_2^\mathrm{init} \lesssim -\theta_2^\mathrm{ss}$ where the order parameter starts rotating clockwise, then reverses direction to rotate counter-clockwise, overshoots $\theta_2^\mathrm{ss}$, then reverses direction again, rotating clockwise to relax back to $\theta_2^\mathrm{ss}$.  The two curves that separate the region where $\theta_2$ relaxes in a purely clockwise fashion from the region where it starts clockwise but then reverses to counter-clockwise, are indicated by thicker lines in the figures.  Since Eq.~(\ref{dth2dt2}) predicts a monotonic increase (i.e., counterclockwise rotation) or monotonic decrease (i.e., clockwise rotation) of $\theta_2$ as the system relaxes, it cannot be describing the system well for such $\theta_2^\mathrm{init}$.
Moreover, being a first order differential equation, Eq.~(\ref{dth2dt2}) would predict that $\theta_2(\gamma)$ would follow a fixed trajectory determined solely by the initial value $\theta_2^\mathrm{init}$.  However, in Figs.~\ref{relax-a4}(a) and \ref{relax-a4}(b) we see curves that pass through the same value of $\theta_2$ (for example $\theta_2=0$) but do not then follow the same trajectory as $\gamma$ increases.

The reason for this more complex behavior lies in the behavior of the magnitude of the order parameter, which in Eq.~(\ref{dth2dt2}) is presumed to stay constant.  In contrast, we see in Figs.~\ref{relax-a4}(c) and \ref{relax-a4}(d) that the rapid change in $\theta_2$ at $\gamma=0$ is accompanied by a pronounced drop in the magnitude of the order parameter $S_2$.  The largest drop in $S_2$, almost but not quite to zero, occurs for those $\theta_2^\mathrm{init}$ which give curves that are on the border between a pure clockwise relaxation and where the relaxation reverses from initially clockwise to counter-clockwise (indicated by the thicker curves in the figure).

\begin{figure}
\centering
\includegraphics[width=3.5in]{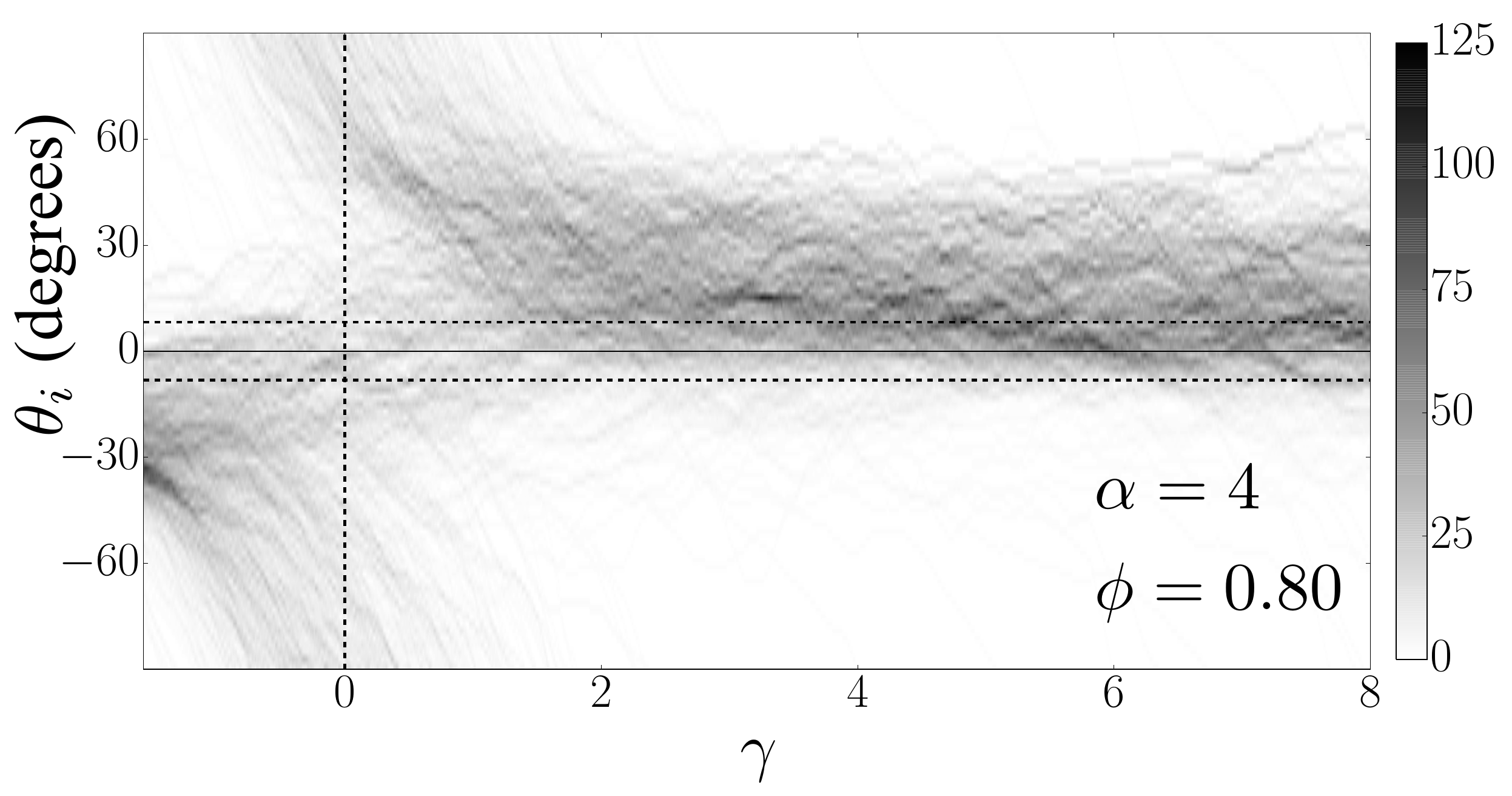}
\caption{For spherocylinders of asphericity $\alpha=4$ at strain rate $\dot\gamma=10^{-5}$ and packing $\phi=0.80$: Intensity plot showing the number of particles oriented at a particular angle $\theta_i$ vs net strain $\gamma=\dot\gamma t$, as the system  relaxes back to steady-state after an initial rotation of a configuration sampled from the steady-state ensemble.  The nematic order parameter  $\mathbf{S}_2$ is rotated to have the value of $\theta_2^\mathrm{init}$ that corresponds to the curve in Fig.~\ref{relax-a4}(c) that has the largest drop in the magnitude $S_2$ at $\gamma=0$.  The strain scale $\gamma$ has been shifted so that the left edge of the figure corresponds to the initial configuration after the rotation, while $\gamma=0$ corresponds to the strain at which $\theta_2=0$.   Horizontal dashed lines indicate the values of $\pm\theta_2^\mathrm{ss}$; the vertical dashed line indicates $\gamma=0$.
}
\label{relax-a4_intensity} 
\end{figure}

\begin{figure}
\centering
\includegraphics[width=3.5in]{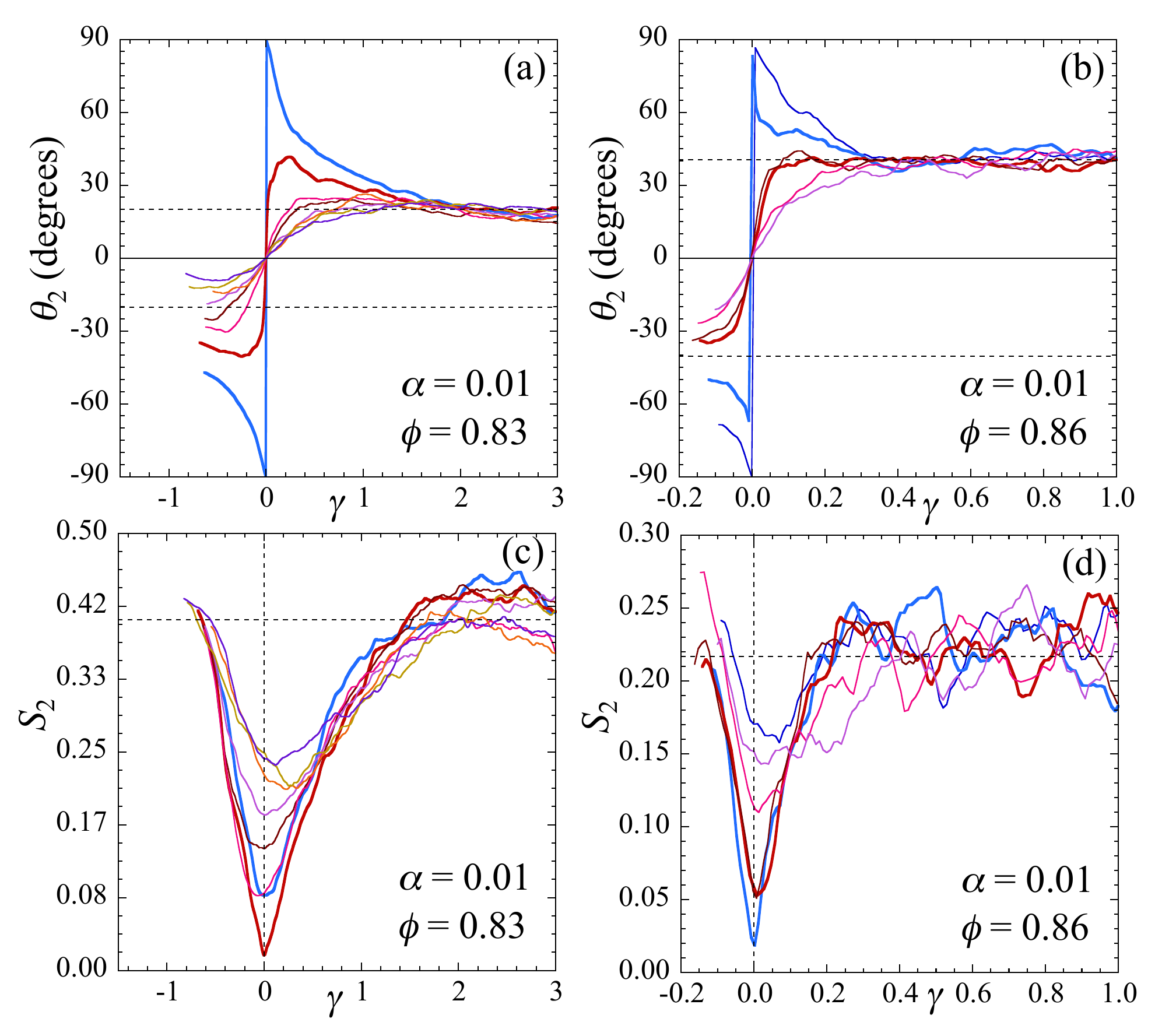}
\caption{For spherocylinders of asphericity $\alpha=0.01$ at strain rate $\dot\gamma=10^{-6}$: (a) and (b) instantaneous angle $\theta_2$, and (c) and (d) instantaneous magnitude $S_2$  of the nematic order parameter $\mathbf{S}_2$, vs shear strain $\gamma=\dot\gamma t$, after a rotation of a configuration in the steady-state  by different angles $\psi$ as illustrated in Fig.~\ref{flipS2}.  (a) and (c) are for $\phi=0.83$ near the minimum in $-\langle\dot\theta_i\rangle/\dot\gamma$, while (b) and (d) are for $\phi=0.86$ above the jamming $\phi_J=0.845$.
In (a) and (b) the left most point on each curve gives the initial value  $\theta_2^\mathrm{init}$ after the system rotation; the horizontal dashed lines give the ensemble averaged steady state values of $\pm\theta_2^\mathrm{ss}$.  In (c) and (d) the horizontal dashed line gives the ensemble averaged steady state value of $S_2$.  For ease of comparison, the strain axis has been shifted for each curve so that the point where $\theta_2=0$ or $90^\circ$ occurs at $\gamma=0$.  The two thicker curves denote (i) the largest of our $\theta_2^\mathrm{init}$ that results in a pure clockwise relaxation to the steady-state, and (ii)  the smallest of our $\theta_2^\mathrm{init}$ that results in a mostly counter-clockwise relaxation.
}
\label{relax-a01} 
\end{figure}

\begin{figure}
\centering
\includegraphics[width=3.5in]{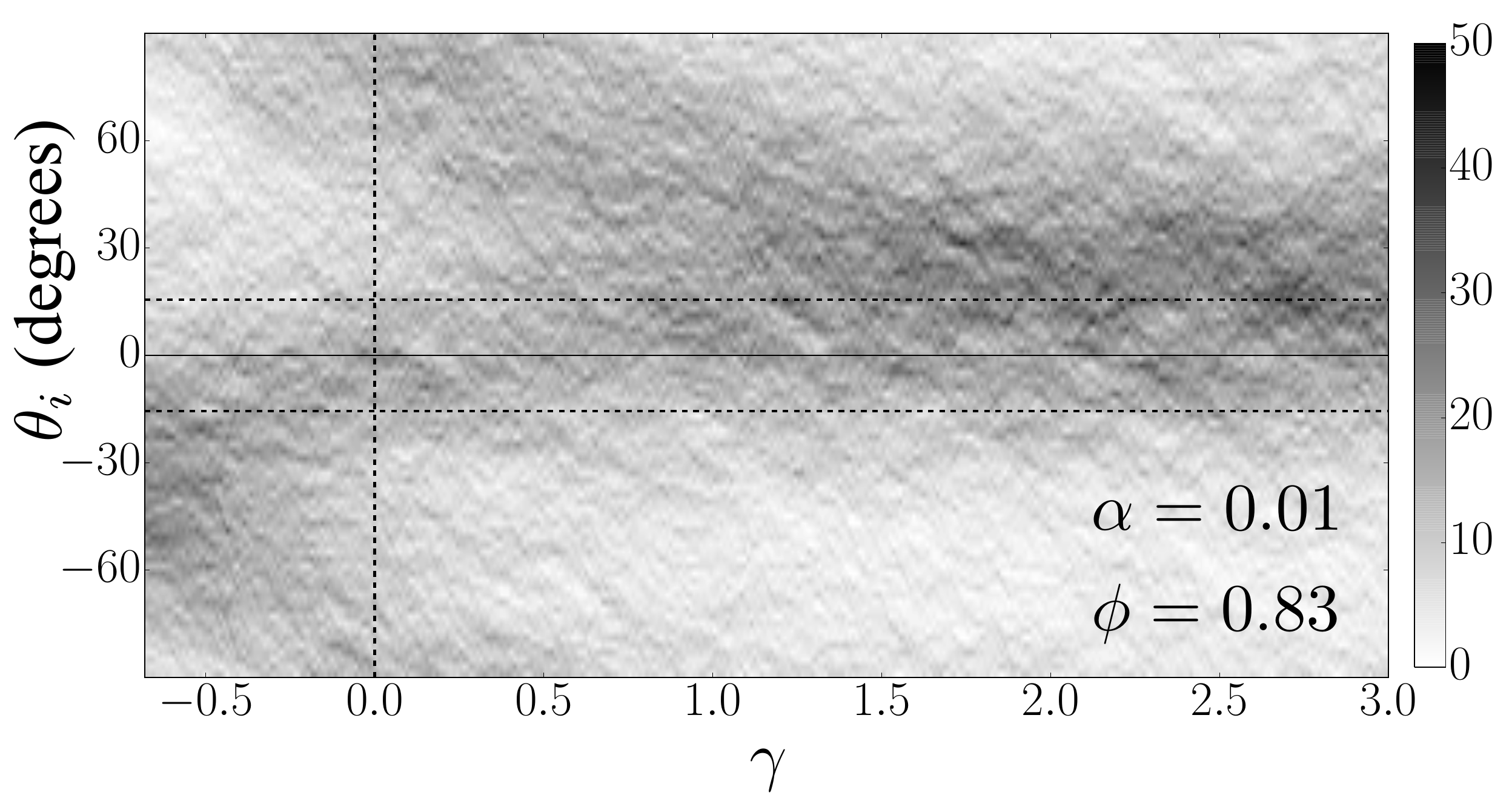}
\caption{For spherocylinders of asphericity $\alpha=0.01$ at strain rate $\dot\gamma=10^{-6}$ and packing $\phi=0.83$: Intensity plot showing the number of particles oriented at a particular angle $\theta_i$ vs net strain $\gamma=\dot\gamma t$, as the system  relaxes back to steady-state after an initial rotation of a configuration sampled from the steady state ensemble.  The nematic order parameter  $\mathbf{S}_2$ is rotated to have the value of $\theta_2^\mathrm{init}$ that corresponds to the curve in Figs.~\ref{relax-a01}(c) that has the largest drop in the magnitude $S_2$ at $\gamma=0$.  The strain scale $\gamma$ has been shifted so that the left edge of the figure corresponds to the initial configuration after the rotation, while $\gamma=0$ corresponds to the strain at which $\theta_2=0$.   Horizontal dashed lines indicate the values of $\pm\theta_2^\mathrm{ss}$; the vertical dashed line indicates $\gamma=0$.}
\label{relax-a01_intensity} 
\end{figure}

To understand this behavior of $S_2$, in Fig.~\ref{relax-a4_intensity} we show an intensity plot of the orientations $\theta_i$ of the individual particles, as a function of the net shear strain $\gamma=\dot\gamma t$, as the system relaxes following the rotation of a configuration sampled from the steady-state.  At each $\gamma$, the range of angles $\theta_i$ is binned into $2^\circ$ intervals and we count the number of particles with orientation $\theta_i$ in each bin; this count is then imaged by the graryscale as shown.
We use the same system as in Figs.~\ref{relax-a4}(a) and \ref{relax-a4}(c), with $\alpha=4$ and $\dot\gamma=10^{-5}$ at packing $\phi=0.80$; a rotation is chosen that corresponds to the curve with the largest drop in $S_2$ seen in Fig.~\ref{relax-a4}(c).  We see that some fraction of the particles relax by rotating clockwise, while the others relax by rotating counter-clockwise.  At $\gamma=0$, corresponding to the smallest value of $S_2$, we see the broadest distribution of values of $\theta_i$.  The sharp drop in $S_2$ as the system relaxes back to steady state is thus due to the lack of coherence in the relaxation of the individual particles.  We find qualitatively the same behavior if we look at other packing fractions near and above jamming.
We note that  similar results as in our Figs.~\ref{relax-a4} and \ref{relax-a4_intensity} have been observed  experimentally by B{\"o}rzs{\"o}nyi et al. for the relaxation of shear-reversed dry granular 3D packings of glass cylinders \cite{Borzsonyi2}.

Finally, in Figs.~\ref{relax-a01} and \ref{relax-a01_intensity} we show similar plots, but now for nearly circular particles with $\alpha=0.01$.  We see the same qualitative features as were found for the more elongated particles with $\alpha=4$.
We thus conclude from these relaxation simulations that the nematic ordering $\mathbf{S}_2$ in our simple sheared system is a consequence of the shearing acting as an ordering field, and not due to large scale cooperative behavior among the particles.
The sharp drop in the magnitude $S_2$ to small values, as the system relaxes back to steady-state, demonstrates that the relaxation takes place through the incoherent rotation of individual particles, not a coherent rotation of many particles that would preserve the magnitude of the ordering.  We  will confirm the absence of long range coherence in particle orientations in a separate work \cite{MTStructure} where we directly compute the spatial correlation function of $\mathbf{S}_2$ and find it to be short ranged.


\subsection{A Numerical Mean-Field Model}
\label{sec:MF}

In the preceding section we have argued that, although there is a finite nematic ordering in the system, there is no macroscopic coherence among the particles.  
In this section we therefore explore whether one can make a mean-field-like model for the rotation  of a particle, that depends only on the state of the individual particle itself, but reproduces reasonably the observed ensemble averages for the nematic order parameter $\mathbf{S}_2$ and the angular velocity $-\langle\dot\theta_i\rangle/\dot\gamma$, as time averages of the single particle.

The rotational motion of a particle is governed by Eq.~(\ref{eq:theta_eom}), which we can rewrite as,
\begin{equation}
\dfrac{\dot\theta_i}{\dot\gamma}=\dfrac{d\theta_i}{d\gamma}=-f(\theta_i)+g_i,\quad\text{where}\quad g_i=\dfrac{\tau_i^\mathrm{el}}{k_d \mathcal{A}_i I_i\dot\gamma}
\end{equation}
gives the interaction with other particles due to the  torques from elastic collisions.  We  consider  four different approximations to $g_i$, replacing the term from the  fluctuating collisional torques by
\begin{flalign}
&\text{(i)}\qquad\qquad\qquad g_i\to \bar g=\langle g_i\rangle&
\end{flalign}
where we average over both different particles in a given configuration, and over different configurations in the steady-state ensemble, and
\begin{flalign}
&\text{(ii)}\qquad\qquad\qquad g_i\to\bar g+\delta g(\gamma)&
\end{flalign}
where $\delta g(\gamma)$ is an uncorrelated Gaussian white noise with
\begin{align}
&\langle \delta g(\gamma)\rangle=0 \\ 
&\langle\delta g(\gamma)\,\delta g(\gamma^\prime)\rangle = [\delta g]^2\delta(\gamma-\gamma^\prime), 
\end{align}
with $[\delta g]^2=\mathrm{var}[g_i]$, where the variance is computed from the steady-state ensemble. 

In the mean-field models (i) and (ii) the elastic torque that the particle experiences is independent of the orientation of the particle.  As a next level of approximation, we 
consider mean-field models in which the elastic torque will be a function of the particle's orientation $\theta$.
\begin{flalign}
&\text{(iii)}\qquad\qquad\qquad g_i\to\bar g(\theta)=\langle g_i\rangle_\theta,&
\end{flalign}
where now the average is restricted to particles oriented at a particular angle $\theta$.
\begin{flalign}
&\text{(iv)}\qquad\qquad\qquad g_i\to\bar g(\theta)+\delta g(\theta;\gamma)&
\end{flalign}
where $\delta g(\theta;\gamma)$ is an uncorrelated Gaussian white noise with 
\begin{align}
&\langle \delta g(\theta;\gamma)\rangle=0\\
&\langle \delta g(\theta;\gamma)\,\delta g(\theta;\gamma^\prime)\rangle=[\delta g(\theta)]^2\delta(\gamma-\gamma^\prime), 
\end{align}
with $[\delta g(\theta)]^2 =\mathrm{var}[g_i]_\theta$, where the variance is taken only over particles with orientation $\theta$.
These different approximations allow us to examine  the relative importance of average torque vs torque noise, and the sensitivity of behavior to the variation of elastic torque with  particle orientation.

\begin{figure}
\centering
\includegraphics[width=3.5in]{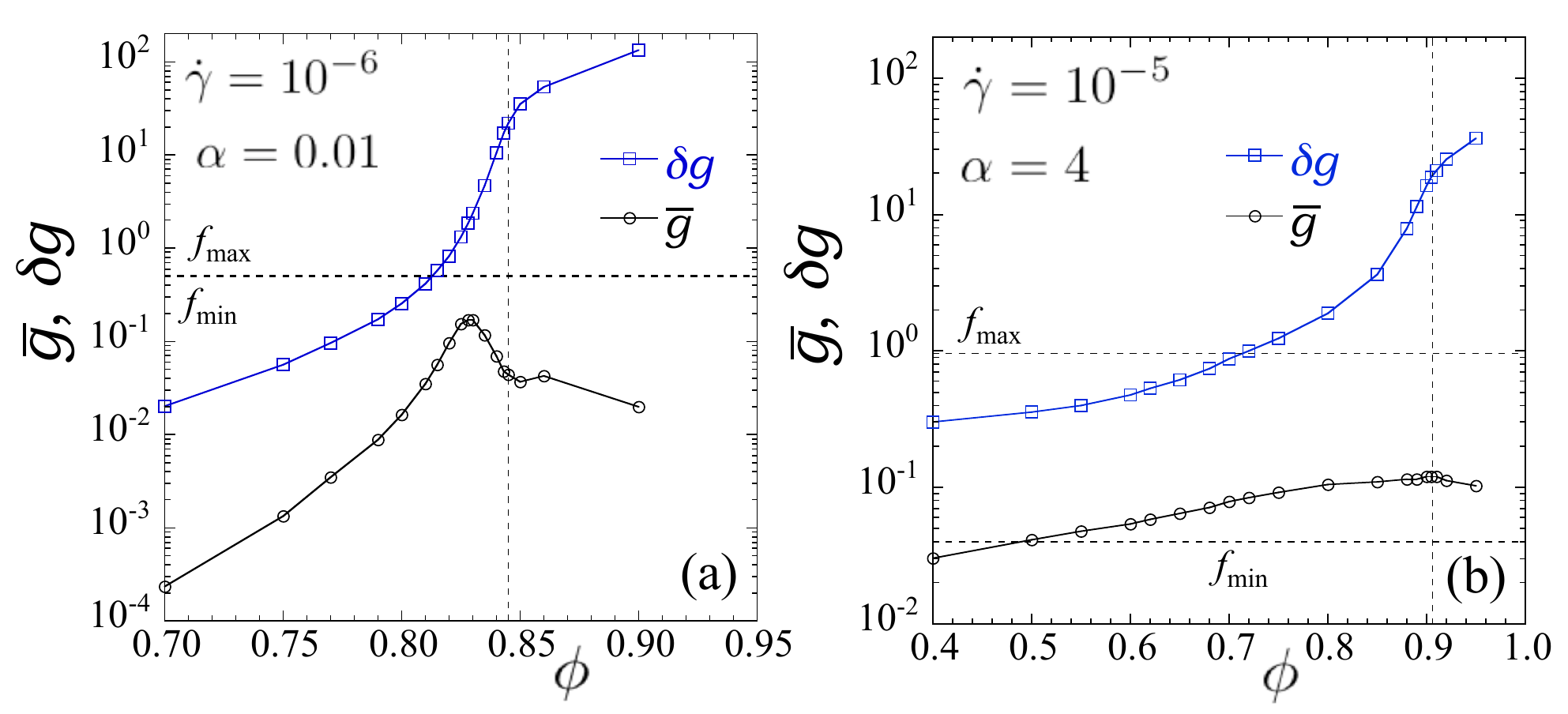}
\caption{For mean-field models (i) and (ii): average elastic torque $\bar g=  \langle \tau^\mathrm{el}_i/k_d \mathcal{A}_i I_i\dot\gamma\rangle$ and associated noise magnitude $\delta g$ vs packing $\phi$ for (a) $\alpha=0.01$ at $\dot\gamma=10^{-6}$, and (b) $\alpha=4$ at $\dot\gamma=10^{-5}$.  Horizontal dashed lines $f_\mathrm{min}$ and $f_\mathrm{max}$ denote the minimum  $f(0)$ and maximum  $f(\pi/2)$ values of $f(\theta)=(1-[\Delta I_i/I_i]\cos 2\theta)/2$ in Eq.~(\ref{eftheta}); note that for $\alpha=0.01$ these two are nearly indistinguishable since $\Delta I_i/I_i=0.00847$ is so small.  Vertical dashed lines locate the jamming packings, $\phi_J=0.845$ for $\alpha=0.01$ and $\phi_J=0.906$ for $\alpha=4$.
}
\label{MF1-2} 
\end{figure}

\begin{figure}
\centering
\includegraphics[width=3.5in]{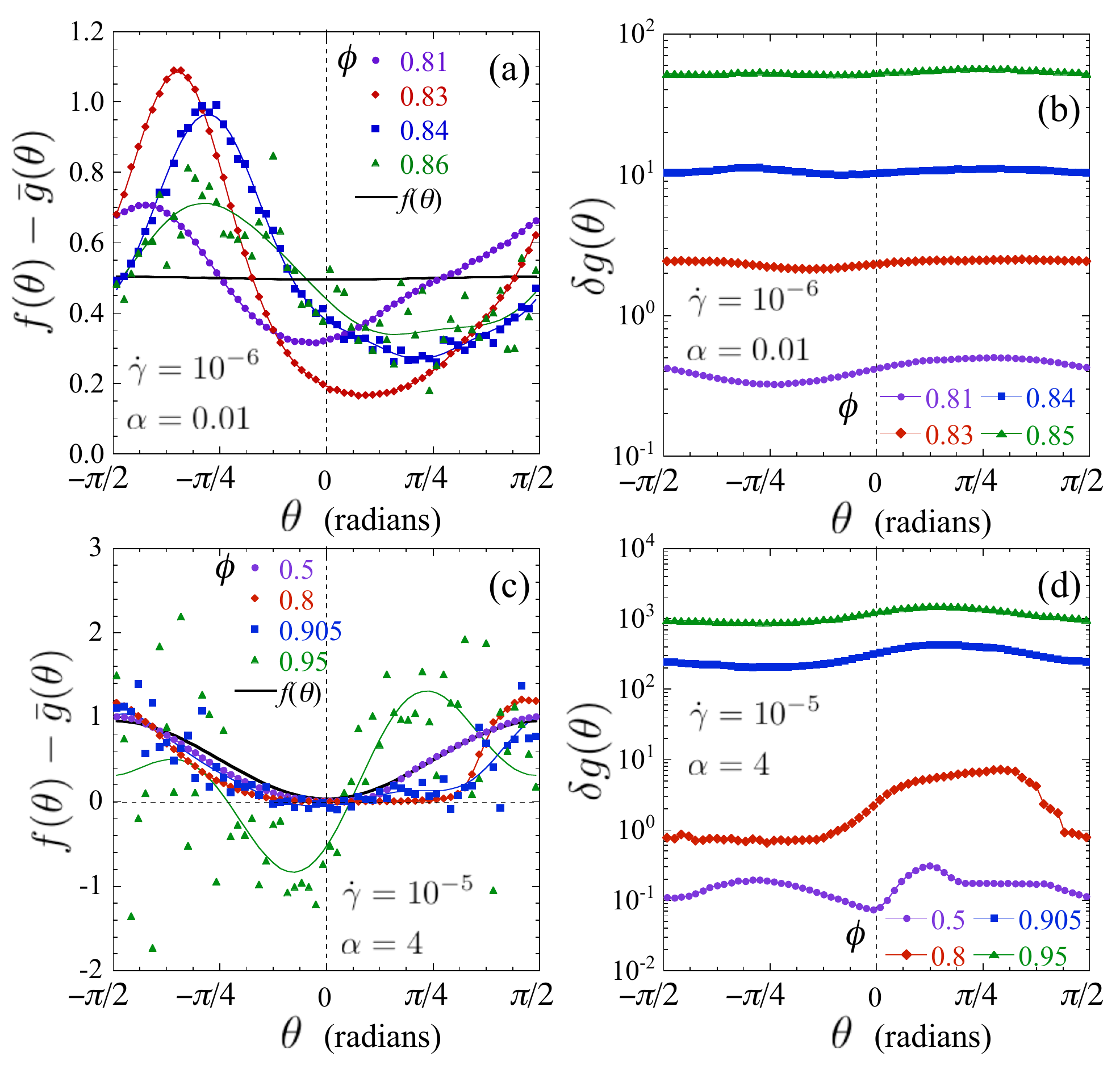}
\caption{For mean-field models (iii) and (iv): average elastic torque $\bar g(\theta)=  \langle \tau^\mathrm{el}_i/k_d \mathcal{A}_i I_i\dot\gamma\rangle_\theta$ and associated noise $\delta g(\theta)$ for particles oriented at angle $\theta$.
Top row (a) and (b) is for $\alpha=0.01$ at $\dot\gamma=10^{-6}$, with $\phi_J=0.845$; bottom row (c) and (d) is for $\alpha=4$ at $\dot\gamma=10^{-5}$, with $\phi_J=0.906$.
(a) and (c): $f(\theta)-\bar g(\theta)$  vs $\theta$ at different packings $\phi$, where $f(\theta)=(1-[\Delta I_i/I_i]\cos 2\theta)/2$  as in Eq.~(\ref{eftheta}).   The thick solid black line is just $f(\theta)$, corresponding to $\phi\to0$ where $\bar g(\theta)=0$.  Thin colored lines are the Fourier series approximation to the data at each $\phi$, as given by Eq.~(\ref{eFourier2}).  (b) and (d): magnitude of the the noise $\delta g(\theta)$ vs $\theta$ at different packings $\phi$.  Note the logarithmic vertical scale.  
}
\label{MF3-4} 
\end{figure}

In Fig.~\ref{MF1-2} we plot our results for $\bar g$ and $\delta g$ vs $\phi$, which are used in constructing the mean-field (MF) models (i) and (ii).  In Fig.~\ref{MF1-2}(a) we show results for nearly circular particles with $\alpha=0.01$ at strain rate $\dot\gamma=10^{-6}$; in Fig.~\ref{MF1-2}(b) we show results for elongated particles with $\alpha=4$ at $\dot\gamma=10^{-5}$.  The horizontal black dashed lines in each panel are  the values of $f_\mathrm{min}\equiv f(0)=(1-\Delta I_i/I_i)/2$ and $f_\mathrm{max}\equiv f(\pi/2)=(1+\Delta I_i/I_i)/2$, which are the minimum and maximum values of $f(\theta)=(1-[\Delta I_i/I_i]\cos 2\theta)/2$ given in Eq.~(\ref{eftheta}).  If ever we have $f_\mathrm{min}<\bar g<f_\mathrm{max}$, then in MF model (i) the direction $\theta_i$ such that $f(\theta_i)=\bar g$ is a stationary point where $\dot\theta_i/\dot\gamma =0$.  From Fig.~\ref{MF1-2} we see that this situation never arises for $\alpha=0.01$, however it does occur for $\alpha=4$ when $\phi>0.5$.  Note that in both cases the average elastic torque $\bar g = \langle \tau^\mathrm{el}_i/k_d \mathcal{A}_i I_i \dot\gamma\rangle$ is positive, showing that, on average, the elastic torques serve to slow down the clockwise rotation of the particles.
Note also that in both cases the magnitude of the noise $\delta g$ is one or more orders of magnitude larger than the average $\bar g$ for the range of $\phi$ considered.

In Fig.~\ref{MF3-4} we show results for $\bar g(\theta)$ and $\delta g(\theta)$ vs $\theta$, which are used for constructing the models MF (iii) and MF (iv).  In Figs.~\ref{MF3-4}(a) and \ref{MF3-4}(b) we show results for $\alpha=0.01$ at $\dot\gamma=10^{-6}$, while in Figs.~\ref{MF3-4}(c) and \ref{MF3-4}(d) we show results for $\alpha=4$ at $\dot\gamma=10^{-5}$.  In each case we show results at four different typical values of $\phi$: below $\phi_{S_2\,\mathrm{max}}$, near $\phi_{S_2\,\mathrm{max}}$, near $\phi_J$ and above $\phi_J$.  Rather than show $\bar g(\theta)$ directly, in Figs.~\ref{MF3-4}(a) and \ref{MF3-4}(c) we instead plot $f(\theta)-\bar g(\theta)=-\dot\theta_i/\dot\gamma$, since this more directly gives the rotational motion of the particle.  A positive value of $f(\theta)-\bar g(\theta)$ indicates a clockwise rotation.  A value of $\theta$ such that $f(\theta)-\bar g(\theta)=0$ indicates a stationary point in MF (iii), where $\dot\theta_i/\dot\gamma=0$; if $d[f(\theta)-\bar g(\theta)]/d\theta >0$ this is a stable stationary point.  

At the larger values of $\phi$ our data for $f(\theta)-\bar g(\theta)$ become quite scattered, particularly for $\alpha=4$. 
To get a smooth $\bar g(\theta)$ for integrating  our mean-field single  particle equation of motion we therefore approximate $\bar g(\theta)$ by expanding our data as a Fourier series and keeping only the lowest several terms,
\begin{align}
\bar g(\theta)&=\dfrac{a_0}{\pi}+\frac{2}{\pi}\sum_{n=1}\left[a_n\cos 2n\theta + b_n\sin 2n\theta\right],\label{eFourier2}\\
a_n&=\int_{-\pi/2}^{\pi/2}\!\!d\theta\,\bar g(\theta)\cos 2n\theta,\\
b_n&=\int_{-\pi/2}^{\pi/2}\!\!d\theta\,\bar g(\theta)\sin 2n\theta.
\end{align}
For the largest $\phi$, where the data are most scattered, we use up to $n=3$ terms for our approximate $\bar g(\theta)$; for smaller $\phi$, where the data are smoother but where there are regions of $\theta$ where $\bar g(\theta)$ is rather flat, we use  up to $n=16$ terms.  This Fourier approximation gives the solid lines in Figs.~\ref{MF3-4}(a) and \ref{MF3-4}(c).

\begin{figure}
\centering
\includegraphics[width=3.5in]{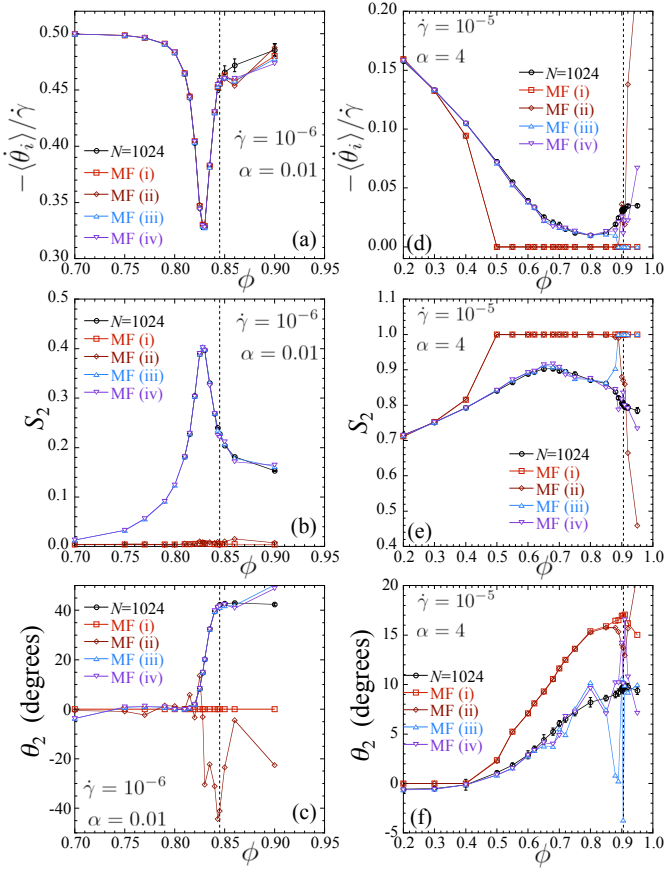}
\caption{Comparison of $-\langle\dot\theta_i\rangle/\dot\gamma$, $S_2$, and $\theta_2$ vs $\phi$ (top, middle, and bottom rows respectively) between our $N=1024$ interacting particle simulations and the single-particle mean-field approximations MF (i), (ii), (iii) and (iv).  The left column is for nearly circular particles of $\alpha=0.01$ at $\dot\gamma=10^{-6}$, while the right column is for elongated particles of $\alpha=4$ at $\dot\gamma=10^{-5}$.  The vertical dashed lines locate the jamming packings, $\phi_J=0.845$ and 0.906 for $\alpha=0.01$ and 4, respectively.
}
\label{MF-av-S2-th2} 
\end{figure}

We now consider how well these mean-field models do in describing the behavior of our interacting many particle system.  In Fig.~\ref{MF-av-S2-th2} we show our results for $-\langle\dot\theta_i\rangle/\dot\gamma$, $S_2$, and $\theta_2$ (top, middle, and bottom rows respectively) vs the packing $\phi$, comparing our $N=1024$ particle simulations against that of the single particle mean-field models MF (i), (ii), (iii), and (iv).  The left column is for nearly circular particles with $\alpha=0.01$ at $\dot\gamma=10^{-6}$, while the right column is for elongated particles with $\alpha=4$ at $\dot\gamma=10^{-5}$.

We discuss $\alpha=0.01$ first.  We see in Fig.~\ref{MF-av-S2-th2}(a) that all the models MF (i) -- (iv) do a good job in predicting the angular velocity $-\langle\dot\theta_i\rangle/\dot\gamma$.  This is not surprising.  For $\alpha=0.01$, the term $\Delta I_i/I_i=0.00847$ is so small that the variation in $f(\theta)$ is exceedingly slight, and so to good approximation one can take $f(\theta)\approx 1/2$;  an isolated particle is essentially rotating uniformly.  The   elastic torque  of MF (i), modeled by the $\theta$-independent $\bar g$, with $\bar g<f_\mathrm{min}$ at all $\phi$ (see Fig.~\ref{MF1-2}(a)), then just subtracts from this average drive $f\approx 1/2$ to give the correct average angular velocity.  Adding the noise $\delta g$ in MF (ii), or using an orientationally dependent $\bar g(\theta)$ in MF (iii) and corresponding noise $\delta g(\theta)$ in MF (iv), does not change this average rotational behavior.  Only as one goes above $\phi_J$, and correlations between particles become longer ranged, do we see a difference between the interacting many particle system and our single particle mean-field models.

In contrast, if we consider $S_2$, we see in Fig.~\ref{MF-av-S2-th2}(b) that the simple MF (i) does an exceedingly poor job.  Again, this is not surprising.  As discussed above, since for $\alpha=0.01$ the model MF (i) results in a particle that rotates almost uniformly, there is no mechanism for $S_2$ to grow above the very small value $S_2=0.0042$ that is found for an isolated particle.  Similarly, as seen in Fig.~\ref{MF-av-S2-th2}(c), MF (i) gives $\theta_2=0$, just as for an isolated particle.  Adding noise, as in MF (ii), does nothing to improve the results for $S_2$ or $\theta_2$.  However, using the orientationally dependent average elastic torque $\bar g(\theta)$ of MF (iii) results in excellent agreement for both $S_2$ and $\theta_2$.  The strong variation of $\bar g(\theta)$ with $\theta$, as seen in Fig.~\ref{MF3-4}(a), results in the non-uniform rotation of the particle that is essential to dramatically increase $S_2$ over the isolated particle limit.
No further improvement is found by adding the orientationally dependent noise $\delta g(\theta)$ of MF (iv).

Turning to elongated particles with $\alpha=4$, we see in Fig.~\ref{MF-av-S2-th2}(d) that now MF (i) fails dramatically even when considering $-\langle\dot\theta_i\rangle/\dot\gamma$. While agreement is not bad at the smallest $\phi$, once $\phi$ increases above 0.5 and $\bar g$ increases above $f_\mathrm{min}=f(0)$ (see Fig.~\ref{MF1-2}(b)), the particle locks into a stationary state where $\dot\theta_i/\dot\gamma=0$, and consequently one has $S_2=1$, as seen in Fig.~\ref{MF-av-S2-th2}(e).  The orientation $\theta_2$, shown in Fig.~\ref{MF-av-S2-th2}(f), then increases with $\phi$ so as to obey $f(\theta_2)=\bar g$.  Adding the noise $\delta g$ of MF (ii) is not sufficient to allow the particle to escape from this stationary state, until $\phi$ gets close to and goes above jamming. 

To get good agreement for $\alpha=4$ it is thus necessary, as we found for $\alpha=0.01$,  to consider the orientational dependence of the average elastic torque.  Using the $\bar g(\theta)$ of MF (iii) we see that we get excellent agreement for all three quantities, $-\langle\dot\theta_i\rangle/\dot\gamma$, $S_2$, and $\theta_2$, for all $\phi$ {\em except} upon approaching close to the jamming $\phi_J$. Close to $\phi_J$,  Fig.~\ref{MF3-4}(c) shows that $f(\theta)-\bar g(\theta)$ can go negative, giving rise to a stationary state when $f(\theta)-\bar g(\theta)=0$.  Thus we see in Fig.~\ref{MF-av-S2-th2}(d) that as $\phi$ approaches $\phi_J$, $-\langle\dot\theta_i\rangle/\dot\gamma$ drops to zero, while in Fig.~\ref{MF-av-S2-th2}(e) we see that $S_2$ jumps to unity.  However adding the noise $\delta g(\theta)$ of MF (iv) is sufficient to allow the particle to escape this stationary state, and restore good agreement with the many particle simulation, until one goes above $\phi_J$.  

We thus conclude that our single-particle mean-field model gives an excellent  description of the rotational motion of our particles, over a wide range of asphericities $\alpha$ and packings $\phi$, provided one includes the proper orientational dependence to the average torque from the elastic interactions, as in MF (iii).  Agreement at large $\phi$ approaching jamming is further improved by adding the noise term of MF (iv).  However our mean-field model seems to do less well as $\phi$ increases above $\phi_J$.  Whether this is an effect of increasing correlations between particles as they jam, or whether it is due to poor accuracy in our estimate of $\bar g(\theta)$, due to poor statistics, remains unclear.

\section{Summary}
\label{sec:discus}

In this work we have considered a  model of sheared, athermal, frictionless two dimensional spherocylinders in suspension at constant volume.  The simplicity of our model, in which the only interactions are pairwise repulsive elastic forces and a viscous damping with respect to the suspending host medium, allows us to shear to very long total strains and completely characterize the behavior of the system over a wide range of packing fractions $\phi$, strain rates $\dot\gamma$, and particle asphericities $\alpha$.  In a prior work we focused on the rheological properties of  this model and the variation of the jamming transition $\phi_J$ with particle asphericity \cite{MT1}.  In the present work we have focused on the shear-induced rotation of particles and their nematic orientational ordering.

We found that, under simple shearing, particles continue to rotate at all packings, even above jamming, and that the nematic order parameter $\mathbf{S}_2$ has a constant, time-independent, value in the sheared steady-state.
We have found that the average angular velocity of particles $-\langle\dot\theta_i\rangle/\dot\gamma$ and the magnitude of the nematic order parameter $S_2$ are non-montonic as the packing $\phi$ increases, with the minimum of $-\langle\dot\theta_i\rangle/\dot\gamma$ and the maximum of  $S_2$ occurring below the jamming transition.  By considering the distribution of strain intervals $\Delta\gamma$ between successive rotations of a particle by $\pi$ in Sec.~\ref{stimedep}, and by comparing the response of the system under pure shear as opposed to simple shear in Sec.~\ref{sPure}, the following scenario emerges.  At the smaller packings $\phi$, behavior is qualitatively similar to that of an isolated particle.  The rotational drive implicit in simple shearing (but absent in pure shearing) causes particles to rotate with a non-uniform angular velocity that depends on the particle's orientation.  As $\phi$ increases, the rate of collisions between particles increases, leading to a broadening of the distribution of rotation times, but still with a typical rotation time comparable to the average.  The average $S_2$ is dominated by the average particle rotation, as evidenced by the observed difference in $S_2$ between simple and pure shearing; in contrast to the increase in $S_2$ as $\phi$ increases under simple shearing, under pure shearing, which has no rotational driving term, $S_2$ shows perfect ordering at small $\phi$ and is monotonically decreasing as $\phi$ increases.

At larger $\phi$, however, the system becomes so dense that the decreasing free volume inhibits rotations.  Particles tend to lock into the local configuration, with rotational rattling about a particular orientation, until a  shear-induced fluctuation in the local particle structure allows a rotation to take place.  Particle rotations become a Poisson-like process in which  the time until the next particle rotation is largely independent of the time since the last rotation.  The average $S_2$ is now dominated by the local structure of the   dense packing, rather than the particle rotations, as evidenced by the qualitative agreement now found  for  the behavior of $S_2$ comparing simple and pure shear (see Fig.~\ref{S2-pure-simple}).

The above scenario helps to explain our surprising result of Ref.~\cite{MKOT}, further discussed in Appendix.~\ref{sAtoZ},  that the $\alpha\to 0$ limit, approaching perfectly circular particles, is singular.  As particles approach the rotationally invariant circular shape, one would naively expect that the nematic orientational order parameter $\mathbf{S}_2$ should vanish.  However, in the limit of finite $\alpha\to 0$, we found that $S_2$ vanishes below $\phi_J$, but remains finite at $\phi_J$ and above.  To explain this,  consider first the behavior under pure shear, where we have argued that particles of any finite $\alpha$, no matter how small, will exponentially relax their orientation  to the minimal stress direction, and so eventually order with $S_2\approx 1$, at sufficiently small packings $\phi$.  As $\phi$ increases, the decreasing free volume inhibits particle rotation, limiting the extent of ordering, and leading to an $S_2$ that decreases monotonically as $\phi$ increases; we found numerically that $S_2$, under pure shear, remains finite above jamming even for very small $\alpha$.  Consider now the behavior under simple shear.  According to the above scenario, above the peak in $S_2$ under simple shear, behavior is dominated by the local structure of the dense configuration, and simple and pure shear result in qualitatively similar ordering.  As $\alpha\to 0$ the location of the peak in $S_2$ moves to the jamming transition.  Hence we expect that, even as $\alpha\to 0$, the simple sheared system will order with finite $S_2$ for $\phi\ge\phi_J$. However for $\phi<\phi_J$, the rotational drive of the simple shear, absent for pure shear, will dominate and cause the particles to rotate with an increasingly uniform (i.e., independent of the particle orientation) angular velocity as $\alpha$ gets small.  As $\alpha\to 0$ this uniform rotation will drive $S_2\to 0$.  Hence our scenario leads one to expect that, as $\alpha\to 0$, one will have $S_2=0$ for $\phi<\phi_J$ but $S_2>0$ for $\phi\ge\phi_J$, just as we found to be the case.

Finally, although our sheared system of aspherical particles displays finite nematic orientational ordering at any packing $\phi$, this ordering is not due to  long range coherence between particles as in an equilibrium liquid crystal, but rather is due to the shearing acting as an ordering field.  
This conclusion is  supported by our results in Sec.~\ref{sRSS}, where we investigated the relaxation of $\mathbf{S}_2$ upon being rotated away from its steady-state direction.  The sharp drop in the magnitude $S_2$ to small values, as the system relaxes back to steady-state, indicates that the relaxation takes place through the incoherent rotation of individual particles, not a coherent rotation of many partices that would preserve the magnitude of the ordering.  Additionally, the success of our numerical mean-field model of Sec.~\ref{sec:MF}, in which we modeled the system by an isolated particle being acted upon by an orientation dependent average elastic torque and  random incoherent torque noise, indicates that correlations between particles are not important to describe the behavior of the system.  We will give further evidence for this conclusion in a separate paper \cite{MTStructure}, where we directly compute the spatial correlation function for $\mathbf{S}_2$ and show that is it short ranged.

\section*{Acknowledgements}

This work was supported in part by National Science Foundation Grants  No. CBET-1435861 and No. DMR-1809318. Computations were carried out at the Center for Integrated Research Computing at the University of Rochester. 

\appendix

\section{Distribution of Particle Orientations}
\label{aOrientations}

Several early works \cite{Campbell,Guo1,Borzsonyi2,Azema2010}     considered the orientational ordering of particles in a shear flow by computing the probability density $\mathcal{P}(\theta)$ for a given particle to be oriented at a particular angle $\theta_i=\theta$.  It is interesting to relate this $\mathcal{P}(\theta)$ to the nematic order parameter $\mathbf{S}_2$, and in particular ask whether the orientation angle $\theta_2$ of the nematic order parameter coincides with the most probable orientation, as determined by the maximum of $\mathcal{P}(\theta)$.  Here we will compute $\mathcal{P}(\theta)$  by sampling both over different particles $i$ within an individual configuration, and over different configurations within our steady-state sheared ensemble.

The ensemble averages defining $S_2$ and $\theta_2$ in Eqs.~(\ref{eS2g0})-(\ref{eS2g2}) can be expressed in terms of  $\mathcal{P}(\theta)$ by considering the Fourier series expansion of the distribution.  
In this context, $S_2$ and $\theta_2$ can viewed as giving the first term in this expansion, 
\begin{equation}
\mathcal{P}(\theta)=\frac{1}{2\pi}+\frac{1}{\pi}\sum_{m\,\mathrm{even}} S_m \cos [m(\theta-\theta_m)],
\label{eFourier}
\end{equation}
where only even integer $m$ terms appear in the sum because $\mathcal{P}(\theta)$ has a periodicity of $\pi$, and the normalization is taken such that $\int_0^{2\pi}\!\!d\theta\,\mathcal{P}(\theta)=1$.

\begin{figure}
\centering
\includegraphics[width=3.5in]{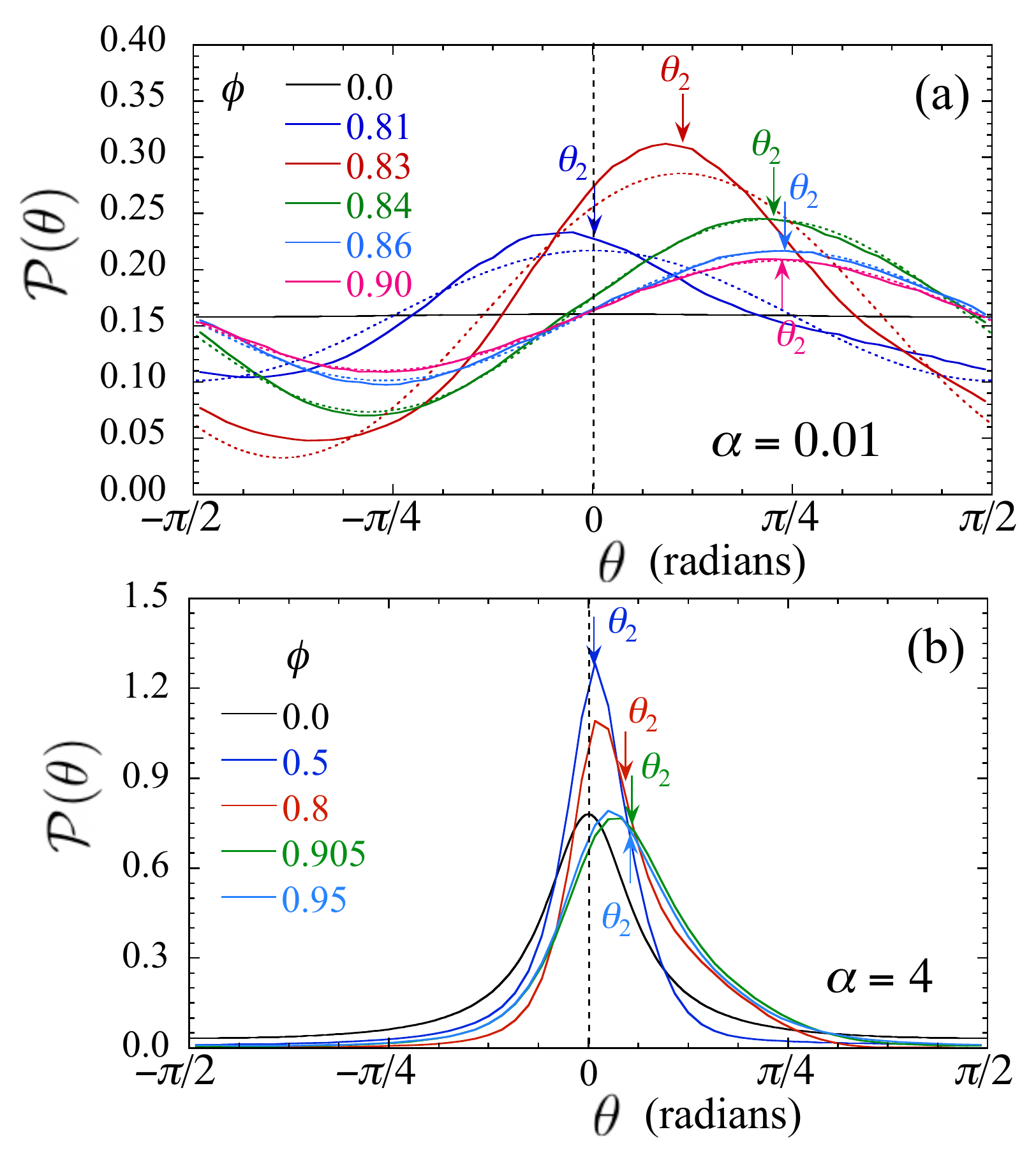}
\caption{Probability distribution $\mathcal{P}(\theta)$ for a particle to be oriented at angle $\theta$, for different packing fractions $\phi$. (a) Nearly circular particles with asphericity $\alpha=0.01$ at $\dot\gamma=10^{-6}$; dotted curves show the approximation to $\mathcal{P}(\theta)$ obtained from the Fourier series expansion of Eq.~(\ref{eFourier}) keeping only the lowest $m=2$ term. (b) Elongated particles with $\alpha=4$ at $\dot\gamma=10^{-5}$.  In both (a) and (b), the curve labeled $\phi=0.0$ is the distribution for an isolated particle given by Eq.~(\ref{Piso}); arrows for each curve of different $\phi$ denote the location of the angle $\theta_2$ of the nematic director.
}
\label{Ptheta-a01-a4} 
\end{figure}

In Fig.~\ref{Ptheta-a01-a4}(a) we plot $\mathcal{P}(\theta)$ vs $\theta$, at several different packings $\phi$, for nearly circular particles with $\alpha=0.01$ at strain rate $\dot\gamma=10^{-6}$.  We show only the range $-\pi/2< \theta \le \pi/2$ because $\mathcal{P}(\theta)$ has a periodicity of $\pi$.
The solid, nearly horizontal, line labeled $\phi=0.0$ is the distribution for an isolated particle, computed using Eq.~(\ref{Piso}); since $\Delta I_i/\Delta I_i=0.0085$ for $\alpha=0.01$, this isolated particle distribution is essentially flat on the scale of the figure.  As $\phi$ increases, and $S_2$ correspondingly increases (see Fig.~\ref{S2_v_phi_a4-01}(b)),   $\mathcal{P}(\theta)$ develops a strong $\theta$ dependence.  The curves for $\phi \ge 0.81$ in Fig.~\ref{Ptheta-a01-a4}(a) show a roughly sinusoidal variation in $\theta$, with an amplitude that varies non-monotonically as $\phi$ increases through the value $\phi_{S_2\,\mathrm{max}}\approx 0.83$ where $S_2$ has its maximum.  The dotted curves in Fig.~\ref{Ptheta-a01-a4}(a)  show the approximation to $\mathcal{P}(\theta)$ obtained from the Fourier series expansion of Eq.~(\ref{eFourier}) keeping only the lowest $m=2$ term, determined by the nematic order parameter.   For the denser packings $\phi\gtrsim 0.84$, near and above the jamming $\phi_J\approx 0.845$, this gives an excellent approximation to $\mathcal{P}(\theta)$; for smaller $\phi<0.84$ we see noticeable deviations.  The direction $\theta_2$ of the nematic order parameter, which always lies at the peak of the dotted curves,  is thus very close to  the most probable particle orientation $\theta_\mathrm{max}$ for the dense cases $\phi\gtrsim 0.84$, but we see that $\theta_2$ is slightly larger than $\theta_\mathrm{max}$ for the more dilute cases.

In Fig.~\ref{Ptheta-a01-a4}(b) we show similar plots of $\mathcal{P}(\theta)$ vs $\theta$ at different $\phi$, but now for elongated particles with $\alpha=4$ at $\dot\gamma=10^{-5}$.   The localized shape of $\mathcal{P}(\theta)$ at all $\phi$ indicates that one would have to take many terms $m$ in the expansion of Eq.~(\ref{eFourier}) to get a good approximation.  Nevertheless one can still ask where $\theta_2$ (indicated by the arrows in Fig.~\ref{Ptheta-a01-a4}) lies with respect to the most probable value $\theta_\mathrm{max}$.  At the smallest $\phi=0.5$, the distribution $\mathcal{P}(\theta)$ is largely symmetric about its maximum and $\theta_2\approx \theta_\mathrm{max}$.  As $\phi$ increases, the location of the maximum $\theta_\mathrm{max}$ increases slightly, but the distribution also becomes noticeably skewed towards the large $\theta$ side of the peak.  Thus we find that $\theta_2$ shifts to the right of the peak and $\theta_2>\theta_\mathrm{max}$.  This difference seems to be at its largest near the $\phi_{\dot\theta\,\mathrm{min}}\approx 0.80$ where $-\langle\dot\theta_i\rangle/\dot\gamma$ is at its smallest. 

\begin{figure}
\centering
\includegraphics[width=3.5in]{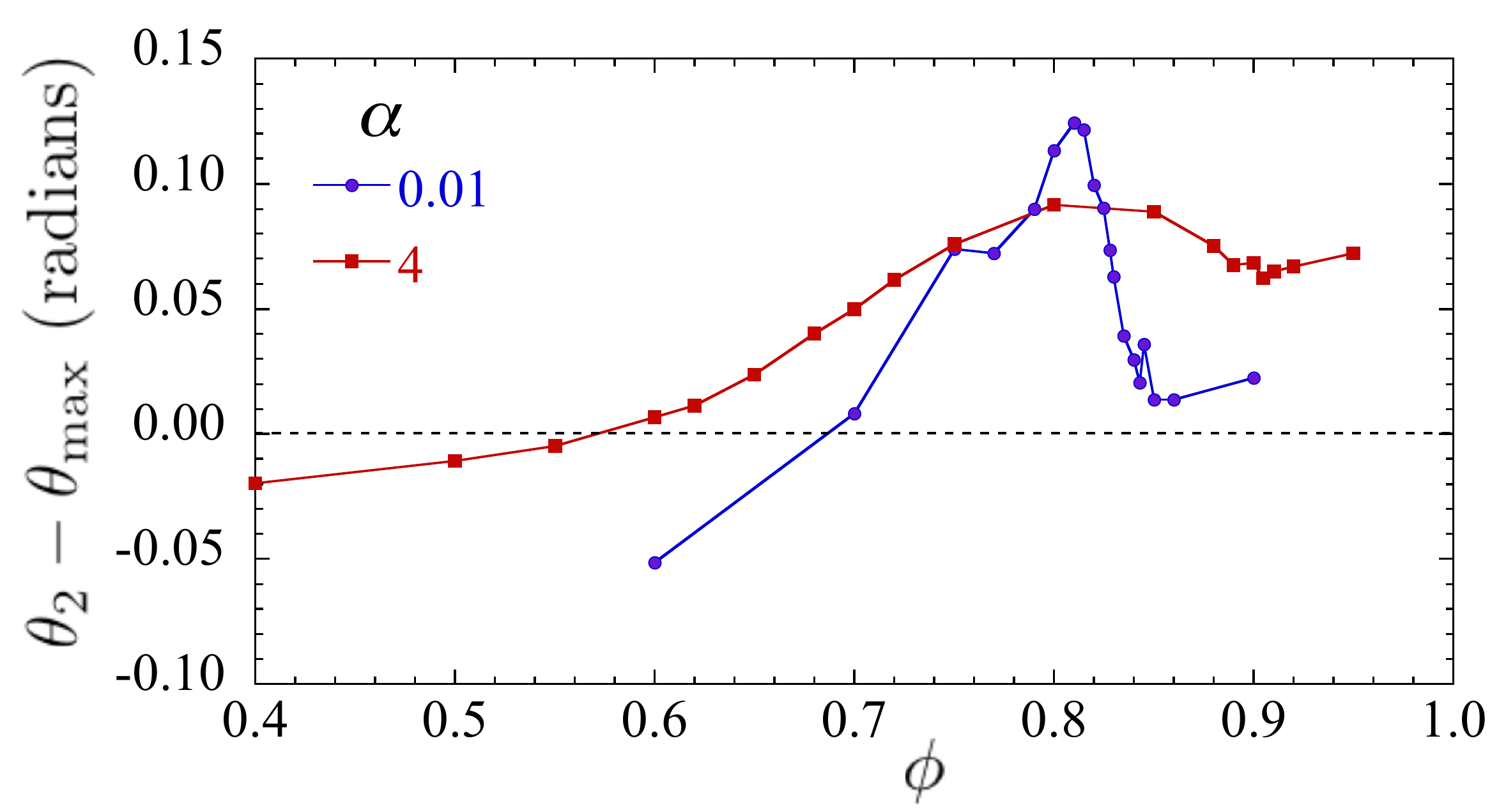}
\caption{Difference between the angle $\theta_2$ of the nematic order parameter and the angle $\theta_\mathrm{max}$ that gives the most probable particle orientation, vs packing $\phi$.  For $\alpha=0.01$ results are from a strain rate $\dot\gamma=10^{-6}$; for $\alpha=4$ results are from $\dot\gamma=10^{-5}$.
}
\label{thetamax-theta2} 
\end{figure}

In Fig.~\ref{thetamax-theta2} we plot the difference between the angle of the nematic director $\theta_2$ and the most probable angle of particle orientation $\theta_\mathrm{max}$ vs the packing fraction $\phi$ for $\alpha=0.01$ and 4.  In both cases $\theta_2-\theta_\mathrm{max}$ is negative at small $\phi$ and then increases as $\phi$ increases, becoming positive and reaching a maximum near (though not exactly equal to)  the packing $\phi_{\dot\theta\,\mathrm{min}}$ where $-\langle \dot\theta_i\rangle/\dot\gamma$ has its minimum, and then decreasing again until $\phi\approx \phi_J$, at which point it increases again as $\phi$ goes above jamming.

\section{The $\alpha\to 0$ limit}
\label{sAtoZ}

For perfectly circular particles with $\alpha=0$, the rotational invariance of the particles implies that there can be no nematic ordering, and so $S_2=0$.  Moreover, for perfectly circular particles the elastic forces are directed radially inwards to the center of the particle and so the torque from the elastic particle collisions necessarily  vanishes, $\tau_i^\mathrm{el}=0$.  Since our model has no Coulomb frictional forces, the rotation of circular particles is thus determined solely by the dissipative torque $\tau_i^\mathrm{dis}$ due to the drag with respect to the background, affinely sheared, host medium. Since by symmetry the moment of inertia has equal eigenvalues, then $\Delta I_i=0$   and Eq.~(\ref{eq:theta_eom}) gives a fixed uniform rotational motion for each particle, $\dot\theta_i=-\dot\gamma/2$.  One might therefore expect that, for spherocylinders of asphericity $\alpha>0$, one would find that $S_2\to 0$ and $-\langle\dot\theta_i\rangle/\dot\gamma\to 1/2$ continuously as $\alpha\to 0$.

However, as we have already noted in connection with Fig.~\ref{dthdg-vs-phiophiJ} for $-\langle\dot\theta_i\rangle/\dot\gamma$ and Fig.~\ref{S2-vs-phiophiJ} for $S_2$, we see a sizable value for $S_2$ and a sizable difference between $-\langle\dot\theta_i\rangle/\dot\gamma$ and $1/2$, even for very nearly circular particles with $\alpha=0.001$, for which the flat sides of the spherocylinder comprise only 0.064\% of the total perimeter.  Here we will argue that the $\alpha\to 0$ limit is singular, and that if one sits at the jamming transition then $\lim_{\alpha\to 0} S_2$ and $\lim_{\alpha\to 0}[1/2-\langle\dot\theta_i\rangle/\dot\gamma]$ stay finite.  We have previously reported on this effect in Ref.~\cite{MKOT}, here we provide  further details.

\begin{figure}
\centering
\includegraphics[width=3.5in]{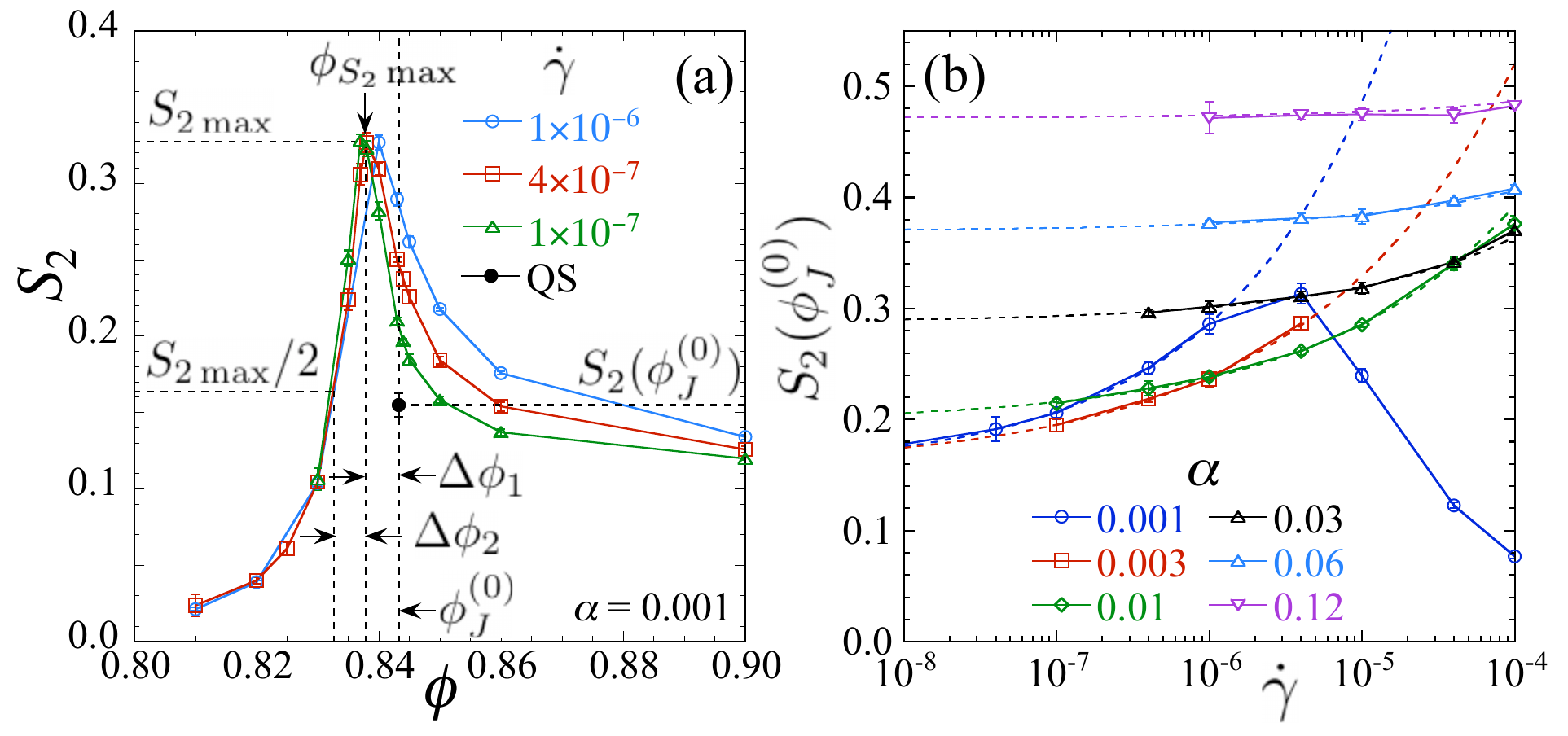}
\caption{(a) For spherocylinders with $\alpha=0.001$, $S_2$ vs $\phi$ at strain rates $\dot\gamma=10^{-6}, 4\times 10^{-7}$ and $10^{-7}$.  The black dot labeled ``QS" represents the extrapolated $\dot\gamma\to 0$ value of $S_2$ at $\phi_J^{(0)}=0.8433$. Widths 
$\Delta\phi_1\equiv \phi_J^{(0)}-\phi_{S_2\,\mathrm{max}}$, and $\Delta\phi_2=\phi_{S_2\,\mathrm{max}}-\phi_{S_2\,\mathrm{half}}$ are denoted in the figure.
(b) Plot of $S_2$ vs $\dot\gamma$ at $\phi_J^{(0)}$ for $\alpha\le 0.12$.  Solid lines connect the data points, dashed lines are fits of the small $\dot\gamma$ points to the form $a+b\dot\gamma^c$ and are used to extrapolate to the $\dot\gamma\to 0$ limit.
}
\label{S2-a001} 
\end{figure}

We are interested in the quasistatic $\dot\gamma\to 0$ limit of $S_2(\phi)$ as $\alpha\to 0$.  To determine this limit, we  define several benchmarks.  The first is the height of the peak in $S_2$ as $\phi$ varies, which we denote by $S_{2\,\mathrm{max}}$, occurring at $\phi_{S_2\,\mathrm{max}}$.  Next is the value $S_2(\phi_J^{(0)})$ at the $\alpha\to 0$ jamming transition of circular particles, $\phi_J^{(0)}=0.8433$.  To characterize the location of the peak in $S_2(\phi)$ we define 
\begin{equation}
\Delta\phi_1=\phi_J^{(0)}-\phi_{S_2\,\mathrm{max}}, 
\end{equation}
the distance of the peak to $\phi_J^{(0)}$.  To characterize the width of the peak we define 
\begin{equation}
\Delta\phi_2=\phi_{S_2\,\mathrm{max}}-\phi_{S_2\,\mathrm{half}}, 
\end{equation}
where $\phi_{S_2\,\mathrm{half}}<\phi_{S_2\,\mathrm{max}}$ is the packing at which $S_2$ takes half the value at its peak, $S_2(\phi_{S_2\,\mathrm{half}})=S_{2\,\mathrm{max}}/2$.  

These parameters are all indicated in Fig.~\ref{S2-a001}(a) where we plot $S_2$ vs $\phi$ for our smallest asphericity $\alpha=0.001$, at the three smallest strain rates $\dot\gamma$.  We see that our smallest $\dot\gamma=10^{-7}$ has reached the desired  quasistatic limit for all $\phi$ up to, and including, the peak.  However above the peak, in particular at $\phi_J^{(0)}$, there remains a noticeable dependence on $\dot\gamma$.  To obtain the quasistatic limit in this case, in Fig.~\ref{S2-a001}(b) we plot $S_2(\phi_J^{(0)})$ vs $\dot\gamma$ for our smallest $\alpha\le 0.12$ (for larger $\alpha$, our smallest $\dot\gamma$ has reached the quasistatic limit at $\phi_J^{(0)}$).  We fit the small $\dot\gamma$ data points to the empirical form $a+b\dot\gamma^c$, shown as the dashed lines, to estimate the quasistatic  $\dot\gamma\to 0$ limit.  For $\alpha=0.001$, this quasistatic value is shown as the black dot in Fig.~\ref{S2-a001}(a).  

Note,  to improve our estimate for $\alpha=0.001$ we have included  in Fig.~\ref{S2-a001}(b) results from a simulation at $\phi_J^{(0)}$ with $\dot\gamma=4\times 10^{-8}$.
Due to the empirical nature of our fits in Fig.~\ref{S2-a001}(b), and the limited range of small $\dot\gamma$ for which we have data, one may question the precision of our estimate for the quasistatic limit of $S_2(\phi_J^{(0)})$.  However, we believe our results are sufficiently accurate to assert that, for $\alpha=0.001$, $S_2$ remains finite at $\phi_J^{(0)}$ and above.

\begin{figure}
\centering
\includegraphics[width=3.5in]{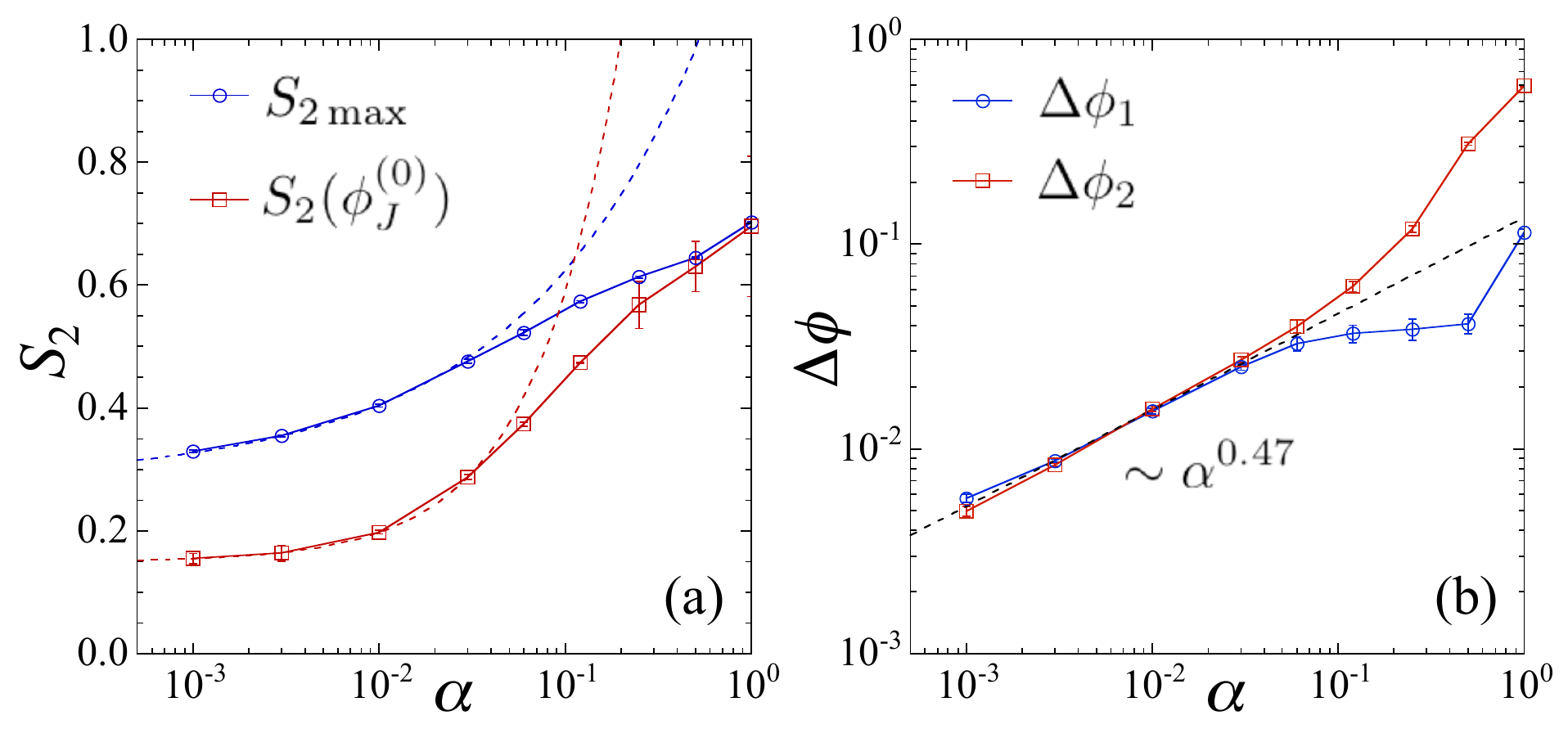}
\caption{(a) Value of the nematic order parameter at its peak, $S_{2\,\mathrm{max}}$, and at the jamming point for circles, $S_2(\phi_J^{(0)})$, vs particle asphericity $\alpha$ in the quasistatic $\dot\gamma\to 0$ limit.  Dashed lines represent fits of the four smallest $\alpha$ data points to the empirical form $a+b\alpha^c$. (b) Distance of the peak from the jamming point for circles, $\Delta\phi_1=\phi_J^{(0)}-\phi_{S_2\,\mathrm{max}}$, and low side half-width of the peak, $\Delta\phi_2=\phi_{S_2\,\mathrm{max}}-\phi_{S_2\,\mathrm{half}}$, vs $\alpha$.  The small $\alpha$ data indicate a vanishing $\Delta\phi_{1,2}\sim\alpha^{0.47}$.
}
\label{S2-heights} 
\end{figure}

In Fig.~\ref{S2-heights}(a) we plot  $S_{2\,\mathrm{max}}$ and the extrapolated quasistatic values of $S_2(\phi_J^{(0)})$ vs $\alpha$.  We see that both appear to stay finite as $\alpha\to 0$.  Fitting the four smallest $\alpha$ data points to the empirical form $a+b\alpha^c$, shown as the dashed lines, we estimate $\lim_{\alpha\to 0} S_{2\,\mathrm{max}}=0.28$ and $\lim_{\alpha\to 0}S_2(\phi_J^{(0)})=0.15$.  In Fig.~\ref{S2-heights}(b) we plot $\Delta\phi_1$ and $\Delta\phi_2$ vs $\alpha$.  From the straight line formed by the smallest data points on this log-log plot, we conclude that both $\Delta\phi_1$ and $\Delta\phi_2$ are vanishing algebraically as $\alpha\to 0$.  Fitting to this algebraic decay we find $\Delta\phi_{1,2}\sim\alpha^{0.47}$.  From Fig.~\ref{S2-heights}(b) we thus conclude that, as $\alpha\to 0$, the location of the peak in $S_2$ moves to $\phi_J^{(0)}$ and the width of the small $\phi$ side of this peak shrinks to zero, so $S_2\to 0$ for $\phi<\phi_J^{(0)}$.  However, from Fig.~\ref{S2-heights}(a) we conclude that $S_2$ stays finite at and above $\phi_J^{(0)}$, though there is a  discontinuous drop in $S_2$ as $\phi$ increases above $\phi_J^{(0)}$.

\begin{figure}
\centering
\includegraphics[width=3.5in]{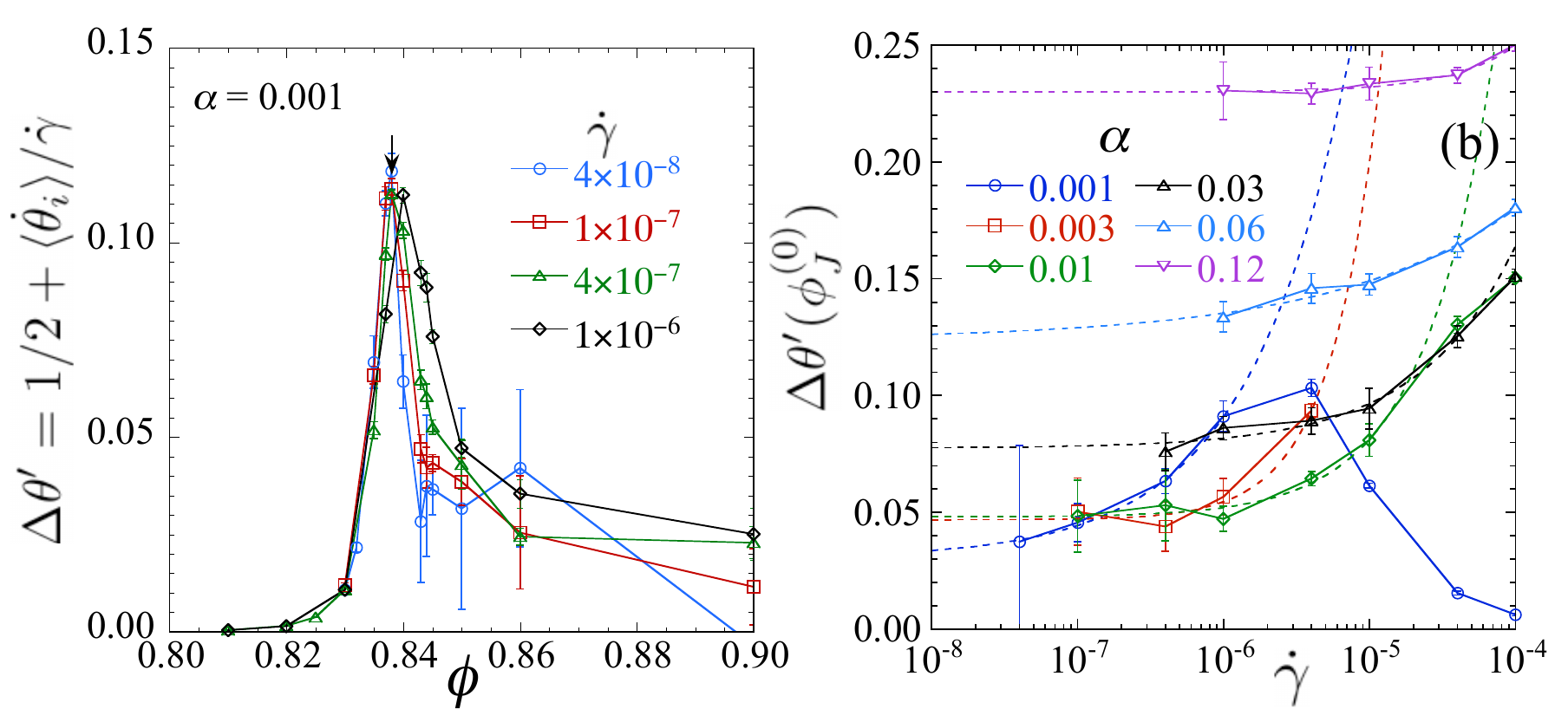}
\caption{(a) For spherocylinders with $\alpha=0.001$, $\Delta\theta^\prime=1/2-\langle\dot\theta_i\rangle/\dot\gamma$ vs $\phi$ at strain rates $\dot\gamma=10^{-6}, 4\times 10^{-7}$ and $10^{-7}$.  The black dot labeled ``QS" represents the extrapolated $\dot\gamma\to 0$ value of $\Delta\theta^\prime$ at $\phi_J^{(0)}=0.8433$. Widths 
$\Delta\phi_1^\prime\equiv \phi_J^{(0)}-\phi_{\Delta\theta^\prime\,\mathrm{max}}$, and $\Delta\phi_2^\prime=\phi_{\Delta\theta^\prime\,\mathrm{max}}-\phi_{\Delta\theta^\prime\,\mathrm{half}}$ are denoted in the figure.
(b) Plot of $\Delta\theta^\prime$ vs $\dot\gamma$ at $\phi_J^{(0)}$ for $\alpha\le 0.12$.  Solid lines connect the data points, dashed lines are fits of the small $\dot\gamma$ points to the form $a+b\dot\gamma^c$ and are used to extrapolate to the $\dot\gamma\to 0$ limit.
}
\label{av-a001} 
\end{figure}

We next consider the average particle angular velocity.  
Since at $\alpha =0$ we expect  particles to have $-\langle\dot\theta_i\rangle/\dot\gamma=1/2$ at all $\phi$, we consider here the deviation from that value.  With $\theta_i^\prime\equiv d\theta_i/d\gamma=\dot\theta_i/\dot\gamma$, we define
\begin{equation}
\Delta\theta^\prime=1/2+ \langle\dot\theta_i/\dot\gamma\rangle.
\end{equation}
In Fig.~\ref{av-a001}(a) we plot $\Delta\theta^\prime$ vs $\phi$ for $\alpha=0.001$, showing results for our three smallest strain rates $\dot\gamma$.  Similar to our analysis of $S_2$ we denote  the height of the peak value in $\Delta\theta^\prime$ by $\phi$ varies as $\Delta\theta_\mathrm{max}^\prime$, occurring at $\phi_{\Delta\theta^\prime\,\mathrm{max}}$, the value at the $\alpha=0$ jamming point by $\Delta\theta^\prime(\phi_J^{(0)})$, and the value $\phi_{\Delta\theta^\prime\,\mathrm{half}}$ as the packing where $\Delta\theta^\prime$ takes half the value at its peak, $\Delta\theta^\prime(\phi_{\Delta\theta^\prime\,\mathrm{half}})=\Delta\theta^\prime_\mathrm{max}/2$.  We similarly define the location of the peak in $\Delta\theta^\prime$ with respect to the jamming transition of circles as
\begin{equation}
\Delta\phi_1^\prime=\phi_J^{(0)}-\phi_{\Delta\theta^\prime\,\mathrm{max}},
\end{equation}
and the half width of the peak as
\begin{equation}
\Delta\phi_2^\prime=\phi_{\Delta\theta^\prime\,\mathrm{max}} - \phi_{\Delta\theta^\prime\,\mathrm{half}}.
\end{equation}
These are indicated in Fig.~\ref{av-a001}(a).  

As seen with $S_2$, we see  in Fig.~\ref{av-a001}(a) that our smallest $\dot\gamma=10^{-7}$ has reached the quasistatic limit for all $\phi$ up to, and including the peak.  However at $\phi_J^{(0)}$ we see that there remains a noticeable dependence on $\dot\gamma$.  Proceeding as was done similarly for $S_2$, in Fig.~\ref{av-a001}(b) we plot $\Delta\theta^\prime(\phi_J^{(0)})$ vs $\dot\gamma$ for the smaller $\alpha$, and fit to the form $a+b\dot\gamma^c$ to extrapolate to the $\dot\gamma\to 0$ limit.  This extrapolated value for $\alpha=0.001$ is indicated by the black dot in Fig.~\ref{av-a001}(a).

\begin{figure}
\centering
\includegraphics[width=3.5in]{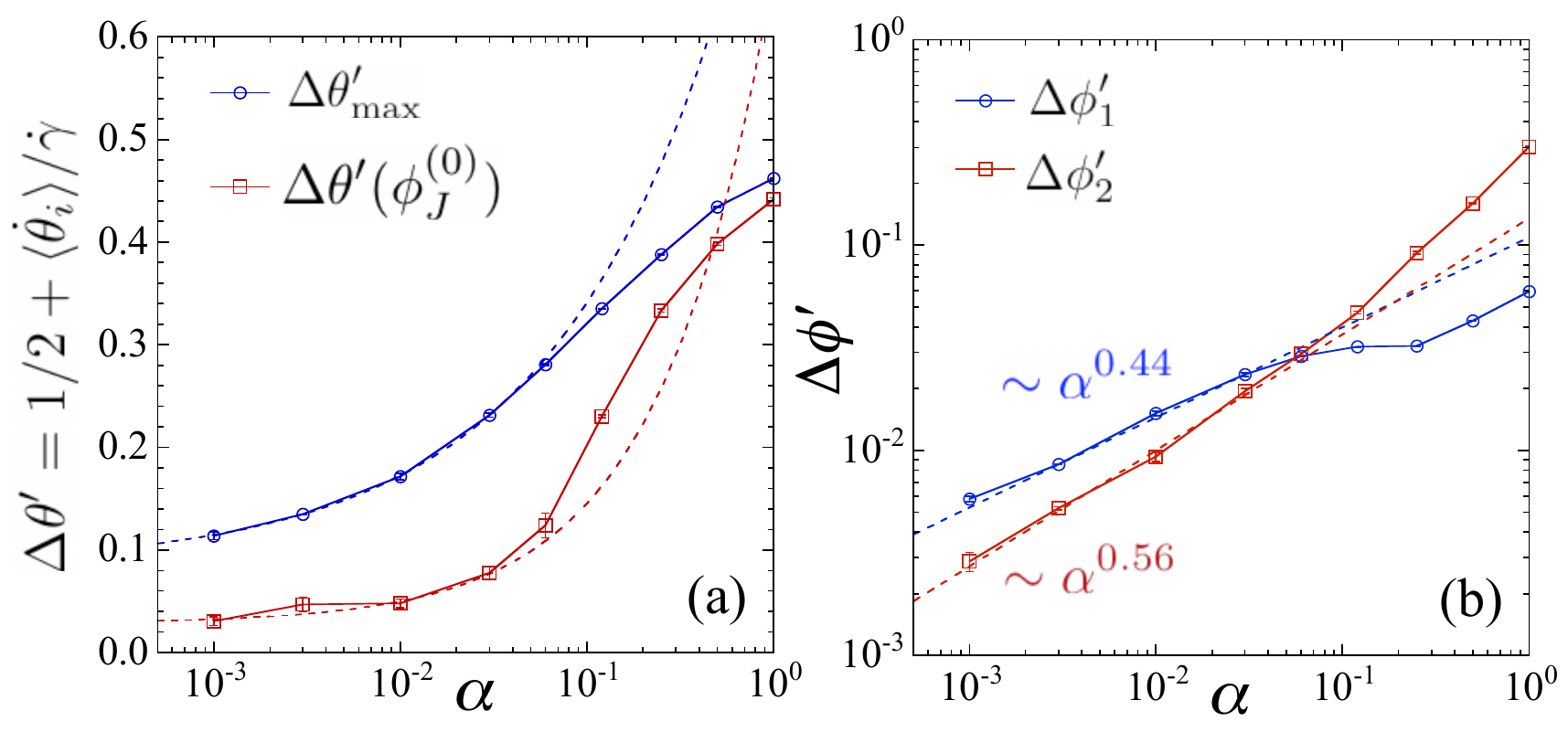}
\caption{(a) Value of $\Delta\theta^\prime=1/2-\langle\dot\theta_i\rangle/\dot\gamma$ at its peak, $\Delta\theta^\prime_\mathrm{max}$, and at the jamming point for circles, $\Delta\theta^\prime(\phi_J^{(0)})$, vs particle asphericity $\alpha$ in the quasistatic $\dot\gamma\to 0$ limit.
Dashed lines represent fits of the four smallest $\alpha$ data points to the empirical form $a+b\alpha^c$. (b) Distance of the peak from the jamming point for circles, $\Delta\phi_1^\prime=\phi_J^{(0)}-\phi_{\Delta\theta^\prime\,\mathrm{max}}$, and low side half-width of the peak, $\Delta\phi_2^\prime=\phi_{\Delta\theta^\prime\,\mathrm{max}}-\phi_{\Delta\theta^\prime\,\mathrm{half}}$, vs $\alpha$.  The small $\alpha$ data indicate an algebraic vanishing $\Delta\phi_{1}^\prime\sim\alpha^{0.44}$ and $\Delta\phi_2^\prime\sim\alpha^{0.56}$.
}
\label{av-heights} 
\end{figure}

\begin{figure}
\centering
\includegraphics[width=3.5in]{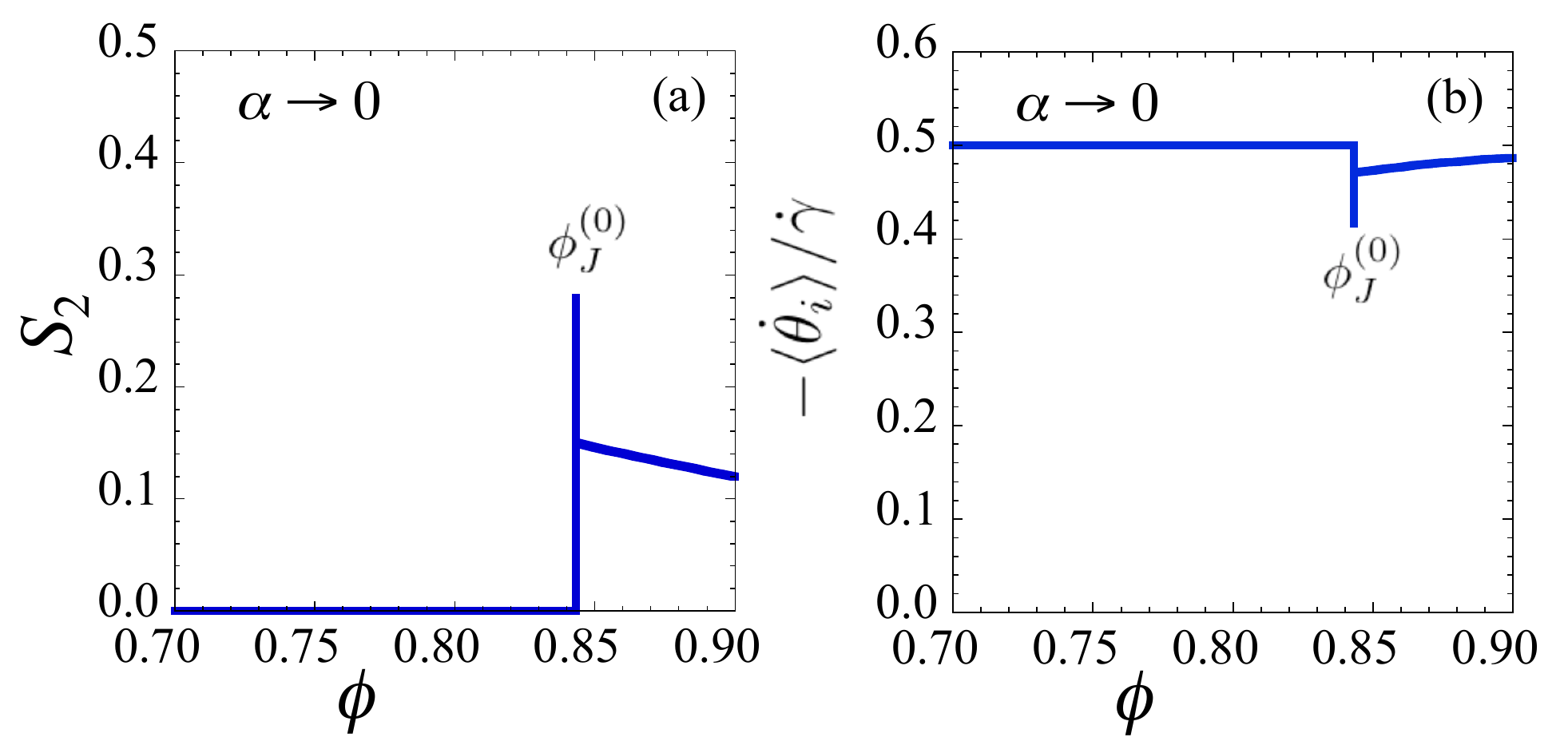}
\caption{Sketch of the quasistatic $\alpha\to 0$ limiting behavior of (a) nematic order parameter magnitude $S_2$ and (b) average  particle angular velocity $-\langle\dot\theta_i\rangle/\dot\gamma$ vs $\phi$, as determined by the results of Figs.~\ref{S2-heights} and \ref{av-heights}.
}
\label{S2-av-alpha-to-0} 
\end{figure}

In Fig.~\ref{av-heights}(a) we plot  $\Delta\theta^\prime_{\mathrm{max}}$ and the extrapolated quasistatic values of $\Delta\theta^\prime(\phi_J^{(0)})$ vs $\alpha$.  As with the corresponding quantities for $S_2$, we see that both appear to stay finite as $\alpha\to 0$.  Fitting the four smallest $\alpha$ data points to the empirical form $a+b\alpha^c$, shown as the dashed lines, we estimate $\lim_{\alpha\to 0} \Delta\theta^\prime_{\mathrm{max}}=0.084$ and $\lim_{\alpha\to 0}\Delta\theta^\prime(\phi_J^{(0)})=0.029$.  In Fig.~\ref{av-heights}(b) we plot $\Delta\phi_1^\prime$ and $\Delta\phi_2^\prime$ vs $\alpha$.  From the straight line formed by the smallest data points on this log-log plot, we conclude that both $\Delta\phi_1^\prime$ and $\Delta\phi_2^\prime$ are vanishing algebraically as $\alpha\to 0$.  Fitting to this algebraic decay we find $\Delta\phi_{1}^\prime\sim\alpha^{0.44}$ and $\Delta\phi_2^\prime\sim\alpha^{0.56}$.  Thus we find for $\Delta\theta^\prime$ qualitatively similar behavior as we found for $S_2$: From Fig.~\ref{av-heights}(b) we  conclude that, as $\alpha\to 0$, the location of the peak in $\Delta\theta^\prime$ moves to $\phi_J^{(0)}$ and the width of the small $\phi$ side of this peak shrinks to zero, so $\Delta\theta^\prime\to 0$ for $\phi<\phi_J^{(0)}$;  however, from Fig.~\ref{av-heights}(a) we conclude that $\Delta\theta^\prime$ stays finite at and above $\phi_J^{(0)}$, though there is a  discontinuous drop as $\phi$ increases above $\phi_J^{(0)}$.
In Fig.~\ref{S2-av-alpha-to-0} we sketch  the  quasistatic ($\dot\gamma\to 0$) $\alpha\to 0$ limiting behavior of the nematic order parameter magnitude $S_2$ and angular velocity $-\langle\dot\theta_i\rangle/\dot\gamma$ vs $\phi$, which follows from the  results of Figs.~\ref{S2-heights} and \ref{av-heights}.

The above analysis is for a system in which particles are bidisperse in size (specifically half big and half small in the ratio $R_b/R_s=1.4$), but monodisperse in shape (all of the same asphericity $\alpha$).  It is interesting to consider what happens to the $\alpha\to 0$ limit if one considers a system of particles that is polydisperse in shape.  We thus now consider a system in which particles are still bidisperse in size, but where the  asphericity $\alpha_i$ of particle $i$ is sampled 
from a distribution of finite width.  Here we use a gamma distribution, determined by two parameters $k$ and $\vartheta$,
\begin{equation}
P(\alpha) = \dfrac{1}{\Gamma(k)\vartheta^k}\alpha^{k-1}\mathrm{e}^{-\alpha/\vartheta}.
\end{equation}
The average and variance of $\alpha$ are  given by,
\begin{equation}
\langle \alpha\rangle=k\vartheta,\quad \mathrm{var}[\alpha]=k\vartheta^2.  
\end{equation}
The relative width of the distribution is,
\begin{equation}
\sigma_\alpha\equiv\sqrt{\mathrm{var}[\alpha]}/\langle\alpha\rangle = 1/\sqrt{k}.  
\end{equation}
We choose $k$ to get the desired relative width, and then choose $\vartheta$ to get the desired $\langle \alpha\rangle$.  
We consider two cases:  $k=100$ corresponding to $\sigma_\alpha=0.1$, and $k=1$ corresponding to $\sigma_\alpha=1$. The later case is just an ordinary exponential distribution with a finite probability density at $\alpha=0$ (circles).  

We have done simulations with $N=1024$ particles at a single slow strain rate $\dot\gamma=4\times 10^{-7}$, for a range of packings $\phi$ near the peak in $S_2$.  We choose $\vartheta$ so that the average $\langle\alpha\rangle$ is equal to the four smallest values of $\alpha=0.001$, 0.003, 0.01, and 0.03 used in the main part of this paper.  In particular, we are interested to see if the singular behavior we found as $\alpha\to 0$ persists once there is polydispersity in the particle shape.

\begin{figure}
\centering
\includegraphics[width=3.5in]{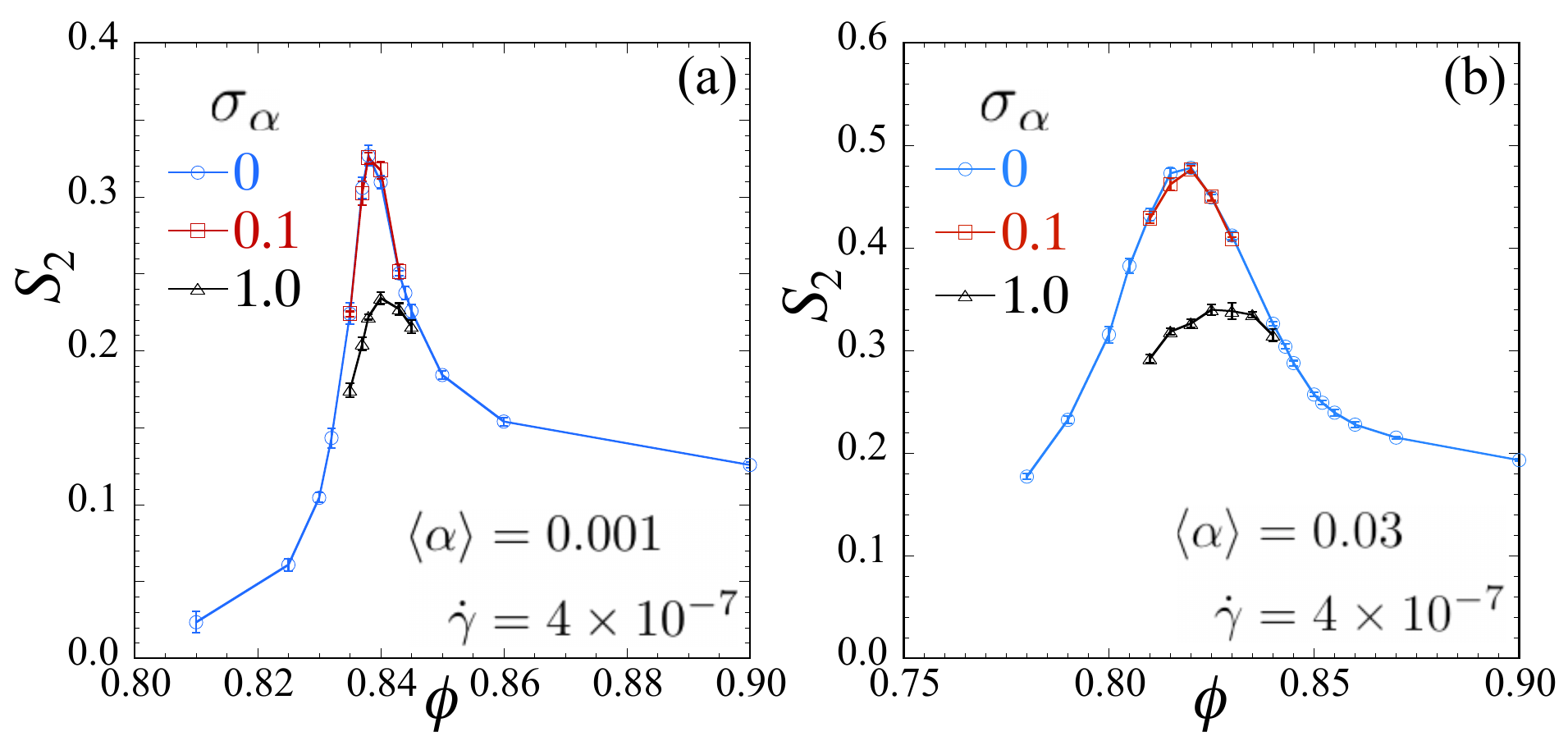}
\caption{Magnitude of nematic order parameter $S_2$ vs packing $\phi$, comparing shape polydisperse distributions of relative widths $\sigma_\alpha=0$ (monodisperse), $\sigma_\alpha=0.1$ and 1.0, at strain rate $\dot\gamma=4\times 10^{-7}$ for (a) $\langle\alpha\rangle=0.001$ and (b) $\langle\alpha\rangle=0.03$.
}
\label{S2-vs-phi-poly} 
\end{figure}

In Fig.~\ref{S2-vs-phi-poly} we show the resulting nematic order parameter $S_2$ vs $\phi$ for our smallest $\langle\alpha\rangle=0.001$ as well as $\langle\alpha\rangle=0.03$.  We compare  results from the distributions with $\sigma_\alpha=0.1$ and $\sigma_\alpha=1$ to  our original shape-monodisperse simulations, i.e., $\sigma_\alpha=0$.   We see that a relative width of $\sigma_\alpha=0.1$ (i.e., 10\% dispersity) produces no noticeable change from the monodisperse case.  For the exponential distribution with $\sigma_\alpha=1$, the peak height $S_{2\,\mathrm{max}}$ decreases, and the location of the peak $\phi_{S_2\,\mathrm{max}}$ slightly increases.

\begin{figure}
\centering
\includegraphics[width=3.5in]{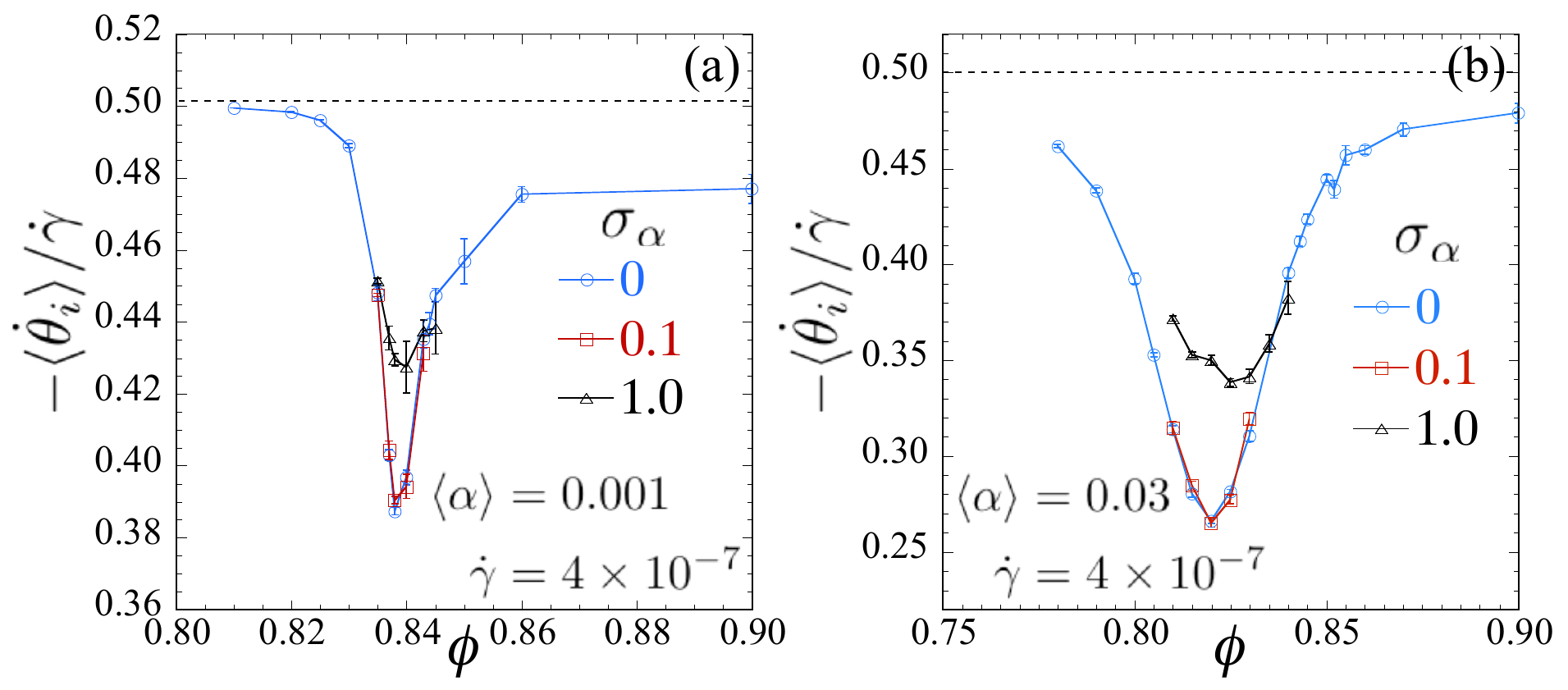}
\caption{Average angular velocity $-\langle\dot\theta_i\rangle/\dot\gamma$ vs packing $\phi$, comparing shape polydisperse distributions of relative widths $\sigma_\alpha=0$ (monodisperse), $\sigma_\alpha=0.1$ and 1.0, at strain rate $\dot\gamma=4\times 10^{-7}$ for (a) $\langle\alpha\rangle=0.001$ and (b) $\langle\alpha\rangle=0.03$.
}
\label{av-vs-phi-poly} 
\end{figure}

In Fig.~\ref{av-vs-phi-poly} we show the average angular velocity $-\langle\dot\theta_i\rangle/\dot\gamma$ vs $\phi$.  We see  results similar to those found for $S_2$.  For $\sigma_\alpha=0.1$ there is no noticeable change from the monodisperse case $\sigma_\alpha=0$.   For $\sigma_\alpha=1$ we see that the depth of the minimum decreases while the location of the minimum shifts to slightly larger $\phi$.

\begin{figure}
\centering
\includegraphics[width=3.5in]{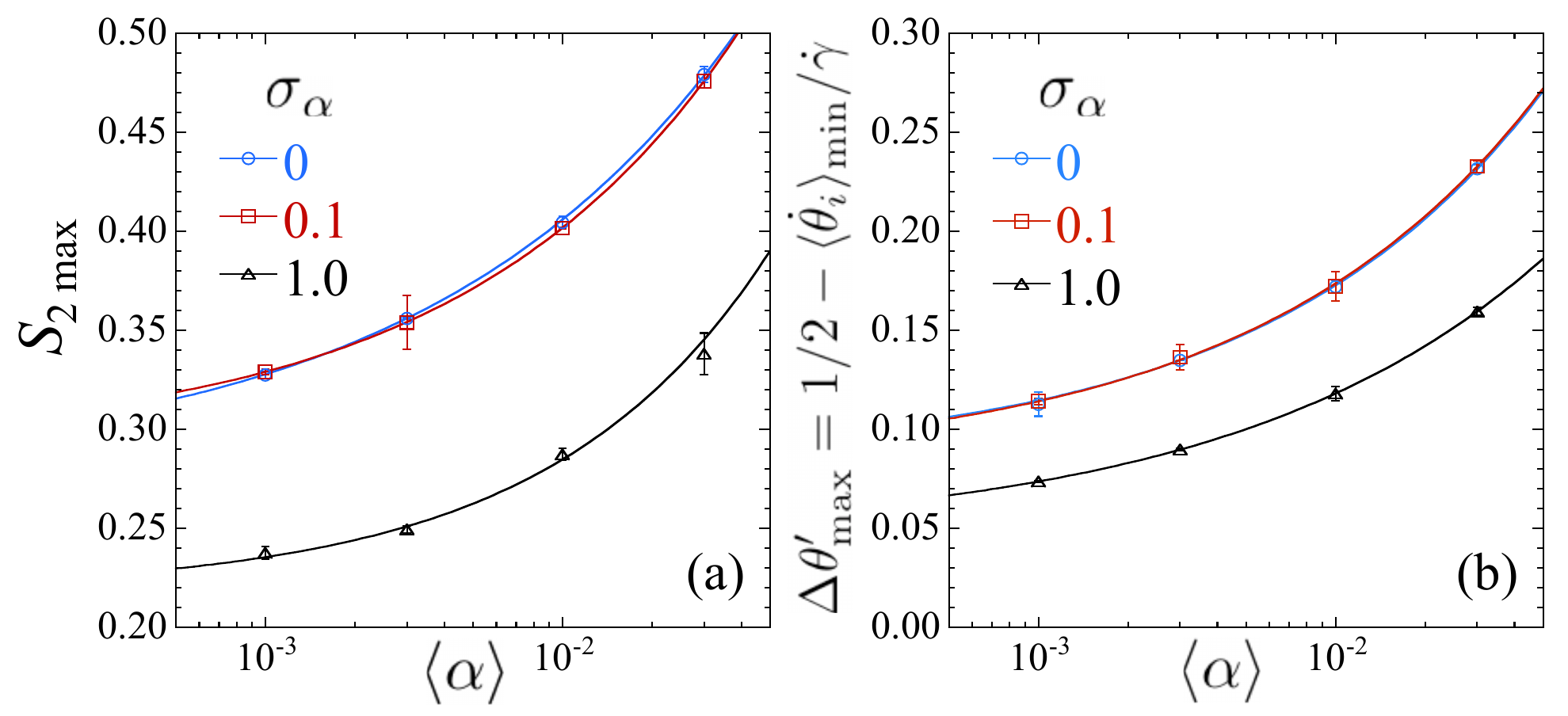}
\caption{(a) Maximum value of the magnitude of the nematic order parameter $S_{2\,\mathrm{max}}$ and (b) minimum average angular velocity $\Delta\theta^\prime_\mathrm{max}\equiv 1/2-\langle\dot\theta_i\rangle_\mathrm{min}/\dot\gamma$  vs average asphericity $\langle\alpha\rangle$, for polydisperse distributions of relative width $\sigma_\alpha=0$ (monodisperse), $\sigma_\alpha=0.1$ and 1.0, at strain rate $\dot\gamma=4\times 10^{-7}$.  Solid lines are fits to the form $a+b\alpha^c$.
}
\label{S2-av-v-alpha-poly} 
\end{figure}

In Fig.~\ref{S2-av-v-alpha-poly}(a) we plot $S_{2\,\mathrm{max}}$ vs $\langle\alpha\rangle$ for these same three values of $\sigma_\alpha=0$, 0.1, and 1.   Again we see that there is no difference between $\sigma_\alpha=0$ and $\sigma_\alpha=0.1$.  A 10\% dispersity results in no noticeable change.  For $\sigma_\alpha=1$ we see that $S_{2\,\mathrm{max}}$ is smaller than for the other two cases, but still we find that $S_{2\,\mathrm{max}}$ seems to be approaching a finite constant as $\langle\alpha\rangle\to 0$.  In contrast to $\sigma_\alpha=0$, for which we found $\lim_{\alpha\to 0}[S_{2\,\mathrm{max}}]=0.28$, fitting to the form $a+b\alpha^c$ for $\sigma_\alpha=1$ gives $\lim_{\alpha\to 0}[S_{2\,\mathrm{max}}]=0.22$.

In Fig.~\ref{S2-av-v-alpha-poly}(b) we plot $\Delta\theta^\prime_\mathrm{max}\equiv 1/2-\langle\dot\theta_i\rangle_\mathrm{min}/\dot\gamma$ vs $\langle\alpha\rangle$.  As with $S_2$ we see no difference between $\sigma_\alpha=0$ and $\sigma_\alpha=0.1$, while results for $\sigma_\alpha=1$ are somewhat smaller, but still appear to be approaching a finite constant as $\langle\alpha\rangle\to 0$.  In contrast to $\sigma_\alpha=0$, for which we found $\lim_{\alpha\to 0}[\Delta\theta^\prime_\mathrm{max}]=0.084$, fitting to the form $a+b\alpha^c$ for $\sigma_\alpha=1$ gives $\lim_{\alpha\to 0}[\Delta\theta^\prime_\mathrm{max}]=0.046$.
Thus, even with considerable dispersity in particle shape,  our conclusion that $\lim_{\alpha\to 0} [S_{2\,\mathrm{max}} ]$  and $\lim_{\alpha\to 0}[\Delta\theta^\prime_\mathrm{max}]$ remain finite appears to remain valid, and so the $\alpha\to 0$ limit continues to be singular.

\bibliographystyle{apsrev4-1}

\end{document}